\documentclass[11pt,notitlepage,a4paper]{article}
\usepackage[utf8]{inputenc}
\usepackage[round]{natbib}
\usepackage{amssymb}
\usepackage{algorithm}
\usepackage{algpseudocode}
\usepackage{amsfonts}
\usepackage{amsmath, amsthm, amssymb}
\usepackage[most]{tcolorbox}
\usepackage{soul}      
\soulregister\ref7
\usepackage{geometry}
\usepackage{tabularx}
\usepackage{longtable}
\usepackage{pdflscape} 
\usepackage{xcolor} 

\definecolor{linkcolor}{rgb}{0, 0, 0.35}
\definecolor{linkbordercolor}{rgb}{.9, .9, .9}
\definecolor{block-gray}{gray}{0.85}
\newtcolorbox{blockquote}{colback=block-gray,grow to right by=-1mm,grow to left by=-1mm,boxrule=0pt,boxsep=0pt,breakable}

\usepackage{graphicx}
\usepackage{listings} 
\usepackage{amsmath}
\usepackage{geometry}

\usepackage{colortbl}
\usepackage[doublespacing]{setspace}

\usepackage{graphicx}

\usepackage{subfig}

\usepackage{fullpage}
\usepackage{booktabs}

\usepackage{color,soul} 
\usepackage{verbatim}
\usepackage{adjustbox}

\definecolor{bg}{rgb}{0.95,0.95,0.95} 
\definecolor{codegreen}{rgb}{0,0.6,0}
\definecolor{codegray}{rgb}{0.5,0.5,0.5}
\definecolor{codepurple}{rgb}{0.58,0,0.82}
\definecolor{backcolour}{rgb}{0.95,0.95,0.95}

\lstdefinestyle{mystyle}{
    backgroundcolor=\color{bg}, 
    commentstyle=\color{codegreen},
    keywordstyle=\color{blue},
    numberstyle=\tiny\color{codegray},
    stringstyle=\color{codepurple},
    basicstyle=\ttfamily\normalsize, 
    breakatwhitespace=false, 
    breaklines=true, 
    captionpos=b, 
    keepspaces=true, 
    numbers=none, 
    numbersep=5pt, 
    showspaces=false, 
    showstringspaces=false, 
    showtabs=false, 
    tabsize=4, 
    frame=none, 
    rulecolor=\color{black}, 
    escapeinside={||}, 
    morekeywords={*,...} 
}
\usepackage[smartEllipses,hybrid]{markdown}

\usepackage{lmodern}

\usepackage{bbm}

\usepackage{thmtools, thm-restate}

\usepackage[draft,textsize=small]{todonotes}   

\usepackage[normalem]{ulem}

\newtheorem{lemma}{Lemma}

\newtheorem{definition}{Definition}

\DeclareRobustCommand\iff{\;\Longleftrightarrow\;}

\usepackage{titlesec}
\usepackage{appendix}
\usepackage{enumitem}

\setlist{noitemsep,topsep=0pt,parsep=0pt,partopsep=0pt}

\usepackage{hyperref}
\hypersetup{
    colorlinks=true,
    linkcolor=linkcolor,
    anchorcolor=linkcolor,
    citecolor=linkcolor,
    filecolor=linkcolor,
    urlcolor=linkcolor,
    citebordercolor=linkbordercolor,
    filebordercolor=linkbordercolor,
    linkbordercolor=linkbordercolor,
}

\begin{document}

\begin{titlepage}
\begin{spacing}{1}
\title{\textsc{Who is More Bayesian: Humans or ChatGPT?}\thanks{We dedicate this study to the memory
of David Grether and Daniel Kahneman for their important contributions to economics and psychology, and
 particularly their pathbreaking work in experimental and behavioral economics and the
study of individual decision making that inspired this paper. 
We thank Mahmoud El-Gamal and David Grether as well as Charles Holt and Angela Smith for providing
the human experimental data that were reanalyzed in this paper. Rust is  grateful for financial support from his University Professorship at Georgetown University.}}
\author{
\textsc{Tianshi Mu, Tsinghua University}\\
\textsc{Pranjal Rawat, Georgetown University}\\
\textsc{John Rust, Georgetown University} \thanks{{\bf Corresponding Author:} Department of Economics, Georgetown University, {\tt jr1393@georgetown.edu}.} \\
\textsc{Chengjun Zhang, Georgetown University} \\
\textsc{Qixuan Zhong, Georgetown University} 
}
\date{\textsc{February 2025}}

\maketitle
\thispagestyle{empty}

\begin{abstract}
\noindent
We compare the performance of human and artificially intelligent (AI) decision makers in simple binary classification
tasks where the optimal decision rule is given by Bayes Rule. We reanalyze choices of human subjects gathered from 
laboratory experiments conducted by El-Gamal and Grether and Holt and Smith. We confirm that while overall,
Bayes Rule represents the single best model for predicting human choices, subjects are heterogeneous and
a significant share of them make suboptimal choices that reflect judgement biases described by Kahneman and  Tversky 
that include the ``representativeness heuristic'' (excessive weight on the evidence from the sample
relative to the prior) and ``conservatism'' (excessive weight on the prior relative to the sample). We compare the 
performance of AI subjects gathered from recent versions of large language models (LLMs)  including 
several versions of ChatGPT. These general-purpose generative AI chatbots are not specifically trained to do well in 
narrow decision making tasks, but are trained instead as ``language predictors'' using a large corpus of textual data from the web.
We show that ChatGPT is also subject to biases that result in suboptimal decisions. However we
document a rapid evolution in the performance of ChatGPT from sub-human
performance for early versions (ChatGPT 3.5) to superhuman and nearly perfect Bayesian classifications in the latest
versions (ChatGPT 4o).

\bigskip


\end{abstract}

\end{spacing}
\end{titlepage}


\large

\section{Introduction}
\label{section:intro}
We compare the performance of human and artificially intelligent (AI) decision makers in simple binary classification
tasks where the optimal decision rule is given by Bayes Rule. AI algorithms such as support vector machines
or neural networks can be trained to closely approximate optimal Bayesian decision rules 
given the relative simplicity of this narrow domain classification task. Machine learning methods
have been extended to more difficult real world classification problems where the covariates
used to classify outcomes can be very high dimensional (e.g. using mammograms to detect breast
cancer). A number of studies have shown these specially 
trained classifiers can perform at superhuman levels, see e.g. \citet{JBC2023}.

It is not surprising that humans, whose brains consume only about 20 watts of power,  do not
outperform special-purpose machine learning algorithms that are trained with large volumes of data to approximate
optimal decision rules for specific tasks. \citet{KuhlEtAl2022} show that humans possess {\it general intelligence\/}
whereas most  machine learning  algorithms are designed to work in narrow domains and will not necessarily make 
sensible decisions in a huge variety of different (and often unexpected) situations as humans do.
As \citet{HM1994} noted, \textit{``From a broader perspective, however, one can argue that optimal solutions are known for a relatively small number of similar, well-specified problems whereas humans evolved to survive in a world filled with a large and diverse set of ill-specified problems. 
Our `suboptimality' may be a small price to pay for the flexibility and adaptiveness of our intuitive decision processes.''}

Rapid recent improvements in large language models (LLMs) and
generative AI suggest that we may be close to the advent of {\it Artificial general intelligence\/}
(AGI) where general purpose algorithms equal or exceed human performance in solving a wide range
of problems even though the algorithms were not specifically trained to do well in specific narrow domain tasks.   
Generative AI models such as ChatGPT are deep neural networks with billions of parameters that have been trained
to predict text, sounds and images using vast databases obtained from the web and other sources.
The progress in this area has been breathtaking, and now a variety of LLMs have demonstrated
a capability to compete with humans on a  wide range of intellectual tasks.\footnote{\footnotesize In the paper we will use interchangeably the abbreviations LLM and GPT (for Generative Pretrained Transformer), though the latter is a subset of the former. See section~\ref{section:conclusion} for
further discussion and comments on the rapidly growing
literature that compares human and AI performance over a wide range of domains and problems.}

Despite the amazingly rapid improvements, the consensus is that LLMs still lack full rationality, including the 
capability to reason and think creatively the way humans do, and other features associated with intelligence
including ``consciousness''. The review by \citet{AIIndex2024} concludes that \textit{``AI has surpassed human performance on several
benchmarks, including some in image classification, visual reasoning, and English understanding. Yet it trails
behind on more complex tasks like competition-level mathematics, visual commonsense reasoning and planning.''} 


Our study focuses on  statistical decision making,  building on a large experimental literature designed to assess the rationality
of human subjects in a simple binary
classification problem used in dozens of previous experimental
studies in economics and psychology. The decision problem is simple enough to have an optimal solution
that is easily characterized using statistical decision theory and Bayes Rule. Numerous studies in psychology and behavioral economics conclude that humans make suboptimal
decisions due to systematic biases, including ``framing''  and contextual effects
that might be caused by reliance on heuristics to reduce cognitive burden, see e.g.  \citet{KT1974}.

However this conclusion is controversial due to the use of ``real world'' situations 
to test decision making (e.g. providing a description of a person and asking them to choose whether
they are more likely to be an engineer or a lawyer), since it provides extraneous information that
 amplifies the potential for framing effects  and stereotyping to  mislead subjects. 
\citet{Grether1978} noted that Kahneman and Tversky's experiments had ``features that make the applicability of the 
findings to economic decisions doubtful'' and their framing of scenarios created  ``difficulty of controlling the information 
given when verbal descriptions or situations are presented. Both of these difficulties could be taken care of 
by the use of actual balls in urns or book-bag poker­chip setups.'' (p. 71-72).
\citet{CT1996} argued that experiments framed in frequentist terms are more likely
to generate behavior that conforms to Bayes Rule since ``our inductive reasoning mechanisms do embody 
aspects of a calculus of probability, but they are designed to take frequency information as input and produce frequencies as output.'' 
(p. 3). 

We find this logic compelling, and build on the experimental
design of \citet{EGG1995} where subjects were shown the outcomes of samples of balls drawn with replacement from one
of two bingo cages, A or B, where each contains different proportions of red and blue balls. The cage used to draw the sample
was selected by a random procedure, such as choosing cage A if the throw of 6-sided die is 1 or 2. This induces a credible
objective prior probability of selecting cage A. The experimental controls involve different choices of these prior probabilities,
the proportions of red and blue balls in the two cages, and number of balls drawn with replacement. 
Subjects were informed of these parameters and the outcome of the sample drawn from the selected cage. Based on this information
subjects chose the cage they believed was more likely to have been selected.
The experiments by \citet{HS2009} used a similar design but asked subjects to directly report subjective
posterior probabilities of cage A.

The ``textbook'' solution to this problem is to use Bayes Rule to compute the posterior probability
of A and select it if and only if its posterior probability exceeds 1/2. However the experiments did not presume that
subjects actually know Bayes Rule and did not train them or tip them off about how to make ``correct'' responses.
The objective of these studies was to assess the extent to which untrained subjects behave as ``intuitive Bayesians'' and make instinctive 
choices that are consistent with Bayes Rule. This should be the case if subjects are {\it intelligent and greedy\/} since subjects received bonus payments for choosing the correct bingo cage used to draw the sample.
We hypothesize that their behavior should be governed by an {\it optimal decision rule,\/} i.e. one that maximizes the probability
of selecting the correct bingo cage. It is straightforward to show that choosing according to Bayes Rule is the optimal
decision rule for this problem. Thus, a comparison of whether humans or GPT is more ``Bayesian'' is equivalent
to a test of their ability to conceptualize probabilities and optimize.\footnote{\footnotesize In the experiments where subjects 
reported their subjective posterior probability for cage A a more complicated incentive scheme
known as a ``BDM mechanism'' was used that requires subjects to solve an even more difficult continuous optimization problem
that we describe in section 4 of the paper.}

\begin{figure}[ht]
\caption{Which group of subjects, 1 or 2, are GPT and which are human?}
\label{fig:humans_vs_robots}
\smallskip
\centerline{\includegraphics[height=3.0in]{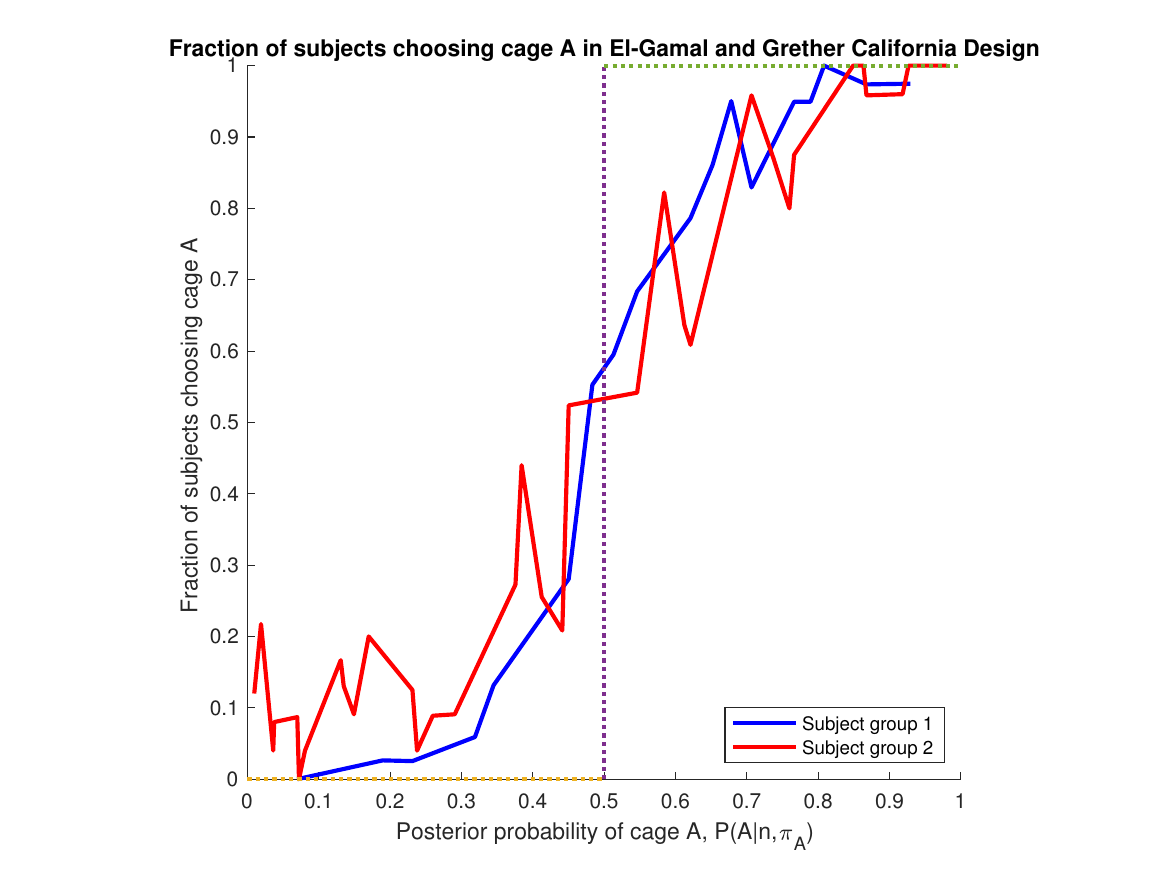}}
\end{figure}

To preview our key findings, and convince the reader that AI subjects are not automatically
superior to human decision makers, consider figure~\ref{fig:humans_vs_robots}. The graph  plots the fraction of subjects 
who chose cage A (y axis) against the true Bayesian posterior probability for cage A (x axis) using the experimental
design of \citet{EGG1995} and \citet{EGG1999} where subjects were shown the outcome of 6 balls drawn with
replacement from one of the randomly selected cages. The thin dotted blue line is the
optimal decision rule of a perfect Bayesian decision maker: choose cage B if the posterior probability of cage A is less than 1/2
and choose cage A otherwise. The solid red and blue lines represent two different groups of subjects: one is a group of 
human subjects, and another is a group of ChatGPT 4 subjects. As a teaser, we won't reveal which group is which at this
point and let the reader try to guess, emphasizing the ``Turing test'' aspect of our analysis and the possibility
that at least the  early generations of GPTs may in fact suffer from some of the same foibles as human decision makers.
Further it is not immediately obvious which of the subject groups exhibits ``better'' performance. 

The simple nature of this binary classification problem allows us to obtain an objective measure of 
``performance'' to compare human and GPT subjects. In section 2 we introduce the relevant statistical
decision theory and define the {\it win function,\/} i.e. the conditional probability of selecting the correct cage 
implied by any given decision rule. The win function is the complement of the {\it loss function\/} 
in statistical decision theory, so any decision rule that maximizes the win function necessarily minimizes the
loss function. Using the win function, we define a simple measure of {\it efficiency\/} of a human or GPT
decision maker: the ratio of the expected win probability of the decision maker to the optimal win probability
implied by Bayes Rule. This is a superior 
measure of performance than the commonly used notion of {\it accuracy,\/} i.e. the fraction of a subject's
 choices that coincide with Bayesian choices. Our efficiency measure 
differentiates between ``hard cases'' (where the Bayesian posterior probability is
close to 1/2) and ``easy cases'' (those where the posterior probability is close to 0 or 1). Among two
subjects facing the same set of trials and whose choices agree with Bayes Rule on the same total number of 
these trials (so the subjects have equal accuracy), the subject who has a greater fraction
of choices that disagree with Bayes Rule on the easy cases 
will have a lower expected payoff and thus lower efficiency.

Our re-analysis of the human experiments confirms the conclusions of \citet{EGG1995} and \citet{HS2009} that humans, on average, closely resemble Bayesian decision makers. The decision efficiency reached above 96\% in both experiments. We detect considerable unobserved heterogeneity in human subjects, with the most efficient type nearly matching Bayesian decision makers.

Next, we document a rapid improvement in the decision efficiency of GPTs over a few generations. Initially, GPT-3.5's decision efficiency is 12.1\% lower than that of humans in the binary choice Wisconsin experiment and 21.9\% lower in the more challenging Holt-Smith experiment. However, with a few updates, GPT-4's efficiency quickly reaches a level comparable to humans, and GPT-4o surpasses human efficiency by 1\% and 3.5\% in the respective experiments. We also conducted a limited analysis of the most recently released version of ChatGPT o1, and found that its efficiency is close to 100\%, with behavior that  closely approximates a perfect Bayesian decision maker.\footnote{\footnotesize So as not to keep the reader in suspense, in figure~\ref{fig:humans_vs_robots} subject group 1 (blue line) are human subjects and subject group 2 (red line) are chatGPT 4 subjects.}

A key advantage of AI subjects over humans is that AI subjects provide full reasoning behind
their answers, broken down into logical steps. As a result we gain a unique window into their ``thought process'' and can better isolate where they make their errors. Analyzing the textual responses, we find that GPT-3.5 often ignores prior information and therefore lacks a Bayesian conceptual framework. GPT-4 generally understands that Bayes Rule is the relevant principle but makes algebraic errors in the process of computing the Bayesian posterior probability. However, GPT-4o shows remarkable improvement, transitioning from a ``conceptual" to an ``accurate" Bayesian approach.

Why should we care how ``Bayesian'' humans or machines are?
Though the decision problem we study is a simple and somewhat abstract ``textbook problem'' we believe it is a reasonable approximation to
and metaphor for a range of real world classification problems involving much higher stakes. One such example
is the problem of making optimal  {\it differential diagnoses\/} (DDx) that involve 
classifying which of several alternative diseases or medical problems most likely caused by a set of observed symptoms in a patient.
Recent studies (e.g. \citet{JAMAOPEN2024} and \citet{GoogleDDX2023})
 have demonstrated that LLMs and GPTs can outperform human physicians in the quality and accuracy
of their differential diagnoses. However unlike our problem, there is no objective measure of what the
``correct'' diagnosis is for most cases. These studies rely on the diagnoses of panels of expert physicians to
score the choices made by GPT and less experienced interns, and use {\it ad hoc\/} scoring rules to evaluate
the reasoning that lead subjects to choose a particular diagnosis, but without a clear notion of the cost or harm from a misdiagnosis. In comparison, we have an objective way of evaluating loss and classifying ``easy'' and ``hard'' problems to identify where subjects make their most costly mistakes.

Our paper contributes to two strands of literature. First, we build on and extend the literature of testing whether human behaviors are consistent with Bayes' rule (\citealp{EGG1995}; \citealp{EGG1999}; \citealp{HS2009}). We formally show that applying Bayes' Rule is optimal in these experiments and develop a novel performance measure superior to widely used accuracy. To infer subjects' posterior subjective beliefs, we propose a new structural logit model, which we show provides a better fit of human behaviors than previously ones in the literature.

Second, we contribute to the growing literature understanding LLM behavior using experimental methods and comparing it with human behavior in a variety of contexts,\footnote{The literature finds that LLMs excel at textual tasks, for example, divergent thinking tasks \citep{Hubert2024}, ophthalmology exams (\citep{Yan2024}; \citep{Taloni2023}), essay writing \citep{Herbold2023}, and uniform bar exams \citep{Martínez2024}. The literature also reports that LLMs can fall short of humans, at least in certain aspects, in some non-textual tasks, including complex mathematical proofs and computation \citep{Frieder2023}, analyzing healthcare data \citep{Li2024}, physics coding \citep{Yeadon2024} and medical board exams \citep{Katz2024}. For additional studies in marketing and finance, see \citealp{goli2024frontiers}, and \citet{zhao2024revolutionizing} and the references there.} among them \citet{GPTstudy2023} and \citet{mei2024turing} are closest to us. \citet{mei2024turing} evaluates the behavioral and personality traits of AI, including trust, fairness, risk-aversion, altruism, and cooperation and find that \textit{``Their behaviors fall within the distribution of behaviors of humans and exhibit patterns consistent with learning''}. \citet{GPTstudy2023} compares chatGPT and a corresponding sample of human subjects in their ability to make rational choices in four different domains involving risk, intertemporal choices, food choices and social choices and concludes that \textit{``GPT’s decisions are largely rational in each domain and demonstrate higher rationality score than those of human subjects in a parallel experiment and in the literature.''} We instead evaluate ChatGPT's Bayesian rationality—a cornerstone of decision theory with broad applications in domains such as disease diagnosis, pricing strategies, financial investment, and beyond. We document a rapid improvement in decision efficiency, moving from varied noisy decision-makers to more uniform, near-Bayesian ones in a few generations. Our estimation of the structural model and novel analysis of textual responses offer deeper insights into how this increase in decision efficiency occurs.

The remainder of this paper is organized as follows. Section 2 reviews the relevant background on statistical decision theory that provides a simple inefficiency index that we use to summarize the degree of irrationality/suboptimality of both human and AI subjects relative to the optimal decisions given by Bayes Rules. Section 3 introduces a structural econometric model of subject responses that we use to infer subject beliefs and summarize their behavior in these experiments. In section 4 we summarize the findings of our reanalysis of human subject data reported in the studies by \citet{EGG1995}, \citet{EGG1999} and \citet{HS2009}. In section 5, we use the
same experimental designs but replace the human subjects with AI subjects from various versions of currently commercially available GPT software, including ChatGPT. In Section 6, we examine the textual responses of GPT subjects to identify where mistakes occur in their reasoning process. Section 7 provides some overall conclusions and discuss the larger significance of our results including revised forecasts of the imminent demise of {\it homo sapiens.\/}



\section{Statistical Decision Theory Background}
\label{section:statistical_decision_theory}
In this section we review some basic concepts of statistical decision theory
that provide a well defined metric for characterizing the degree of irrationality/inefficiency 
in the responses of both human and AI subjects. The human subject data we reanalyze in section~\ref{section:human_subjects} 
were gathered from four separate experiments: 1) 257 student subjects from
four different universities in California reported in \citet{EGG1995}, 2) 
79 student subjects at the University of Wisconsin reported in \citet{EGG1999},
3) 22 subjects at the University of Virginia, and 4) 24 subjects who participated  in web-based
experiments, both reported in \citet{HS2009}. 

All of the experiments analyzed in this paper can be classified as ``binomial designs''
where subjects are presented with two bingo cages, labeled A and B, each containing the same  known number of balls of 
two types (i.e. balls labeled  N or G in experiments 1 and 2,  light and dark marbles in experiment 3, or
red and blue balls in experiment 4). A credible random mechanism was used to select one of the two cages
(e.g. selecting one of the cages based on a toss of a die or a random number generator) 
though subjects were not shown which cage was selected. A random sample of $D$ balls with replacement was drawn 
from the selected cage and shown to the subjects. 

In experiments 1 and 2 subjects were asked to choose the
cage they believed was most likely to have been used to draw the sample. In experiments 3 and 4 subjects were
asked to report the probability that cage A was the one from which the sample was drawn using an incentive-compatible procedure 
introduced by \citet{BDM1964} known 
as the {\it BDM mechanism\/} which involves a second stage lottery whose payoff depends on the
probability the subject reports. This lottery is designed so that the payoff maximizing report 
is the subjective posterior probability of cage A. We will describe the BDM mechanism in more detail
in section~\ref{section:holtsmith} below.
In experiments 1 a subset of subjects were paid a \$10 bonus if they selected the actual cage used
to draw the random sample of balls for a randomly selected trial out of the total trials they participated in.
In experiment 2 all subjects received a \$20 bonus for each correct response in 3 randomly selected trials.

It is important to note that in all experiments incentive payments were made {\it after\/} all trials were completed. 
Beyond an initial description of the bingo cage setup
and a single demonstration of how it works at the start of the experiment, {\it none of the subjects received any feedback on whether they
had selected the correct cage after each trial in the experiment.\/} This was evidently an intentional feature of the experimental
design, to reduce the possibilities of non-stationarity in subjects' decision rules during the experiment due to
``learning-by-doing'' that is enhanced by real-time feedback.\footnote{\footnotesize We tested for the presence of learning by doing
effects simply due to repeated participation even without feedback by comparing performance on the first third of trials with the
last third. We find small learning by doing effects even for the non-incentivized subjects, even in the absence of sequential
feedback about whether they had selected the correct cage after each trial. However the effect is sufficiently small that we
ignore it in the subsequent analysis in this paper.}

The problem of selecting the cage from which
the observed sample was most likely to have been drawn is an elementary {\it statistical decision problem\/} whose
optimal solution is given by Bayes Rule. In the next section we summarize the relevant
statistical decision theory that provides the conceptual framework for our comparison of human and AI behavior.  
Then we introduce a {\it structural logit model\/}
of subject decision making that we use to analyze  human and AI responses and 
compare their degree of suboptimality using  statistical decision theory.

\subsection{Bayes Rule, Decision Rules, Win and Loss Functions}
\label{section:lossfunctions}

First we introduce the notation to more precisely describe these {\it statistical experiments.\/}
Let $d$ denote the number of balls in the sample of $D$ balls that have a designated type (i.e.  balls marked N 
in experiments 1 and 2, light balls in experiment 3 or red balls in experiment 4). Though $d$ is a sufficient
statistic for the full random sample, subjects were shown the full sample outcomes.\footnote{\footnotesize 
In experiments 1 to 3 subjects were shown the results of each draw sequentially,
but in experiment 4 the results of the draws were presented simultaneously ``to mitigate the tendency of subjects
to overweight recent observations'' (p. 129).} Let $p_A$ and $p_B$ be the probabilities of selecting the designated
type of ball from each cage. The probabilities equal the fractions of the total number of balls in each cage of the designated
type.  Let $f(d|p_A,D)$ and $f(d|p_B,D)$ be the probabilities of observing $d$ balls
of the designated type in the random sample of $D$ balls for cages A and B.  These  are
binomial distributions with parameters $(p_A,D)$ and $(p_B,D)$, respectively. Finally, let $\pi \in (0,1)$ denote
the credible {\it objective prior probability\/} that cage A was selected to draw the random sample of $D$ balls.

The behavior of subjects in the experiments
can be summarized by a {\it decision rule\/} which is a function $\delta(d,\pi,p_A,p_B,D)$  mapping
the information provided to subjects in the experiments into a choice of cage A or B. Following 
\citet{EGG1995} we do not assume all subjects use the same decision rule, and our analysis will attempt
to identify different {\it types\/} of subjects who use similar decision rules, using statistical
methods for discovering unobserved heterogeneity including the {\it Estimation-Classification\/} (EC) algorithm
introduced by \citet{EGG1995} as well as a finite mixture model approach introduced by \citet{HS1984}.

Our analysis of human and AI subject data also allows for probabilistic decision rules (i.e. ``mixed strategies'')
as well as pure strategies that appear probabilistic to the experimenter
because the subject's choice depends on additional information or stochastic
psychological ``decision errors'' $\nu$ that are observed only by the subject and not by 
the experimenter. This results in a decision rule of the form $\delta(d,\pi,p_A,p_B,D,\nu)$. For this reason
we define the decision rule for a subject as a conditional probability of selecting cage A that depends
on the public information $(d,\pi,p_A,p_B,D)$.

\begin{definition}{\bf Decision Rule:}
Any conditional probability $P(A|d,\pi,p_A,p_B,D)$ of selecting cage A
as a function of the publicly observable information $(d,\pi,p_A,p_B)$ as well as other
information or influences $\varepsilon$  that are privately observed/experienced by the subject. 
If $\nu$ has CDF $F(\nu)$ that is independent of the public signals, 
then $P$ has the representation \begin{eqnarray}
    P(A|d,\pi,p_A,p_B,D) & = & \mbox{Pr}\{A=\delta(d,\pi,p_A,p_B,D,\nu)\}, \nonumber \\
  & = & \int_\varepsilon I\{A=\delta(d,\pi,p_A,p_B,D,\nu)\}F(\nu).
\label{eq:ccd_def}
\end{eqnarray}
\end{definition}
Note that $P$ is also referred to as a {\it conditional choice probability\/} (CCP) 
and it does {\it not\/} represent the subjective beliefs  about the probability that the sample
came from cage A. In fact a decision rule $\delta$ may have no connection to any well
defined subjective beliefs. For example a variety of machine learning algorithms such as support vector
machines or neural networks can be trained to have nearly optimal decision rules but they do 
not require or depend on well defined subjective posterior beliefs about A.  

However humans, as well as chatGPT, are able to report subjective posterior probabilities. Thus, we would
like a model where the decision rule depends, at least implicitly, on their subjective beliefs. In 
section~\ref{section:structural_model} we introduce a parametric {\it structural logit model\/} of subject choices that
does explicitly depend on subjective posterior beliefs, and which in principle can be used to uncover subjective
beliefs from observed choices. With sufficient data on subject choices in many trials
for different values of $(d,\pi,p_A,p_B,D)$, it is possible to estimate $P$ non-parametrically even
in situations where it is not possible to identify subjective posterior beliefs. 
As we show, knowledge of $P$ is sufficient for our key conclusions about the relative efficiency of human vs AI subjects. 

\begin{definition}{\bf Bayes Rule:} The conditional probability that cage $A$ was selected given
the information $(d,\pi,p_A,p_B,D)$ given by
\begin{equation}
    \Pi(A|d,\pi,p_A,p_B,D) = \frac{\pi f(d|p_A,D)}{\pi f(d|p_A,D) + (1-\pi) f(d|p_B,D)}.
\label{eq:bayes_rule}
\end{equation}
\end{definition}

Bayes Rule is often interpreted as an individual's {\it subjective posterior beliefs\/} since the prior belief $\pi$ is treated
as a subjective prior probability that can differ from subject to subject. However due to the setup of these experiments, 
$\pi$ should be interpreted as an {\it objective prior probability\/} 
that cage A is selected, and we assume that this information has been credibly communicated to subjects.
Thus, it follows that the Bayesian posterior probability $\Pi(A|d,\pi,p_A,p_B,D)$ given in equation (\ref{eq:bayes_rule}) is an 
{\it objective posterior belief\/} i.e. it is the true conditional probability that cage A is selected given $(d,\pi,p_A,p_B,D)$.

Even though some of the subjects were not incentivized via a bonus payment for selecting the correct cage in a randomly
selected trial, it seems reasonable to presume that both incentivized and non-incentivized subjects were trying
to maximize the probability of selecting the correct cage. 
In other words, we presume that subjects in the experiment will want to use
an {\it optimal decision rule\/} if it is not too difficult for them to calculate, or some approximation to it
otherwise.

It is convenient to define two binary random variables, $\tilde W_P$ and $\tilde L_P$, implied by decision rule $P$ by
$\tilde W_P=1$ if the subject selects the correct cage from which the sample was drawn, and 0 otherwise. Thus, $\tilde W_P$
is an indicator for a ``win'' i.e. a correct prediction or classification. $\tilde L_P$ is the indicator for a loss,
i.e. an incorrect prediction. It follows that with probability 1 we have $1=\tilde W_P+\tilde L_P$, and so we can define
an optimal decision rule as one that maximizes the probability of a win or conversely one that minimizes the probability
of a loss. Following the standard terminology from the literature on statistical decision theory, we define

\begin{definition}{\bf Loss Function} The loss function is the conditional  probability of a loss,
\begin{eqnarray}
  L_P(d,\pi,p_A,p_B,D) &=& E\{\tilde L_P|d,\pi,p_A,p_B,D\} \nonumber \\
  & =& P(A|d,\pi,p_A,p_B,D)[1-\Pi(A|d,\pi,p_A,p_B,D)] \nonumber \\
 &+&[1-P(A|d,\pi,p_A,p_B,D)]\Pi(A|d,\pi,p_A,p_B,D).
\label{eq:lossdef}
\end{eqnarray}
\end{definition}

\begin{definition}{\bf Win Function} The win function is the conditional  probability of a win, i.e. selecting the
correct cage,
\begin{equation}
  W_P(d,\pi,p_A,p_B,D) = E\{\tilde W_P|d,\pi,p_A,p_B,D\}  = 1-L_P(d,\pi,p_A,p_B,D).
\label{eq:windef}
\end{equation}
\end{definition}

An {\it optimal decision rule\/} $P$ maximizes the probability of a win, or equivalently it  minimizes
the probability of a loss. Using equation (\ref{eq:lossdef}) or (\ref{eq:windef}) the optimal
decision rule is the pure strategy (\ref{eq:optpure}) defined in terms of Bayes Rule
in Lemma~\ref{lemma:bayes_optimality}.

\begin{lemma}
\label{lemma:bayes_optimality}
The optimal decision rule for a statistical experiment with a binomial design can be defined in terms of  Bayes Rule by
\begin{equation}
    \delta^*(d,\pi,p_A,p_B,D) = \left\{\begin{array}{ll}
            A & \mbox{if} \;\; \Pi(A|d,\pi,p_A,p_B,D) \ge 1/2 \cr
            B & \mbox{otherwise}.
\label{eq:optpure}
\end{array}
\right. 
\end{equation}
\end{lemma}

The win function $W_P$ conditions on the full information set 
$(d,\pi,p_A,p_B,D)$. Of course, $d$ varies stochastically over trials for each subject, 
but the other 4 variables serve as experimental controls.
For example in the California experiments the prior $\pi$ varied across trials taking
on three possible values, $\pi \in \{1/3,1/2,2/3\}$ but the remaining variables were fixed at 
$D=6$, $p_A=2/3$ and $p_B=1/2$. In the Wisconsin experiments $\pi$ also varied over trials
but the remaining variables $(D,p_A,p_B)$ only changed over successive rounds (groups of trials) taking values $(6,2/3,1/2)$ in
round 1 and $(7,.4,.6)$  in round 2.

In our comparisons of the performance of human vs AI decision makers, it is convenient to have a single  overall scalar summary 
measure of {\it decision efficiency\/} which we define as the ratio of the subject's expected win probability to the optimal expected
win probability implied by Bayes Rule, where we compute expectations  over the empirical distributions for the values of the
experimental controls in the experiments. For example, we can define
a {\it ex ante\/} or unconditional expected loss by first taking expectations over the unconditional distribution
over the realized values of $d$ given $(\pi,p_A,p_B,D)$ by
\begin{eqnarray}
    W_P(\pi,p_A,p_B,D) &=& E\{\tilde W_P|\pi,p_A,p_B,D\} \nonumber \\
& =& \sum_{d=0}^D W_P(d,\pi,p_A,p_B,D)[f(d|p_A,D)\pi+f(d|p_B,D)(1-\pi)].
\end{eqnarray}
If $F(\pi,p_A,p_B,D)$ is the empirical distribution of the experimental
control variables in all trials of the experiment, the overall expected loss for a subject with decision rule $P$ in
this  experiment is given by
\begin{equation}
    W_P = E\{\tilde W_P\} = \int_\pi\int_{p_A}\int_{p_B}\int_{D} W_P(\pi,p_A,p_B,D)dF(\pi,p_A,p_B,D).
\end{equation}
We will use $W_P$ as a single summary statistic for the overall performance of decision rule $P$ 
using our econometric estimates of $P$ from our structural logit model of subject choice behavior discussed below.
Similarly, we can define a single summary statistic for the optimal win probability in the same experimental design
under the optimal decision rule implied by Bayes Rule, $W_{\delta^*}$. Thus our overall scalar efficiency metric
is given by $\omega_P$ equal to the ratio of the subject's expected win probability to the optimal win probability of a perfect Bayesian decision maker, $\omega_P =  W_P/W_{\delta^*}$. Clearly
we have $0 \le \omega_P \le 1$.

\section{Structural Econometric Model of Subject Responses}
\label{section:structural_model}

This section introduces an econometric model of subject decision making that we refer to as a {\it structural logit model.\/}
It is related to and subsumes as special cases the ``probability weighting'' model used by \citet{HS2009} to analyze reported posterior and a 
``structural probit'' model introduced by \citet{EGG1999}. It also includes the optimal Bayesian decision rule as a special
case as well. The structural logit model also has an interpretation as a  two layer neural
network where the first input layer uses ``transformed inputs'' equal to the log-likelihood ratio and the log posterior odds ratio
and the second output  layer uses the subjective posterior probability output from the first layer as its input and includes
it in a logistic ``squashing function'' that is a monotonic function of the difference between the subjective posterior and
$1/2$.  A key advantage of the structural logit model is that
it enables us to recover estimates of subjective posterior beliefs even when subjects only make binary choices
of which cage they believe is more likely to have generated the observed sample $d$.

We compare the structural logit model to an alternative structural model introduced 
by  \citet{EGG1995} that assumes that subjects use {\it cutoff rules\/} to make their decisions. Their cutoff rule 
model does not require subjects to have subjective posterior beliefs, and thus bypasses trying to estimate them.
They estimated this model using data from their California experiment cage A where $D=6$ and $p_A=2/3$ and $p_B=1/2$.
Since higher realized values for $d$ provide stronger evidence in favor of cage A, 
a plausible cutoff rule involves choosing cage A if $d > c_\pi$ where $c_\pi$ is a prior-specific
threshold.  In their California design there were three possible priors, $\pi \in \{1/3,1/2,2/3\}$ so there are  512
possible cutoff rules $(c_{1/3},c_{1/2},c_{2/3})$ that subjects could use.  For example, Bayes Rule corresponds to the
cutoff rule $c=(c_{1/3},c_{1/2},c_{2/3})=(4,3,2)$. However the cutoff rule model is {\it statistically degenerate\/}
i.e. for most subjects observed over $T$ successive trials, none of the cutoff rules will be able to perfectly predict the subject's choices.
To deal with this and avoid a ``zero likelihood problem'' \citet{EGG1995} introduced an additional error rate parameter
$\sigma \in (0,1)$ which has the interpretation as the probability that the subject guesses one of the cages
at random rather than using the cutoff rule. This results in a non-degenerate choice probability (decision rule) given by
\begin{equation}
   P(A|d,\pi,p_A,p_B,D) = \left\{\begin{array}{ll} 1-\frac{\sigma}{2} & \mbox{if}\;\; d > c_\pi \cr
                    \frac{\sigma}{2} & \mbox{otherwise.}
\end{array}
\right.
\label{eq:eggprob}
\end{equation}
Using the choice probability $P(A|d,\pi,p_A,p_B,D)$ in equation (\ref{eq:eggprob}), \citet{EGG1995}
were able to write a likelihood function of the sequence of choices made by each subject in the
$T$  trials in the experiment,  and use it to estimate the four unknown
parameters of their model, $\theta=(c_{1/3},c_{1/2},c_{2/3},\sigma)$ by maximum likelihood. 

Though the model is innovative and interpretable,
it has several drawbacks that lead us to abandon it in favor of the structural logit model we introduce
below.  First, the cutoff rule parameters are not ``structural'' i.e. they are not invariant to changes
in the experimental controls $(\pi,p_A,p_B,D)$. For example,  the cutoffs $c_\pi$ are only estimated for the three possible
values of $\pi$ used in their experiment.  Ideally, a structural model should depend on parameters
that are {\it policy invariant\/} and do not change when the experimental design controlling the ``policy'' or ''environment'' is changed.
Second, the nature of subject errors implied by their model is not entirely plausible: the probability
that a subject randomly guesses, $\sigma$, does not depend on the strength of the evidence the subject observes. 
For example, for values $(d,\pi,p_A,p_B,D)$ we should not expect to see much guessing in
the ``easy cases'' where the Bayesian posterior probability is close to 0 or 1.  Random guesses should be 
prevalent mostly in  ambiguous cases where the Bayesian posterior is close to 1/2. 

We now introduce a {\it structural logit model\/} that avoids these limitations. It implies a parametric
family of stochastic decision rules $\delta(d,\pi,p_A,p_B,D,\nu,\varepsilon)$ and stochastic subjective
posterior beliefs $\Pi_s(A|d,\pi,p_A,p_B,D,\nu)$ that includes the optimal
Bayesian decision rule $\delta^*(d,\pi,p_A,p_B,D)$ as a special case.  The structural
logit model depends on 5 unknown parameters $\theta=(\beta_0,\beta_1,\beta_3,\sigma,\eta)$ where the first three parameters
are allowed to take any real value and determine the subject's subjective posterior beliefs. The last parameters $\eta$ and
$\sigma$ are restricted to be positive and determine the scaling/level of idiosyncratic ``noise'' affecting the subject's posterior
beliefs via $\nu$, and final choice via  $\varepsilon$.

Consider first the subject's subjective posterior beliefs $\Pi_s(A|d,\pi,p_A,p_B,D,\nu)$,
where $\nu$ is a random variable that reflects stochastic shocks and ``calculational errors'' that subjects
experience trying to compute their subjective posterior probability. We presume that subjects
attempt to transform their information $(d,\pi,p_A,p_B,D)$ into 
$\mbox{LPR}(\pi)$ and $\mbox{LLR}(d,p_A,p_B,D)$ where $\mbox{LPR}(\pi)$ is the log
posterior odds ratio and $\mbox{LLR}(d,p_A,p_B,D)$ is the log-likelihood ratio given by
\begin{eqnarray} 
     \mbox{LPR}(\pi) & = & \log(\pi/(1-\pi)) \nonumber \\
     \mbox{LLR}(d,p_A,p_B,D) & = &  \log(f(d|p_A,D)/f(d|p_B,D)).
\end{eqnarray}
However we also assume that both human and AI subjects can make algebraic errors trying to evaluate
the quantities  $\mbox{LPR}(\pi)$ and $\mbox{LLR}(d,p_A,p_B,D)$.
Let the scalar random variable $\nu$ equal the sum of these errors, so the log-posterior odds ratio
that the subject actually perceives and reports (or makes their choice on), $\Pi(A)$, is given by
\begin{equation}
\label{eq:slpor}
\log\left(\Pi_s(A)/(1-\Pi_s(A))\right) = \beta_0+\beta_1 \mbox{LLR}(d,p_A,p_B,D)+\beta_2 \mbox{LPR}(\pi)+\nu.
\end{equation}
In our empirical analysis below we assume $\nu \sim N(0,\eta^2)$.
Solving equation~(\ref{eq:slpor}) for $\Pi_s(A|d,\pi,p_A,p_B,D,\nu)$ results in the following logistic specification given by
\begin{equation}
\Pi_s(A|d,\pi,p_A,p_B,D,\nu) = \frac{ \exp\{\beta_0+\beta_1\mathrm{LLR}(d,p_A,p_B,D)+\beta_2 \mathrm{LPR}(\pi)+\nu\} }{ 1 + \exp\{\beta_0+\beta_1\mathrm{LLR}(d,p_A,p_B,D)+\beta_2 \mathrm{LPR}(\pi)+\nu\} }.
\label{eq:subjectiveposterior}
\end{equation}
Notice that the true Bayesian posterior $\Pi(A|d,\pi,p_A,p_B,D)$ given in equation (\ref{eq:bayes_rule}) is a special
case of (\ref{eq:subjectiveposterior}) when $\beta=(0,1,1)$ and $\nu=0$.  For other values
of $\beta$ the subjective posterior can capture a number of well known biases observed in past studies, including
an outright bias for cage A or B if $\beta_0 \ne 0$ as well as {\it overconfidence\/} and {\it underconfidence\/} about the
posterior probability of cage A, {\it base rate bias\/} $(\beta_2 < \beta_1)$ resulting in
behavior consistent with the representativeness heuristic (i.e. excessive weight on the data via $\mbox{LLR}$ relative
to the prior via $\mbox{LPR}(\pi)$), as well as conservatism ($\beta_2 > \beta_1$, i.e. putting excessive weight on prior
information relative to sample information).

Next we  model the subject's binary choice of cage A or B, which we assume depends on whether their perception of the
expected reward from choosing cage A exceeds the expected reward from choosing cage B. Suppose the subject receives  
a reward $R$ if they select the correct cage and 0 otherwise. In experiments where subjects were not paid
for making a correct choice, $R$ can be viewed as a ``psychological reward'' the subject expects from making a
correct choice. However we assume that the subject's expected reward 
for choosing either cage can also be affected by unobserved idiosyncratic preference shocks $\varepsilon=(\varepsilon(A),\varepsilon(B))$ that we
assume are distributed independently of $\nu$. In our empirical analysis below we assume that
$\varepsilon$  has  a bivariate Type 1 extreme value distribution
with location parameter normalized to 0 and a common scale parameter $\sigma$.  Normally we would expect the subject
to select cage A if $\Pi_s(A|d,\pi,p_A,p_B,D,\nu) > 1/2$ and cage $B$ otherwise. However we allow the subject's report
to be affected by the additional preference shocks $\varepsilon$ to capture behavior such as simply guessing one of 
the cages, or some other psychological factors that may cause a subject to report a cage that may not have the higher
posterior probability. This specification implies the following decision rule for the subject
\begin{equation}
      \delta(d,\pi,p_A,p_B,D,\nu,\varepsilon) = 
\left\{ \begin{array}{ll} A & \mbox{if} \quad  R\Pi_s(A) +\sigma\varepsilon(A) \ge  R\Pi_s(B)+\sigma\varepsilon(B) \cr 
    B & \mbox{otherwise}.
\label{eq:dr}
\end{array}
\right.
\end{equation}
As is well known from the discrete choice literature (see e.g. \citet{McFadden1974}) the probability that the subject chooses
cage A is given by the binomial logit formula
\begin{align} 
   P(A|d,\pi,p_A,p_B,D,\nu) &= \Pr\{ \delta(d,\pi,p_A,p_B,D,\nu,\epsilon)=A|d,\pi,p_A,p_B,D,\nu\} \nonumber \\
    &= \frac{ \exp\{R \Pi_s(A|d,\pi,p_A,p_B,D,\nu)/\sigma\} }{ \exp\{R \Pi_s(A|d,\pi,p_A,p_B,D,\nu)/\sigma\} +  \exp\{R \Pi_s(B|d,\pi,p_A,p_B,D,\nu)/\sigma\} } \nonumber \\
    &= \frac{ 1 }{ 1 + \exp\{ R[1-2\Pi_s(A|d,\pi,p_A,p_B,D,\nu)]/\sigma\} }.
\label{eq:structural_logit}
\end{align}
It follows that when $\Pi_s(A|d,\pi,p_A,p_B,D,\nu)=1/2$ the subject is indifferent between choosing cage A or B and the noise
terms $(\varepsilon(A),\varepsilon(B))$ determine the subject's choice, so $P(A|d,\pi,p_A,p_B,D,\nu)$ is also equal to 1/2. 
However as $\Pi_s(A|d,\pi,p_A,p_B,D,\nu)$
approaches 0 or 1, the ``strength of the evidence'' reduces the role of the idiosyncratic shocks $\varepsilon$ on the subject's choice.
Thus, $P(A|d,\pi,p_A,p_B,D,\nu)$ increases to 1 when $R$ is sufficiently large or $\sigma$ is sufficiently small 
and  $P_s(A|d,\pi,p_A,p_B,D,\nu) > 1/2$, and conversely $P(A|d,\pi,p_A,p_B,D,\nu) \to 0$ when
$P_s(A|d,\pi,p_A,p_B,D,\nu) < 1/2$ as $R/\sigma \to \infty$.

Notice that that the structural logit model results in a ``mixed strategy'' for the choice of bingo cage due to the 
effect of the idiosyncratic shocks $\varepsilon$. However as $\sigma \to 0$
it converges to a pure strategy that chooses cage A with probability 1 when $\Pi_s(A|d,\pi,p_A,p_B,D,\nu) > 1/2$ and cage B 
when the subjective posterior is strictly less than 1/2. 
Further if $\beta=(0,1,1)$ and $\eta=0$ (so that subjects do not make random errors calculating
$\mbox{LPR}$ and $\mbox{LLR}$), then
their subjective posterior belief coincides with the Bayesian posterior belief and the structural logit model encompasses
the optimal pure strategy Bayesian decision rule $\delta^*(A|d,\pi,p_A,p_B,D)$ given in equation (\ref{eq:optpure}) as a 
special case.\footnote{\footnotesize \label{footnote:structural_probit} \citet{EGG1999}
introduced a three parameter {\it structural probit model\/} which is a decision rule
of the form $P(A|d,\pi,p_A,p_B,D,\beta)=\Phi(\beta_0+\beta_1\mbox{LLR}(d,p_A,p_B,D)+\beta_2\mbox{LPR}(\pi))$ where
$\Phi$ is the standard normal CDF. The optimal Bayesian decision rule $\delta^*(d,\pi,p_A,p_B,D)$ is not
nested within this class of models, though it can be approximated arbitrarily well in the limit
when $\beta_0=0$ and $\beta_1=\beta_2=\beta > 0$ and $\beta \to \infty$.}  

In experiments 1 and 2 subjects only report which cage is more likely, so the unobserved ``calculational error''
$\nu$ must be integrated out to produce a CCP that can be used for estimation of the model. When $\nu \sim N(0,\eta^2)$ the CCP  
is a mixed logit given by
\begin{equation}
  P(A|d,\pi,p_A,p_B,D) = \int \frac{ 1 }{ 1 + \exp\{ R[1-2\Pi_s(A|d,\pi,p_A,p_B,D,\eta\nu)]/\sigma \} } \phi(\nu) \, d\nu,
\label{eq:ml}
\end{equation}
where $\phi(\nu)$ is the standard normal density.

The structural logit model can also be interpreted  
as a two layer feedforward neural network (or 3 layer in the mixed logit case) that uses transformed inputs
$\mbox{LPR}(\pi)$ and $\mbox{LLR}(d,p_A,p_B,D)$ and produces a single output: the probability of
choosing cage A.  Maximum likelihood of the structural logit  model  can
be interpreted as training the neural network to behave like a human being. The ability of the structural logit model to fit a wide range of
subject behaviors can be ascribed to the flexibility afforded by using a parsimoniously parameterized neural network to
predict subject behavior.\footnote{\footnotesize It is not necessary to pre-transform the inputs
$(d,\pi,p_A,p_B,D)$ into $(\mbox{LPR}(\pi),\mbox{LLR}(d,p_A,p_B,D))$: additional layers can be added to the neural network 
so that the inputs to the deeper neural network can entered without any pre-transformation. 
Then the initial layers of this deeper neural network can be viewed as 
producing approximations to the transformed inputs $(\mbox{LPR}(\pi),\mbox{LLR}(d,p_A,p_B,D))$ that 
then feed into the two layer neural net that used the transformed inputs to compute a subjective posterior
probability and the top layer producing an output equal to the conditional probability of selecting cage A.
These deeper networks require far more parameters, but do not result in substantially better predictions of subject behavior than our
parsimonious 4 parameter two layer neural network specification. Indeed we can ``train'' our 4 parameter neural
network specification to behave nearly identically to a perfect Bayesian decision maker using training samples
with only a few dozen observations.}

Regardless of how it is interpreted, the structural logit implies a CCP that is  
a function of the unknown parameter vector $\theta=(\beta,\eta,\sigma)$ that
can be estimated by maximum likelihood. We use a panel likelihood function since each subject $s$ in the experiment participates in 
a total of $T_s$ independent trials, so we observe a sequence of binary choices $d_{ts}$ and corresponding
controls $(\pi_{ts},D_{ts})$ for each subject $s$ over trials $t=1,\ldots,T_s$ assuming
$(p_A,p_B)$ remain fixed across trials.  Let $y_{ts}$ be a binary indicator of the choice of subject $s$ in trial $t$:
$y_{ts}=1$ if the subject chose A and $y_{ts}=0$ otherwise.  The likelihood $L(\theta)$ is given by
\begin{equation}
     L(\theta) = \prod_{s=1}^S \prod_{t=1}^{T_s} P(A|d_{ts},\pi_{ts},p_A,p_B,D_{ts},\theta)^{y_{ts}}
[1-P(A|d_{ts},\pi_{ts},p_A,p_B,D_{ts},\theta)]^{1-y_{ts}}.
\label{eq:lf}
\end{equation}

\subsection{Identification of  Beliefs}
\label{section:identification}

Both of the unobserved private shocks $\varepsilon$ and $\nu$ imply idiosyncratic responses
by subjects, so even the same subject can provide two different choices to the same information
$(d,\pi,p_A,p_B,D)$. Since these shocks have similar effects on a subject's choice, it is not clear 
whether it is possible separately identify the independent effect of each type of shock 
using data from experiments 1 and 2 where subjects provide only binary responses. 
For these experiments we estimated two different restricted
versions of the mixed structural logit specification (\ref{eq:ml}): 1) a 3 parameter {\it structural probit\/}
that restricts $\sigma=0$ and normalize $\eta=1$, 2) a 4 parameter structural logit model where set 
$\eta=0$ but estimate the scale
parameter $\sigma$ of the extreme value distribution of the preference shocks $\varepsilon$.

In experiments 3 and 4 subjects reported their subjective
posterior probabilities but did not make binary choices. So we assume that subjects would select cage A 
if and only if their reported subjective posterior probability exceeds $1/2$, which is equivalent to
restricting $\sigma=0$. However we estimate the standard deviation parameter  $\eta$ of the ``calculational errors''
in their reported subjective posteriors. It is not hard to see from equation (\ref{eq:slpor}) that the belief
parameters $\beta$ and $\eta$ are parametrically identified given sufficient subject data and sufficient
variation in the experimental controls $(\pi,p_A,p_B,D)$ to avoid multicollinearity in the covariates
$\mbox{LLR}(d,p_A,p_B,D)$ and $\mbox{LPR}(\pi)$.

Identification of the parameters is more challenging in the case where subjects only report binary
choices of cage A or B, even under the restriction that $\eta=0$. 
First, we observe that it is impossible to separately
identify the reward $R$ and the error or noise parameter $\sigma$ since it 
is obvious from formula (\ref{eq:structural_logit}) that these parameters only appear together as a  ``signal to noise ratio''
$R/\sigma$ in the decision rule which in turn is used to form the likelihood for the data.  
Thus,  we assume that the reward $R$ from making a correct
decision is known (e.g. it equals the \$10 bonus for picking the correct cage)
and we normalize the payoff to $R=1$ and estimate only $\sigma$ subject to this normalization. 

Thus, one must keep in mind that a high value of $\sigma$ may not necessarily indicate a high level
of random noise affecting the subject's decision making:
it could also indicate that the subject's evaluation of the reward from a correct
decision, $R$, is sufficiently low. In our empirical reanalysis of \citet{EGG1995} 
the structural logit model estimates of $\sigma$ are lower in the incentivized experiments where subjects received a \$10 bonus if they correctly selected the cage in a randomly selected trial  than in the ``no pay'' experiments. The lower estimated $\sigma$
may reflect the higher reward and not necessarily that the scale of the random preference shocks is lower 
among the incentivized subjects.

An important question is whether the estimated structural logit model provides accurate estimates of
subjects' posterior beliefs, something we do not directly observe in the El-Gamal and Grether experiments.
Ours is an exercise in {\it revealed beliefs\/} where we attempt to uncover subjective posterior beliefs
from observations of subject's binary choices. In fact in the case where $\sigma=0$, we already have an identification
problem in that there are a continuum of non-Bayesian posterior beliefs that are consistent with the optimal
Bayesian decision rule $\delta^*(d,\pi,p_A,p_B,D)$.  To see this consider a family of beliefs indexed by a single parameter $\lambda > 0$
given by $\theta_\lambda=(\beta_\lambda,\sigma,\eta)$ where $\beta_\lambda=(0,\lambda,\lambda)$ and $\eta=0$ and $\sigma=0$. 
For any $\lambda$ and any information  $(d,\pi,p_A,p_B,D)$ we have
\begin{equation}
  \Pi(A|d,\pi,p_A,p_B,D) \le 1/2  \iff  \Pi_s(A|d,\pi,p_A,p_B,D,\theta_\lambda) \le 1/2.
\label{eq:iff}
\end{equation} 
This equivalence implies that the implied decision rule for the subject given in equation (\ref{eq:dr})  coincides with the optimal
Bayesian decision rule $\delta^*(A|d,\pi,p_A,p_B,D)$ given in equation (\ref{eq:optpure}). Thus, even though the subject's posterior
beliefs are not Bayesian, their decision rule is still optimal and achieves the same minimal loss as the one defined
in terms of Bayes Rule.  However we can show that the subject's  posterior beliefs can be identified when $\sigma > 0$ if
the experiment exposes the subject to priors that can be arbitrarily close to $0$ or $1$

\begin{lemma} {\bf Identification of subject beliefs when $\sigma>0$}
\label{lemma:two}
Assume that $\eta=0$. When $\sigma>0$, all four parameters of the structural logit model are identified,
so the subject's subjective beliefs can be identified from knowledge
of their decision rule $P(A|\pi,n)$, where the latter
is identifiable given sufficient experimental data on a subject's choices. 
\end{lemma}
The proof of Lemma~\ref{lemma:two} is in \ref{appendix:A}.
Even though Lemma~\ref{lemma:two} provides a theoretical justification for the
identification of the model when $\sigma > 0$, in practice it can be hard to distinguish the decision rule of a subject with Bayesian
posterior beliefs where $\sigma $ takes on relatively large values (i.e. a ``noisy Bayesian'') from  a decision rule of a
non-Bayesian who has a very small value of $\sigma$ but whose $\beta$ coefficients are also close to zero.

\begin{figure}[ht]
\caption{Example of weak identification of subjective posterior beliefs}
\label{fig:weak_identification}
\centering 
\begin{minipage}[t]{0.48\textwidth} 
    \centering 
    \includegraphics[height=2.5in]{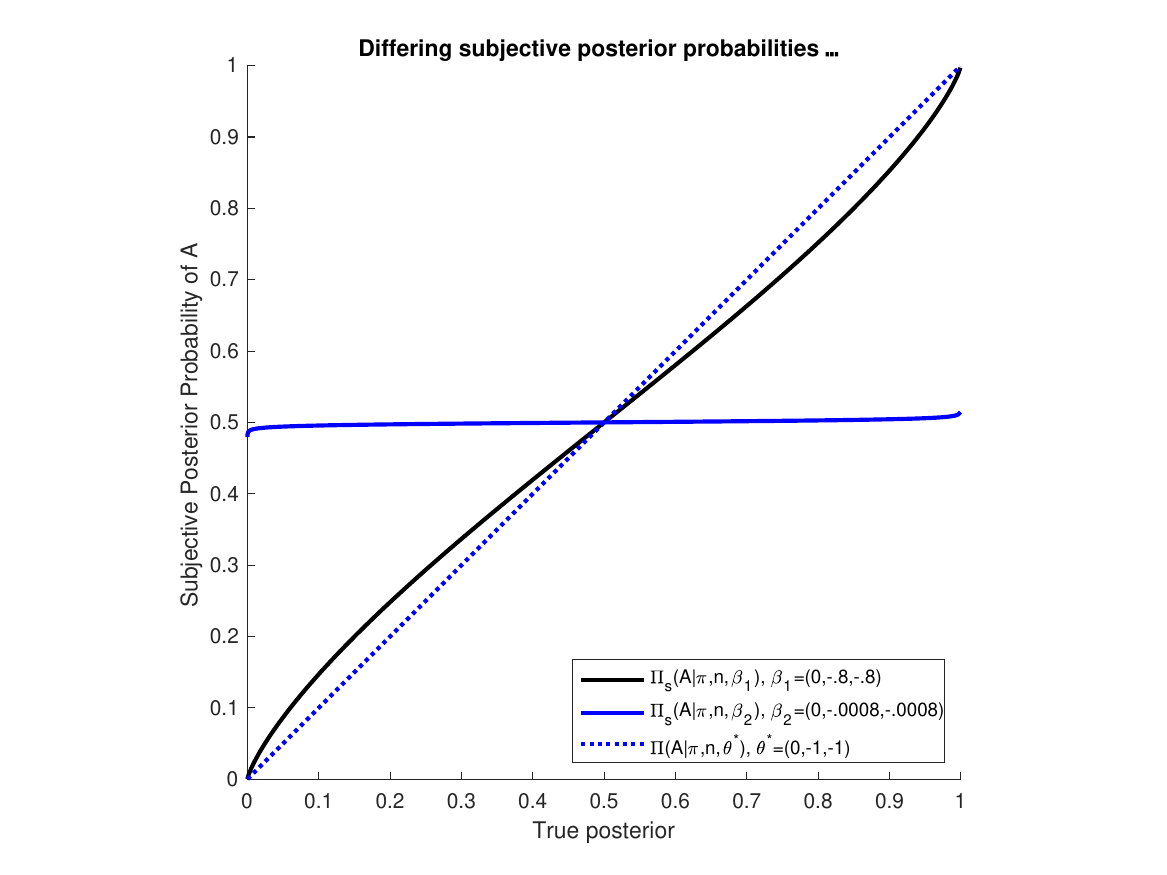}
\end{minipage}
\hfill 
\begin{minipage}[t]{0.48\textwidth} 
    \centering 
    \includegraphics[height=2.5in]{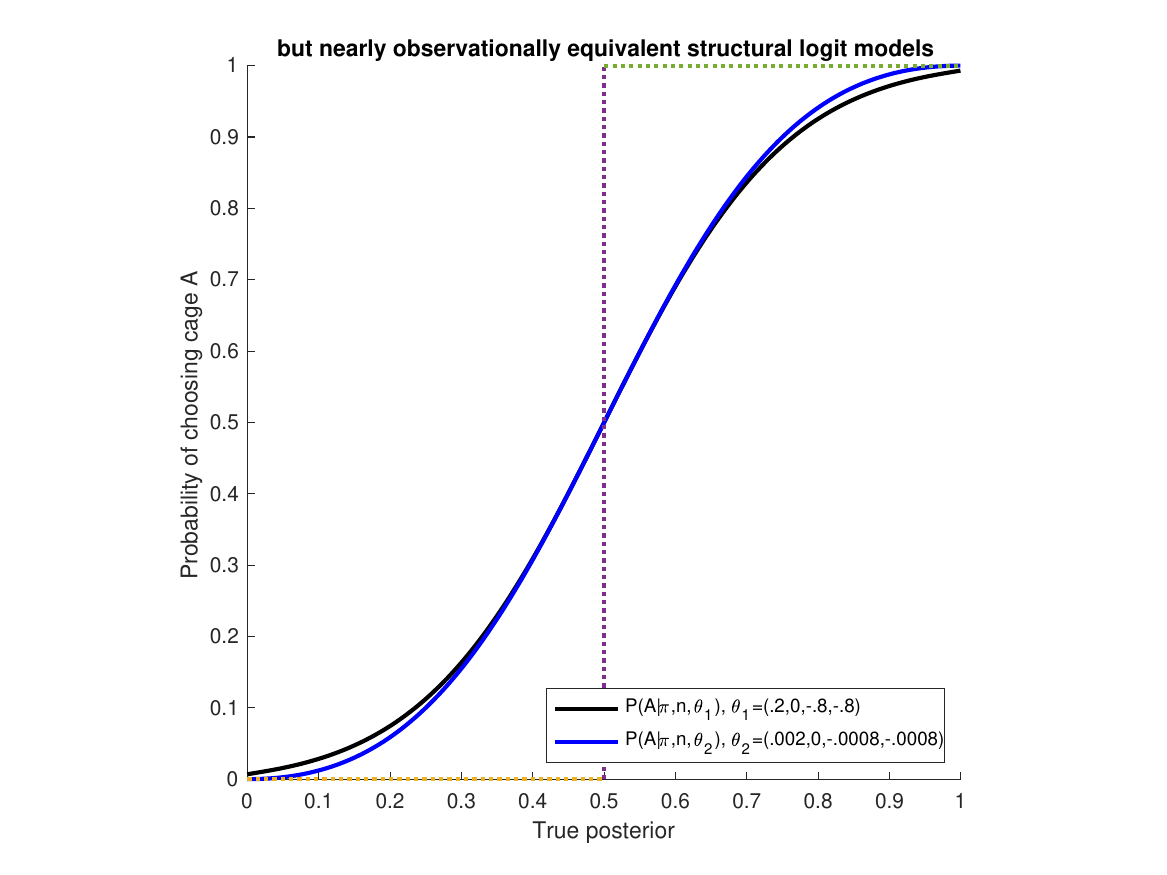}
\end{minipage}
\end{figure}

The weak identification of subjective posterior beliefs is illustrated in figure~\ref{fig:weak_identification}. The figure
plots the CCPs for two very different values of the structural parameters:  $\theta_1=(0,.8,.8,.2,0)$ and
$\theta_2=(0,.008,.008,.002,0)$, i.e. $\theta_2=\theta_1/100$. The left panel plots the subjective posterior
beliefs  for these two different ``subjects'' and we can see that they are very different, both from each other and
from the true Bayesian posterior (the dashed 45 degree line). The right hand panel shows that the 
implied CCPs are nearly identical. This example does not contradict our identification result, Lemma~\ref{lemma:two},
because we can see that $P(A|d,\pi,p_A,p_B,D,\theta_1) \ne P(A|d,\pi,p_A,p_B,D,\theta_2)$ but the differences are really only apparent
at extreme values such as when $\pi$ is close to 0 or 1. So even though the structural logit is ``technically identified''
as a theoretical matter, but in any practical sense it is only weakly identified.\footnote{\footnotesize One way to understand the
identification problem is to note that the structural probit model of \citet{EGG1999} discussed in
footnote~\ref{footnote:structural_probit} must be subject to a location scale normalization on its normally  distributed error term. 
As we discussed above in the structural logit model the extreme value scale parameter $\sigma$ only appears as a ratio
$R/\sigma$ where $R$ is the payoff for a correct classification. Even if we normalize $R=1$ and Lemma~\ref{lemma:two}
assures us that the structural logit CCP is not scale invariant in its 4 parameters $\theta$, it is {\it approximately scale invariant\/}
as shown in figure~\ref{fig:weak_identification}. However imposing an arbitrary normalization such as $\sigma=1$ has behavioral
implications, since it determines the probability of incorrect classifications at the extreme values for the prior
probability of choosing cage A, i.e. $\pi=0$ or $\pi=1$. Indeed, our proof of identification used the knowledge of these classification
error probabilities at $\pi=0$ and $\pi=1$ to identify $\sigma$. But in actual data sets we may not have sufficient
data at $\pi=0$ or $\pi=1$ and in such cases the identification of $\sigma$ and the entire parameter vector $\theta$ will be 
more tenuous.}

As a result, the focus of the empirical work in this section will be on the {\it optimality of subjects'
decision rules\/} (which are non-parametrically identified) not necessarily on whether their posterior beliefs are Bayesian. Condition 
(\ref{eq:iff}) can be viewed as a sufficient condition for optimality of the decision rule of a non-Bayesian subject
and it results in a test for a weaker form of Bayesian rationality: $H_o: \beta_0=0$ and $\beta_1=\beta_2$. If this latter
hypothesis is satisfied, then the subject will still be modeled as behaving as a  ``noisy Bayesian''  even though their posterior
beliefs are not Bayesian. In section~\ref{section:holtsmith}  we return to the question of inferring subjective posterior beliefs
by reanalyzing the experiments of \citet{HS2009}  that directly elicited subjects' beliefs.

\subsection{Accounting for Unobserved Subject Heterogeneity}
\label{section:heterogeneity}

The experimental data we received from \citet{EGG1995} and \citet{EGG1999} do not contain any covariates
or observable characteristics of the subjects in the experiment beyond 
the particular school (out of 4 in California) where the experiment was run. They found that students
at the more selective schools (e.g. UCLA) were more likely to be Bayesian and less likely 
to guess. 

However there is undoubtedly additional heterogeneity in the decision rules that subjects are using, including
different thought processes (and different subjective posterior beliefs) that affect their choices.
One possible way to account for this additional {\it unobserved heterogeneity \/} is to use {\it fixed effects\/} where 
we estimate subject-specific parameter vectors. However with $S$ subjects observed over $T$ time periods, we have
$ST$ total observations but a total of $4S$ parameters in the structural logit model. The time dimension (i.e. number
of trials) is rather limited (in the California experiments  $T \le 20$).  Since both $T$ and $S$  are
relatively small, the  {\it incidental parameters problem\/} of \citet{KW1956} suggests that inferences from
a fixed effects approach are unreliable.

Instead we use a {\it random effects\/} approach to inferring unobserved subject heterogeneity, where following
\citet{KW1956} we posit a  {\it distribution\/} $\mu(\theta)$ of preference parameters in the population and attempt
to estimate it. Treating $\mu$ as an arbitrary
element of the space of all distributions over $\theta$ results in an infinite dimensional ``parameter space'' 
and the estimation problem can be ill-posed unless some restrictions are imposed. Following \citet{HS1984} we estimate
a finite mixture approximation to $\mu$ by maximum likelihood, treating this mixture as a {\it sieve\/} (i.e. an expanding
parametric family that increases with sample size $S$ and can eventually approximate any $\mu$ when $S$ and the
number of mixture components is sufficiently large). 

Let $K$ denote the number of unobserved types to be estimated, and $\theta=(\theta_1,\ldots,\theta_K)$ be the $4K \times 1$
vector of parameters of the ``mixed structural logit model'' and let $\lambda=(\lambda_1,\ldots,\lambda_K)$ be the corresponding
$K \times 1$ vector of population probabilities of each of the $K$ types. Then the mixed logit likelihood function, $L(\theta,\lambda)$
is given by
\begin{equation}
   L(\theta,\lambda) = \prod_{s=1}^S \sum_{k=1}^K \lambda_k  L_s(\theta_k),
\label{eq:hs}
\end{equation}
where $L_s(\theta_k)$ is the likelihood function for subject $s$ evaluated at $\theta_k$.
We estimate a sequence of such models starting with $K=1$ and increasing the number of types $K$ until a log-likelihood ratio test
is unable to reject a model with $K$ types in favor of a model with $K+1$ types, or alternatively,  picking the value of
$K$ for which the Akaike Information Criterion (AIC) equals $2(N_K-L(\hat\theta,\hat\lambda))$ where  $N_K$ is the total number
of parameters in the $K$-type model, is minimized.

We also estimate a closely related random effects approach called the {\it Estimation-Classification\/} (EC) algorithm
by \citet{EGG1995}. This algorithm also maximizes a likelihood function but instead of computing a mixture over
types for each subject as in equation (\ref{eq:hs}) the EC algorithm {\it assigns each subject in the sample their most
likely type.\/} That is, the EC algorithm is the $4K\times 1$ vector $\theta=(\theta_1,\ldots,\theta_K)$ together with the 
$S \times 1$ vector of maximum likelihood subject classifications $\kappa=(k_1,\ldots,k_S)$ where $k_s$
denotes the most likely type for subject $s$. That is, we can write the EC estimator $(\hat\theta,\hat\kappa)$ as the 
pair $(\theta,\kappa)$ that maximizes the following likelihood
\begin{equation}
 L(\theta,\kappa) =  \prod_{s=1}^S \max_{1\le k \le K} L_s(\theta_k),
\label{eq:ec}
\end{equation}
where $L_s(\theta_k)$ is also the likelihood function for subject $s$ evaluated at $\theta_k$.
The Heckman-Singer finite mixture estimator can be conceptualized as a ``mixed strategy'' since it 
treats each subject as a mixture over $K$ possible types, whereas the EC estimator can be regarded as
a ``pure strategy'' since instead of using a mixture over types, it assigns each subject their most likely type. 
The EC algorithm results in the following implied mixture probabilities,
\begin{equation}
         \hat\lambda_k = \frac{1}{S} \sum_{s=1}^S  I\{\hat k_s=k\},
\end{equation}
where $\hat k_s$ is the type that maximizes the likelihood for subject $s$.
Thus $\hat\lambda_k$ is simply the fraction of subjects whose most likely type is $k$. Similar to the finite
mixture estimator, we use the same sequential procedure to identify the number of types $K$ by starting with $K=1$ and
sequentially increasing $K$ until a likelihood ratio test fails to reject the hypothesis that a model with $K$
types fits significantly better than a model with $K+1$ types. Alternatively we can pick the value of $K$ with the smallest AIC.

\section{Optimality and revealed beliefs of human subjects}
\label{section:human_subjects}

\subsection{Reanalysis of California Experiments}

\begin{figure}[ht]
\caption{Comparison of subject behavior and models in the California experiments}
\label{fig:california_experiments}
\centering 
\begin{minipage}[t]{0.48\textwidth} 
    \centering 
    \includegraphics[height=2.5in]{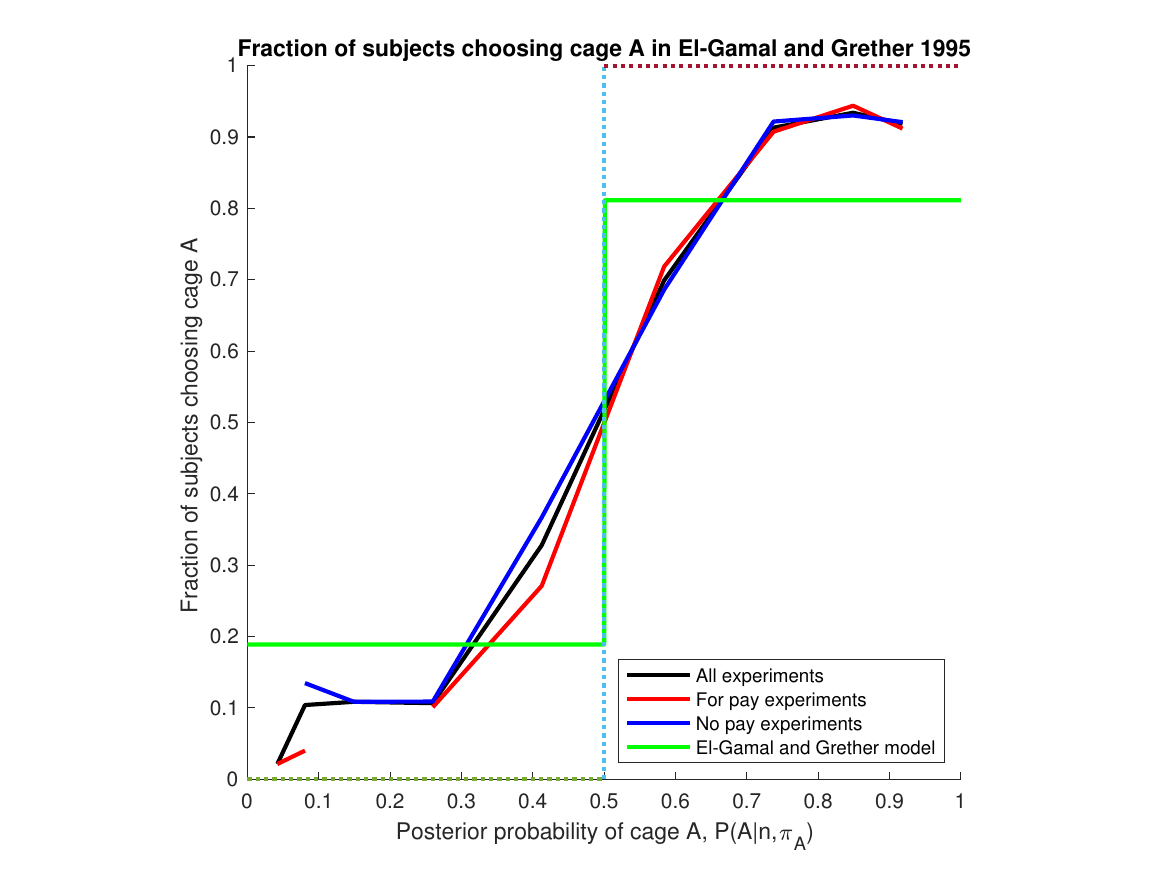}
\end{minipage}
\hfill 
\begin{minipage}[t]{0.48\textwidth} 
    \centering 
    \includegraphics[height=2.5in]{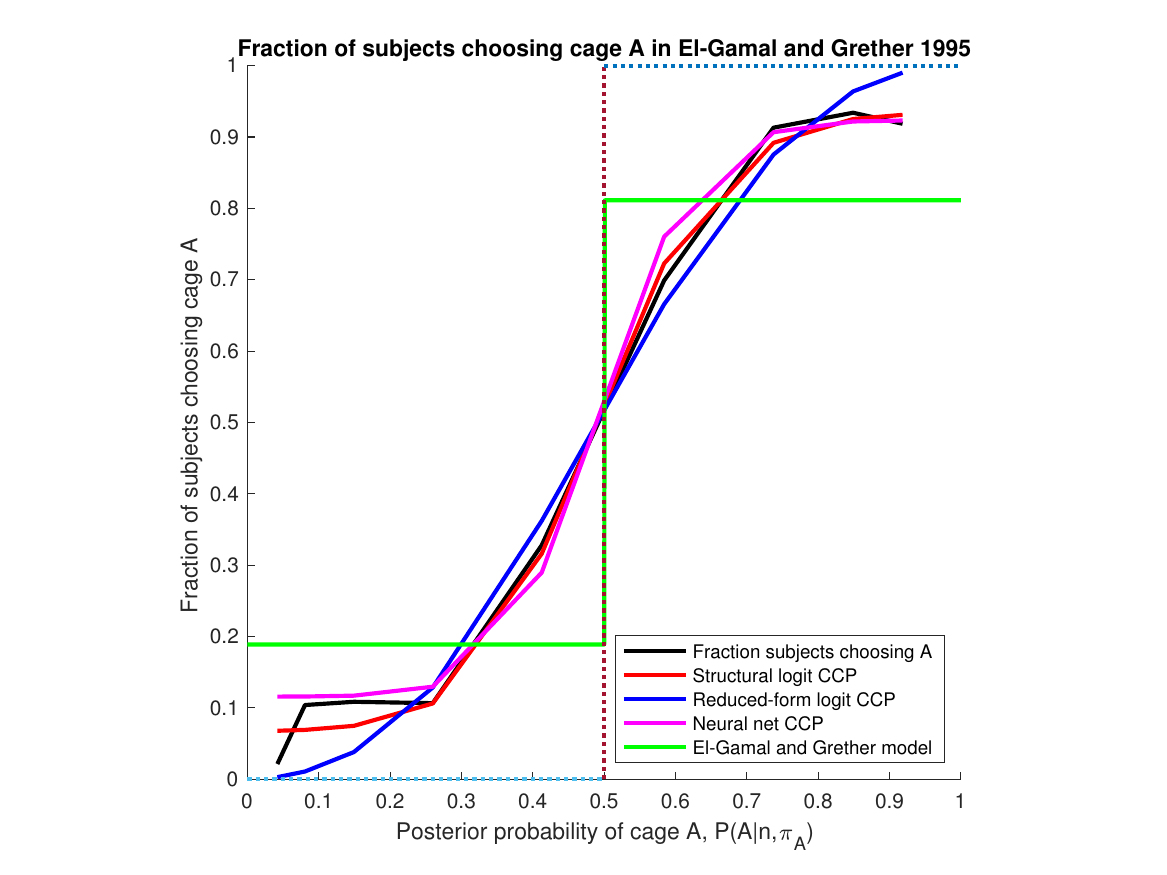}
\end{minipage}
\end{figure}

We estimated the El-Gamal and Grether threshold model and the structural logit model for 221 of the 247 subjects reported
in \citet{EGG1995}.\footnote{Due to a corrupted data file we were unable to include 26 subjects from Pomona Community College
under the incentivized (i.e. for pay) design.} The results are shown graphically
in figure~\ref{fig:california_experiments} which compare predicted subject choices for several different
models and subsamples. In both panels, we plot of the fraction of subjects choosing
cage A (y axis) as a function of the true Bayesian posterior probability of cage A (x axis). The black lines
in both panels are the actual fraction of the 221 subjects who chose cage A in the different trials of the
experiment where the values of the ``treatment variables'' $(\pi,n)$ are binned so we can plot results
on a two dimensional graph with the Bayesian posterior probabilities, $\Pi(A|\pi,n)$, on the x-axis.
Similar to figure~\ref{fig:humans_vs_robots} the dashed blue line represents the optimal decision rule of a perfect
Bayesian decision maker.  

The left panel illustrates the effect of the incentive payments on subject behavior (blue curve for the no-pay
subjects, red for the subjects who were paid) and it is evident that it has negligible effect on overall
behavior.\footnote{The average decision efficiency for the 90 subjects in the incentivized trials 
was 93.5\% (std error $0.5\%$) which is not significantly higher than the 92.3\% efficiency
of the 132 subjects in the non-incentivized trials (std error 1.5\%). 
We also separately analyzed data from the first and last third of the trials 
 see if there were any substantial ``learning by doing'' or ``experience effects'' and these
were also negligible.} The maximum likelihood predictions from El-Gamal and 
Grether's model of subject choice are the green curves in both panels. We see that this model fails to
fit the data well, particularly for the ``easy cases'', i.e. $(\pi,n)$ values where the Bayesian posterior probability is near $0$ or $1$.  But it also misses near the ```hard cases'' where the true posterior is near $1/2$. This
pattern of prediction errors follows from their assumption about subject behavior already discussed, namely
that with probability $\sigma=.38$ subjects randomly guess a cage otherwise with probability $1-\sigma=.62$
they follow Bayes Rule. This implies a discontinuous jump in the predicted probability of selecting cage A
right at $\Pi(A|\pi,n)=1/2$ since at that point the 62\% of subjects who are choosing according to Bayes Rule
jump from choosing cage B to choosing cage A.

The right hand panel of figure~\ref{fig:california_experiments} plots the predictions from
the structural logit model (red curve) as well as several ``reduced form'' models: 
1) a binary logit model with 3 parameters (for a constant, coefficient of $\pi$ and coefficient of $n$),
and 2) a 5 parameter two layer neural network that
includes an additional bias parameter in the upper output layer. We can see visually that the structural logit
model fits the data significantly better than the El-Gamal and Grether model even though both models have
4 parameters.  The El-Gamal and Grether model restricts 3 of the parameters, the cutoffs $c_\pi$, to a finite grid of integers, 
which allows far less flexibility in fitting the data compared to the 4 continuous parameters of the structural logit. The structural logit also
outperforms the 3 parameter reduced form logit (which can be regarded as a single layer feedforward neural network), but
produces approximately the same predictions as a 5 parameter neural network specification. The
structural logit model can be viewed as a restricted 4 parameter version of the 5 parameter neural network
where the bias term in the output layer is restricted to be -1/2  times the value of the input weight parameter, which is
$1/\sigma$ in the notation of the structural logit model. Intuitively, the structural logit model imposes a harmless
restriction that subjects compare their subjective posterior $\Pi_s(A|\pi,n)$ to the value 1/2 as the threshold for choosing  
cage A, and in particular, the restriction implies that subjects are equally likely to choose cage A or B when $\Pi_s(A|\pi,n)=1/2$.

\begin{table}[htbp]
\caption{Log-likelihood values for alternative models of subject choices}                                                                       
\begin{center} 
\label{tab:llf_california1}
\fontsize{11pt}{11}\selectfont
        \setlength{\tabcolsep}{2pt}
        \renewcommand{\arraystretch}{1.15}
                \begin{tabular}{|l|c|c|c|} 
                        \hline
{\bf Model} & {\bf Number of parameters}& {\bf Log-likelihood} & {\bf AIC} \\
\hline
El-Gamal/Grether discrete cutoff rule & 4 &  -1952 & 3912 \\
\hline
Structural probit & 3 & -1847 & 3700 \\
\hline
Reduced-form logit  & 3&  -1821 & 3648 \\
\hline
Noisy Bayesian & 1 & -1801 & 3604 \\
\hline
Structural logit & 4 & -1773 & 3554 \\ 
\hline
Neural network & 5 & -1772 & 3554 \\
\hline
\end{tabular}
\end{center}
\end{table}

Table~\ref{tab:llf_california1} summarizes the fit of the various models of subject behavior. The final column
of the table reports the Akaike Information Criterion (AIC) used for model selection
and  defined as $2(k-LL)$ where $k$ is the number of parameters in the model and $LL$ is the maximized
value of the log-likelihood function for that model. Though the 4 parameter
El-Gamal and Grether model is not nested as a special case of the 4 parameter structural logit model, using the non-nested
likelihood-based specification test of \citet{Vuong1989} we can strongly reject the El-Gamal and Grether model in favor
of the structural logit model (P-value $2.5\times 10^{-4}$). Similarly, we can strongly reject the 3 parameter structural probit
(see footnote~\ref{footnote:structural_probit}) and the 3 parameter reduced form logit model as well (P-values 
$1.1\times 10^{-6}$ and $2 \times 10^{-5}$ respectively).
The noisy Bayesian model is a restricted 1 parameter version of the structural logit model where we allow $\sigma$ to be freely
estimated and restrict $\beta$ to impose Bayesian beliefs, i.e. $(\beta_0,\beta_1,\beta_2)=(0,1,1)$. A likelihood
ratio test strongly rejects the hypothesis that subjects are ``noisy Bayesians'' (P-value $9.2\times 10^{-12}$). 
The structural logit model is a restricted version of the 5 parameter neural network model and a likelihood
ratio test fails to reject the implicit upper-level restriction on the bias term  discussed above. (P-value $.115$).
Overall, the structural logit model achieves the lowest value of the AIC (the same as the 5 parameter neural network
model) and it is another compelling reason why we use it as the preferred model for our subsequent analysis.

\begin{table}[htbp]
\caption{Maximum likelihood estimates of the structural logit/probit model parameters}                                                
\begin{center} 
\label{tab:sl_california_ml_estimates}
\fontsize{11pt}{11}\selectfont
        \setlength{\tabcolsep}{2pt}
        \renewcommand{\arraystretch}{1.15}
                \begin{tabular}{|l|c|c|c|c|c|} 
\multicolumn{6}{c}{Structural logit model} \\
\hline
Parameter & $\sigma$ & $\eta$  & $\beta_0$ &  $\beta_1$ & $\beta_2$ \\
                        \hline
Estimate &  .38  & 0 &  .05  & 2.38   &   1.86     \\
\hline
Standard error &  (.02) & (0)   &  (.05) &  (.28) &  (.19) \\
                 \hline
\multicolumn{6}{c}{Structural probit model} \\
\hline
Parameter & $\sigma$ & $\eta$  & $\beta_0$ &  $\beta_1$ & $\beta_2$ \\
                        \hline
Estimate &  0 & 1   &  .03  & 1.05   &   .97     \\
\hline
Standard error &  (0) & (0)   &  (.02) &  (.04) &  (.04) \\
\hline
                \end{tabular}%
\end{center}
\end{table}

Table~\ref{tab:sl_california_ml_estimates} presents the maximum likelihood coefficient estimates for the
structural logit and probit  models. Since the coefficient on $\mbox{LLR}(d,p_A,p_B,D)$, $\beta_1$, is significantly greater 
than the coefficient on $\mbox{LPR}(\pi)$, $\beta_2$, the estimation results suggest the typical subject
in El-Gamal and Grether's California experiments display the representativeness heuristic, contrary to
their finding using their cutoff rule model that the single best model to predict subject behavior is the ``noisy Bayesian'' model.
On the other hand, structural probit model results support the hypothesis that the single best model is the
noisy Bayesian model. Neither a Wald test nor a Likelihood ratio test rejects the hypothesis that $\beta^*=(0,1,1)$,
the values that imply subjective beliefs $\Pi_s(A|d,\pi,p_A,p_B,D,\nu)$ in equation (\ref{eq:subjectiveposterior}) ($P$-value
of likelihood ratio test, $.13$). However the results in
 table~\ref{tab:llf_california1} show that the data reject the structural probit specification in favor
of the structural logit model. We can see why this is visually in the right hand panel of figure~\ref{fig:california_experiments} where the blue
line for the 3 parameter reduced-form logit  (which fits better than the 3 parameter structural probit but results in a similar
CCP) predicts that subjects make no classification errors when the evidence (as captured by the Bayesian posterior) is
close to 0 or 1. In actuality, subjects do make significant classification errors at these extremes, and this is why $\sigma > 0$
in the structural logit model and it is able to fit the data significantly better.

Now we  show how our conclusions change when we allow for unobserved heterogeneity in subjects' beliefs
and behavior.  We estimated multiple type models using both the EC algorithm and the finite mixture of types (hereafter
abbreviated as FM)
method described in section~\ref{section:heterogeneity}. We found that AIC is smaller for a specification with $K=3$
unobserved types compared to $K=2$ types or the single type specification presented in table~\ref{tab:sl_california_ml_estimates}.
Rather than presenting the coefficient estimates, we illustrate the predictions and key findings in a series of graphs below.

\begin{figure}[ht]
\caption{Inferred Posterior beliefs of California subjects from EC and FM algorithms}
\label{fig:ec_vs_fm_1}
\centering 
\begin{minipage}[t]{0.48\textwidth}
    \centering
    \includegraphics[height=2.5in]{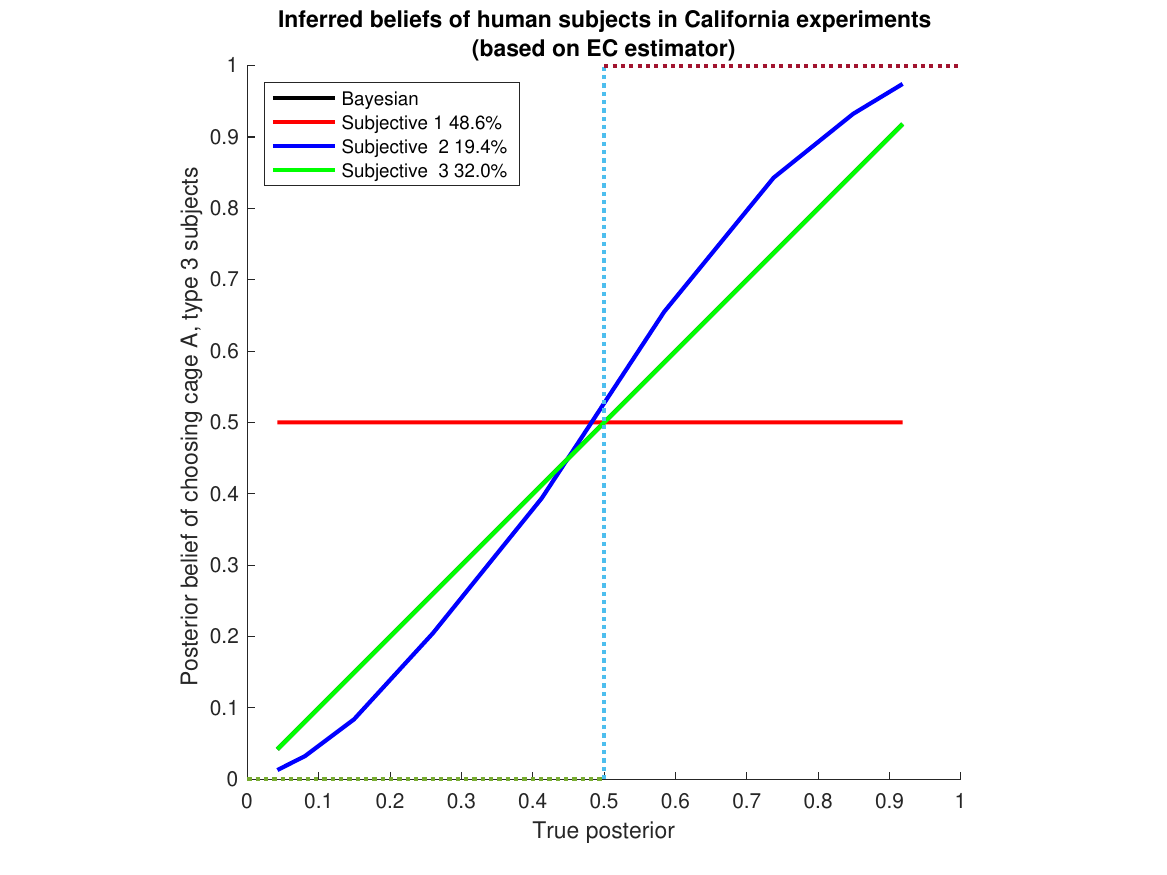}
\end{minipage}%
\hfill
\begin{minipage}[t]{0.48\textwidth}
    \centering
    \includegraphics[height=2.5in]{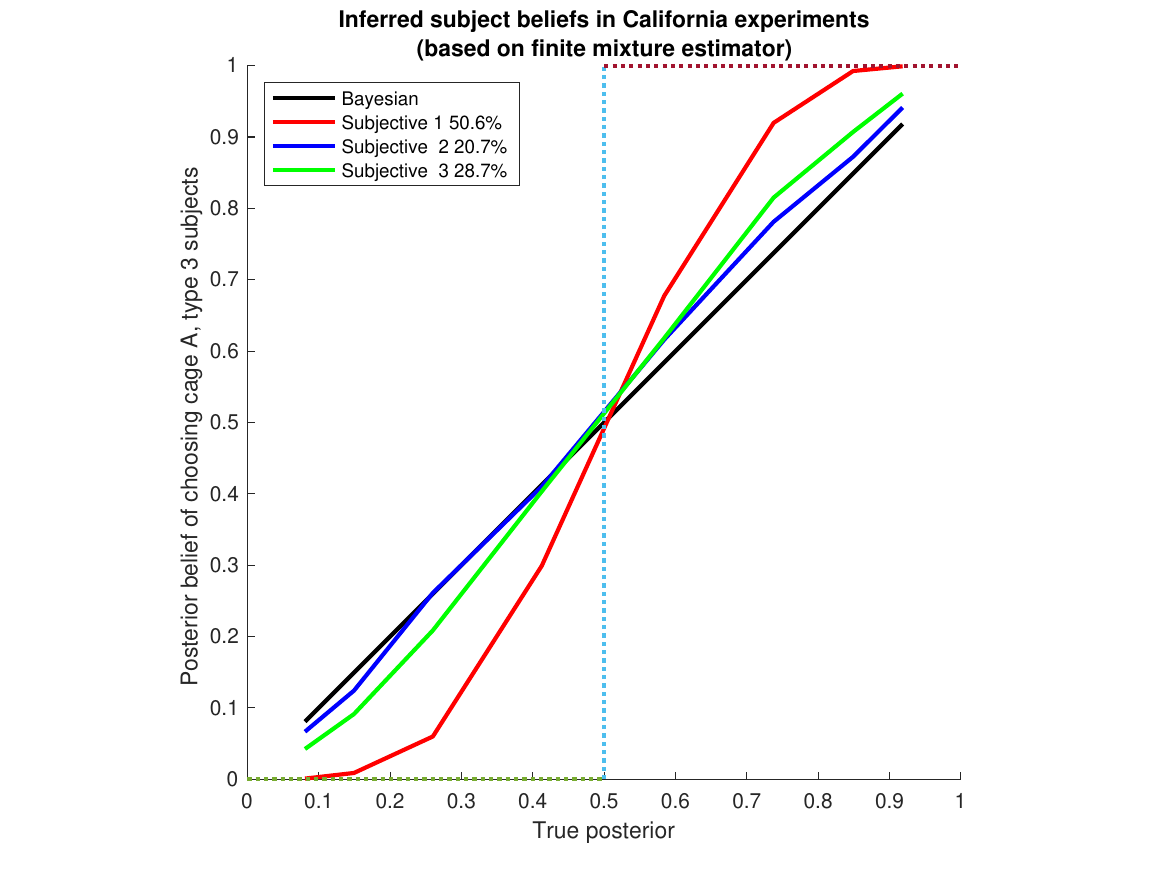}
\end{minipage}
\end{figure}

Figure~\ref{fig:ec_vs_fm_1} compares the types identified by EC and FM in terms of the revealed posterior beliefs.
The methods result in different inferences about the proportions of the three types as well as their beliefs. The left
hand panel shows results from EC algorithm. Type 1 subjects (48.6\% of the sample) 
have a virtually flat posterior. This is due to very small
estimates of $\beta$ and $\sigma$, echoing the example we discussed in section~\ref{section:identification} and the beliefs
are similar to those plotted in the example of weak identification of beliefs in figure~\ref{fig:weak_identification}. 
Type 3 subjects (32\% of the sample) have beliefs that are virtually identical
to the Bayesian posterior (with the two lines appearing as a single solid green line in the figure)
and type 2 subjects (19\% of the sample) have beliefs consistent with conservatism 
 since the coefficient $\beta_1$ on $\mbox{LLR}$ is significantly lower than $\beta_2$, the coefficient of $\mbox{LPR}$. Their
beliefs, plotted by the blue line in the left panel of figure~\ref{fig:ec_vs_fm_1}, display overconfidence, i.e. they believe
cage A is less likely than the true (Bayesian) posterior when the true posterior is less than 1/2, and more likely otherwise.

Though the posterior beliefs of the type 3 subjects are closest to Bayesian posterior beliefs, the degree of idiosyncratic decision noise for these subjects is far higher than for type 1 subjects whose subjective posterior seems nearly flat but is actually strictly increasing as a function of the true posterior. For this reason, we will see below that in terms of decision efficiency, the type 1 subjects are the ``most Bayesian'' of the three types. That is, both type 1 and 3 subjects are ``noisy Bayesians'' but type 3 subjects have substantially more noise in their responses with an estimated $\sigma$ of $\hat\sigma=.3$ for type 3 and $\hat\sigma=1.9\times 10^{-6}$ for the type 1 subjects. 

The right panel of figure~\ref{fig:ec_vs_fm_1} plots the inferred beliefs from the FM approach. The results appear rather 
different than those from the EC algorithm: 
the Type 1 subjects (50.6\% of the sample) conform to the representativeness heuristic, since the estimated
coefficient on $\mbox{LLR}$ is approximately twice as large as the coefficient of $\mbox{LPR}$. The other two types
identified by the FM method can be classified as ``noisy Bayesians'' since the coefficients on
$\mbox{LLR}$ and $\mbox{LPR}$ are not significantly different from each other. However type 2 subjects are substantially
noisier than type 3 subjects: the estimated $\sigma$ for the former is $.75$ compared to only $.16$ for the latter. Thus, we will designate the type 3 subjects as the ``most Bayesian'' type discovered by the finite mixture method. Overall the optimized
likelihood from the FM approach, $-1655$ is considerably below the value yielded by EC algorithm, $-1543$, reflecting the
advantage of EC as a ``pure strategy'' for identifying types compared to FM which is akin to using a ``mixed strategy''.
Despite these differences, we show that both methods provide similar overall results and conclusions.

\begin{figure}[ht]
\caption{CCPs implied by EC and FM models of California subjects}
\label{fig:ec_vs_fm_2}
\centering 
\begin{minipage}[t]{0.48\textwidth}
    \centering
    \includegraphics[height=2.5in]{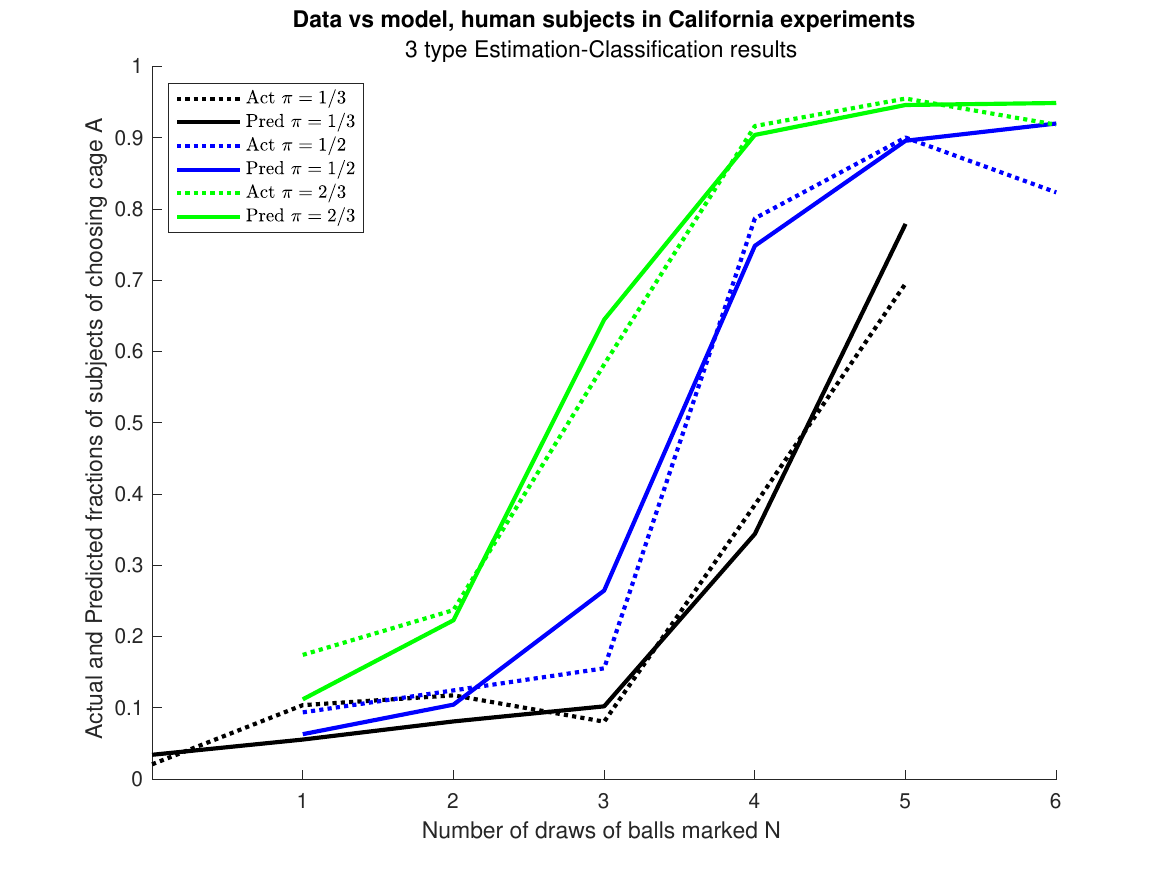}
\end{minipage}%
\hfill
\begin{minipage}[t]{0.48\textwidth}
    \centering
    \includegraphics[height=2.5in]{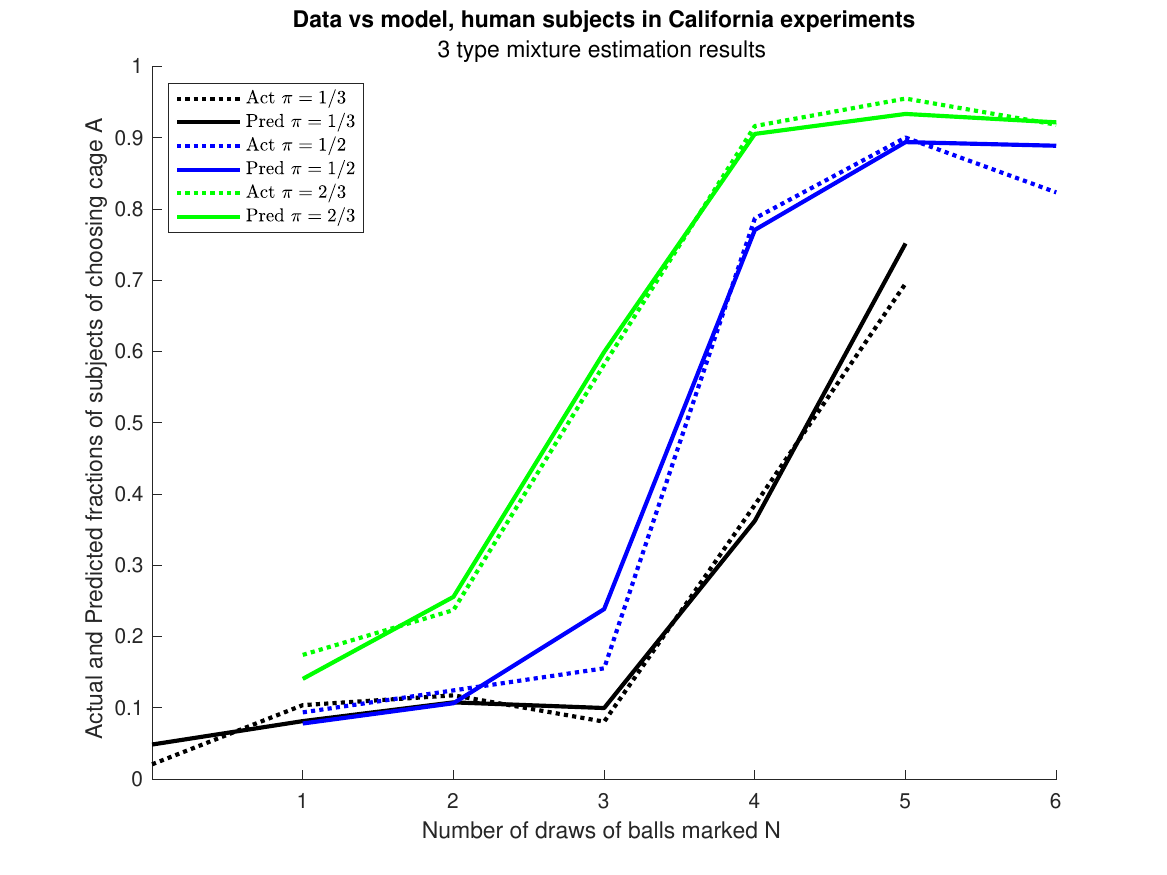}
\end{minipage}
\end{figure}

Figure~\ref{fig:ec_vs_fm_2} compares the overall CCPs (for all subjects, not broken down by type), plotted
for each of the three priors used and for all observed outcomes for $d$. We see that EC and FM give very
similar overall fits to the data.

\begin{figure}[ht]
\caption{Classification thresholds implied by EC and FM models of California subjects}
\label{fig:ec_vs_fm_3}
\centering 
\begin{minipage}[t]{0.48\textwidth}
    \centering
    \includegraphics[height=2.5in]{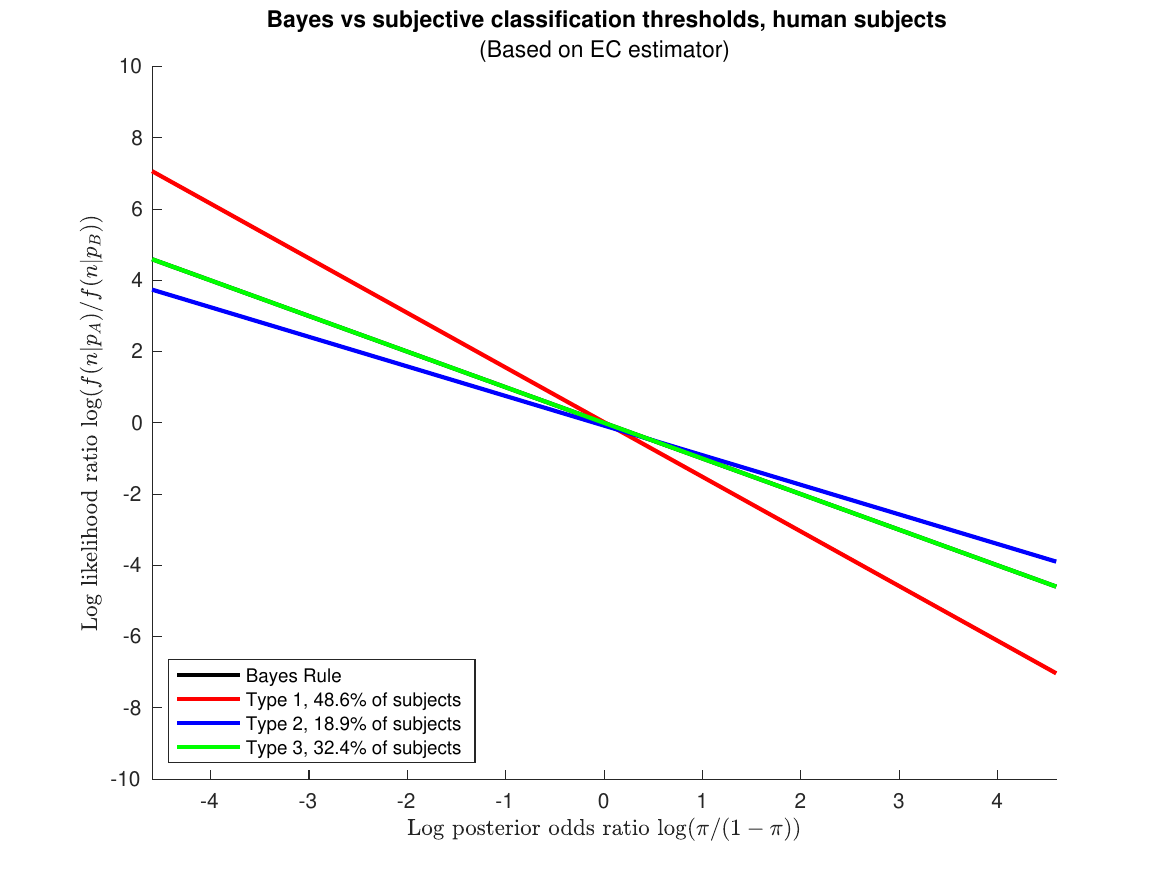}
\end{minipage}%
\hfill
\begin{minipage}[t]{0.48\textwidth}
    \centering
    \includegraphics[height=2.5in]{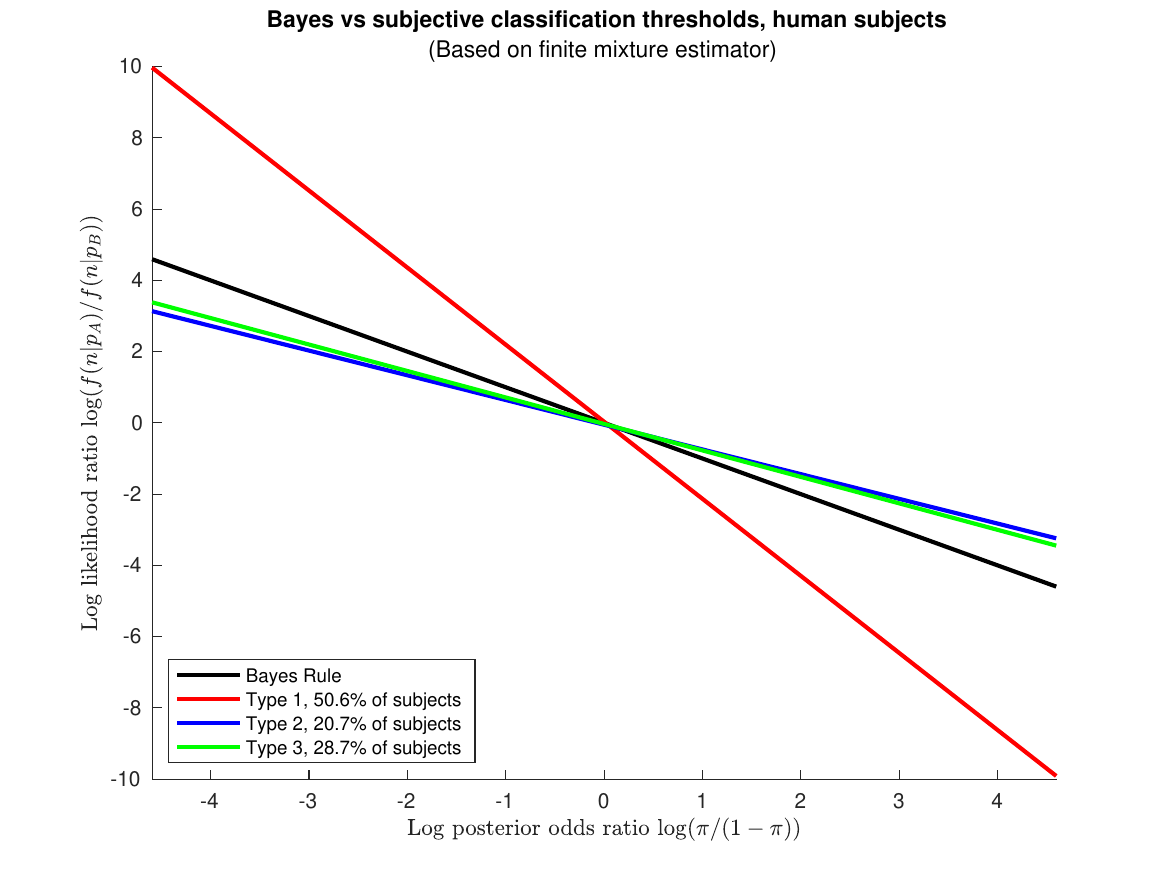}
\end{minipage}
\end{figure}

Figure~\ref{fig:ec_vs_fm_3} plots the classification thresholds for each type of subject. These are
hyperplanes of the form $\hat\beta_{0,k}+\hat\beta_{1,k}\mbox{LLR}+\hat\beta_{2,k}\mbox{LPR}=0$ for each
type $k \in \{1,2,3\}$ where cages A and B are subjectively equally likely. We also plot
the classification threshold for Bayes Rule, the black line with a slope of $-1$. 
Even though figure~\ref{fig:ec_vs_fm_1} creates an impression that EC and FM approaches 
lead to different conclusions about the proportions of subjects with different
subjective posterior beliefs, figure~\ref{fig:ec_vs_fm_3} shows that the implied
classification thresholds are quite similar. In particular the subjects whose beliefs are consistent with Representativeness
are the ones with the steeper classification threshold in each graph and these are the Type 1 subjects that are
estimated to be about 50\% of the subject pool by both methods. The main difference is that the type 3 subjects
identified by the EC method have a classification threshold that is virtually identical to Bayes Rule, whereas there is
a clearer difference for the type 3  subjects identified by the FM method.

\begin{figure}[ht]
\caption{Accuracy and efficiency of California subjects implied by EC and FM algorithms}
\label{fig:accuracy_efficiency_california}
\centering 
\begin{minipage}[t]{0.48\textwidth}
    \centering
    \includegraphics[height=2.5in]{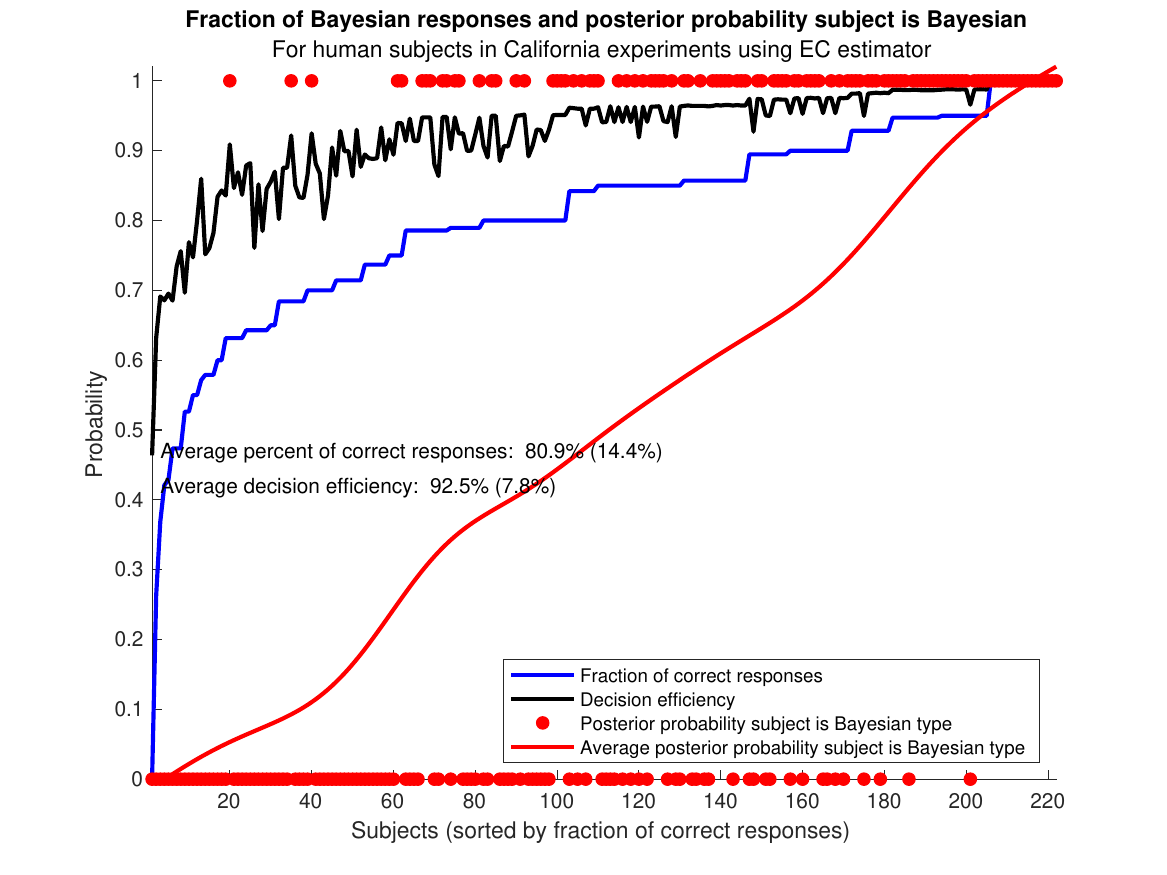}
\end{minipage}%
\hfill
\begin{minipage}[t]{0.48\textwidth}
    \centering
    \includegraphics[height=2.5in]{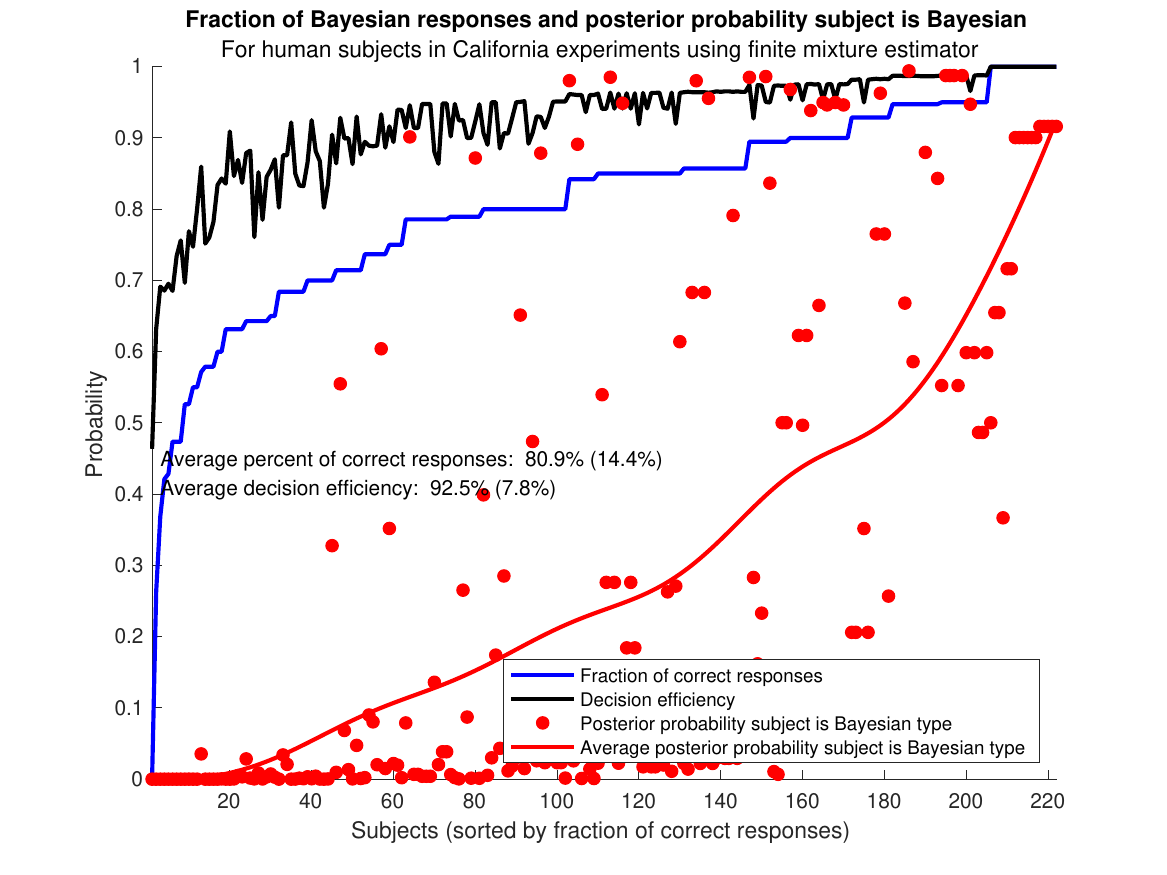}
\end{minipage}
\end{figure}

Figure~\ref{fig:accuracy_efficiency_california} plots subject-specific accuracy and efficiency scores as well as the posterior probability that the subject is the ``Bayesian type'' implied by each subject's choices and the estimated structural logit model, as estimated using the EC and finite mixture methods, respectively. The accuracy score is simply the fraction of each subject's choices that coincide with the choices of 
perfect Bayesian decision maker. The efficiency score is the sum  of expected wins in the $T_s$ trials each subject $s$ participated in to the corresponding wins for a perfect Bayesian, i.e.  the ratio $\omega_s$ given by
\begin{equation}
    \omega_s = \frac{ \sum_{t=1}^{T_s} \left[ \Pi(A|d_{ts},\pi_{ts})^{y_{ts}} + [1-\Pi(A|d_{ts},\pi_{ts})]^{(1-y_{ts})} \right] }{ \sum_{t=1}^{T_s} \left[ \Pi(A|d_{ts},\pi_{ts})^{y^*_{ts}} + [1-\Pi(A|d_{ts},\pi_{ts})]^{(1-y^*_{ts})} \right] }.
\end{equation}
where  $d_{ts}$ and $\pi_{ts}$ are the trial outcomes and priors, respectively, and $y_{ts}$ is an indicator for subject $s$'s choice of cage A in trial $t$, and $y^*_{ts}$ is the choice a perfect Bayesian would make in the same trial.

The red dots in each figure are the posterior probabilities  implied by the structural logit model that each of the 222 subjects is the ``most Bayesian type'' (i.e. type 1 for the EC algorithm and type 3 for the finite mixture model). Since the EC algorithm classifies each subject to be the most likely type, these posterior probabilities are either 0 or 1 depending on whether the most likely type of subject $s$ is type 1. In the case of the finite mixture method, we can use the estimated probabilities of each type as a ``prior probability'' of that type and use the subject-specific likelihood to compute a posterior probability for each type $\tau \in \{1,2,3\}$, denotes by $\Pi(\tau|y_s,d_s,\pi_s)$ and given by
\begin{equation}
\Pi(\tau|y_s,d_s,\pi_s,\hat\theta) = \frac{ \hat P(\tau) L(y_s,d_s,\pi_s|\tau,\hat\theta) }{ \sum_{k=1}^3 \hat P(k) L(y_s,d_s,\pi_s|k,\hat\theta) }.
\end{equation}
where $y_s$ is the sequence of choices by subject $s$ in the $T_s$ trials, and $d_s$ and $\pi_s$ are the corresponding outcomes and priors for these trials, and $\hat P(k)$ is the estimated fraction of type $k$ subjects, and $L(y_s,d_s,\pi_s|k,\hat\theta)$ is the subject-specific likelihood for subject $s$ at the estimated parameter values $\hat\theta$ assuming the subject is type $k$.

The right hand panel of figure~\ref{fig:accuracy_efficiency_california} 
 plots the posterior probabilities, for each subject, that they are the  ``most Bayesian type'' (type 3 per the discussion above). Of course these posterior probabilities are not all zero or one for the finite mixture estimator. To enable better comparison the red line in both panels plots the local average probability that the subject is the most Bayesian type and we can see that this probability is monotonically increasing in the fraction correct responses (accuracy) and is also strongly positively correlated with subject-specific decision efficiency, though variation across subjects in efficiency is not as great as the variation in accuracy. This is a reflection of the observation we made in the introduction that a subject with lower accuracy need not have significantly lower efficiency if the trials where their choices deviate from Bayes Rule are mostly the ``hard cases'' where the Bayesian posterior is close to 1/2.

\begin{figure}[ht]
\caption{Loss functions implied by EC and FM models of California experiments}
\label{fig:ec_vs_fm_4}
\centering 
\begin{minipage}[t]{0.48\textwidth}
    \centering
    \includegraphics[height=2.5in]{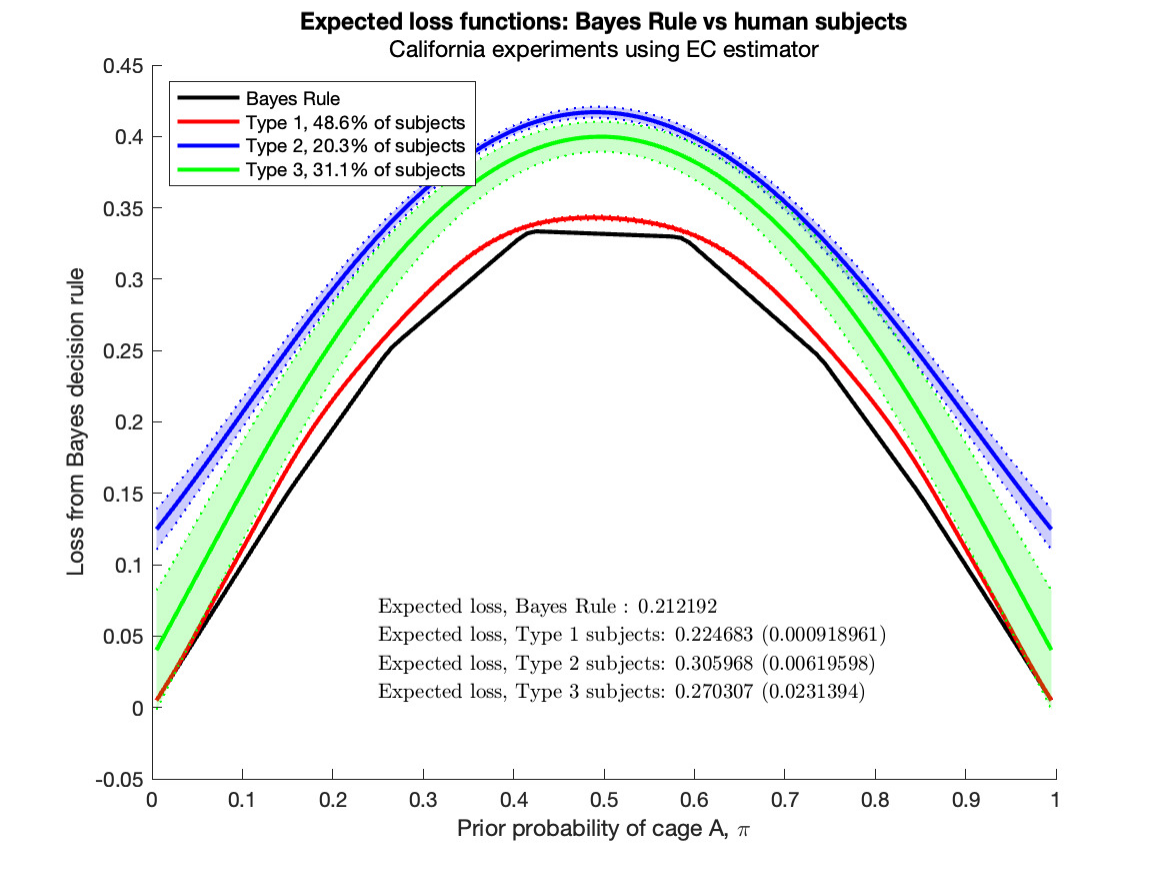}
\end{minipage}%
\hfill
\begin{minipage}[t]{0.48\textwidth}
    \centering
    \includegraphics[height=2.5in]{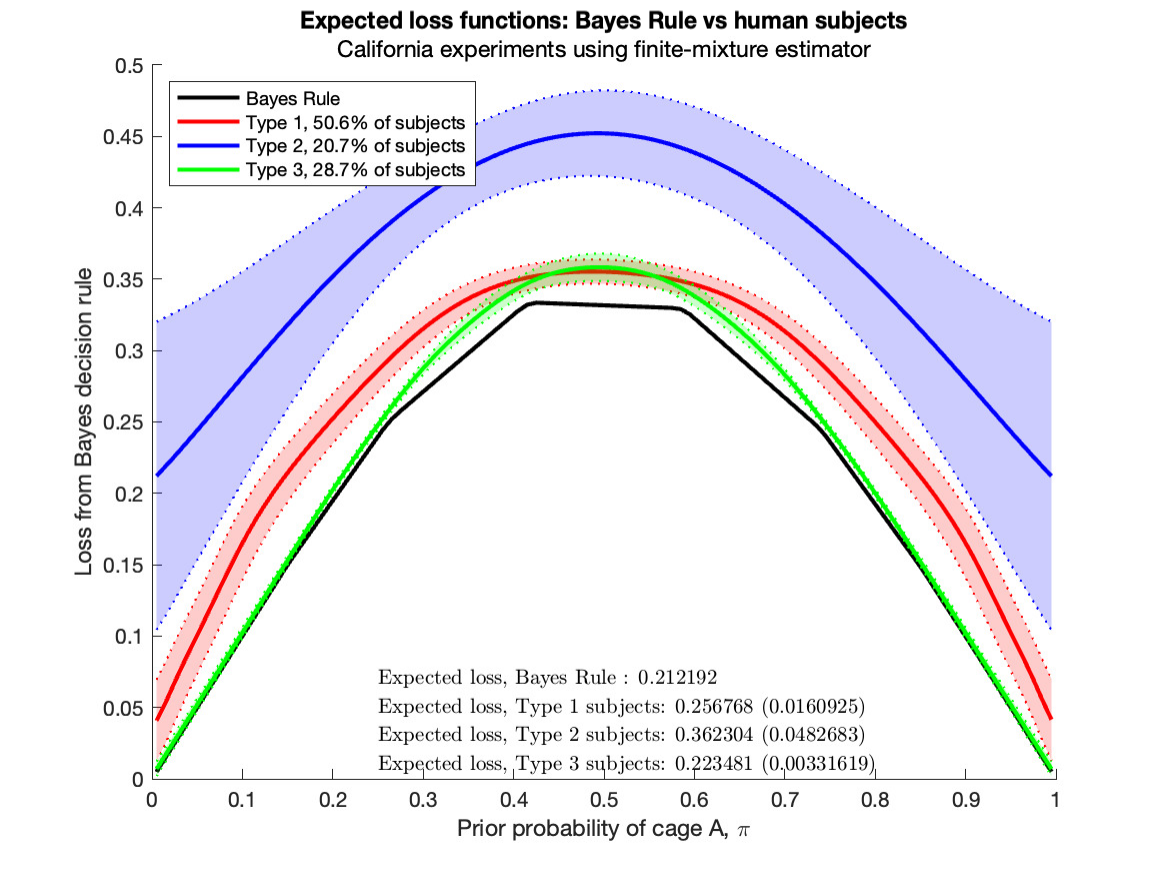}
\end{minipage}
\end{figure}

Figure~\ref{fig:ec_vs_fm_4} plots the loss functions for the three types of subjects implied by the EC and FM models
as a function of the prior $\pi$. We see that again the implied loss functions are similar, though the estimated
standard error bands are larger for the loss functions estimated by FM, especially for the type 2 subjects (blue line, the
noisier subset of noisy Bayesians).  The EC method predicts that the type 1 subjects have the lowest loss function,
whereas the FM method predicts that the type 3 subjects have the lowest loss.

We calculated expected win probabilities using
the empirical distribution of $\pi$ for all three types. For these experiments a Bayesian
would, on average, predict the correct cage with probability 70\%. According to the EC algorithm type 1 subjects
predict the correct cage with an average probability of 68\%, so the estimated average decision efficiency score is $\omega_p=97.5\%$
(standard error, $0.7\%$). The FM algorithm predicts that the less noisy Bayesians (Type 3 subjects) are the most efficient
decision makers, with a decision efficiency score of 94.7\% (1.5\%). Both the EC and FM algorithms predict that the least efficient
decision makers are the type 2 subjects with decision efficiencies of 87\% and 82\%, respectively.
Overall, for all three types of California subjects, both the  EC and FM algorithms predict a decision efficiency score of 
93\% with estimated standard errors of  0.7\% and 1.2\%, respectively.
Wald tests strongly reject the hypothesis that subjects are fully efficient decision makers either individually
for each type or overall as a group.

\subsection{Reanalysis of Wisconsin Experiments}\label{subsection:reanalysis_human_wisconsin}

Now we reanalyze  data from the ``Wisconsin experiments'' conducted
by \citet{EGG1999}. They recruited  79 student subjects from the University of Wisconsin-Madison
and employed a two stage experimental design to test for {\it context effects\/} by altering
the ``California design'' used in \citet{EGG1995} (where $D=6$ and $p_A=2/3$ and $p_B=1/2$) 
to a new ``Wisconsin design'' ($D=7$, $p_A=.4$ and $p_B=.6$).  The experiments were conducted
on two successive days. On the first day approximately half the subjects began with the 6 ball California
design and then switched to the 7 ball Wisconsin design on the second day, whereas for the other half of subjects
this order was reversed to see if the ordering of the designs affects subjects' choices.

The motivation for this design
is twofold: 1) by drawing an odd number of balls, they made it harder for subjects to rely
on the ``representativeness heuristic'' in their choice of cage A or B, and 2) by reversing the proportions of
balls marked N they reversed the classification rules. That is, in the 7 ball Wisconsin design larger values of $d$
constitute stronger evidence for cage B, not cage A as was the case in the California design. These design changes
could potentially change subject behavior, whereas they have no effect on the optimal Bayesian decision rule
which is a function of $\mbox{LLR}(d,p_A,p_B,D)$ and $\mbox{LPR}(\pi)$. In fact, the structural logit
model predicts that the design change is fully accounted
for in the likelihood ratio term entering Bayes Rule. It follows that even if the subject is not Bayesian
(e.g. the $\beta$ coefficients do not equal the values $\beta=(0,1,1)$), as long as the structural coefficients are 
unaffected by the change in experimental design, then so is the implied behavior of these subjects.

From our perspective, the Wisconsin experiments provide an opportunity to treat the switch from the 6 ball California
design to the 7 ball Wisconsin design as a {\it policy change\/} to study how well the structural logit model
is able to predict subjects' choices under a different experimental design (data generating mechanism). To do this we estimate a multi-type
version of the structural logit model for all 79 subjects using the 6 ball California design trials as the
{\it training sample\/} and use the 7 ball Wisconsin design trials as the {\it evaluation sample.\/} This approach
also allows us to conduct a likelihood ratio test for {\it structural stability,\/} i.e. to test whether the coefficients
of the structural logit model are invariant across the two experimental designs.
As we noted in the introduction, \citet{EGG1999} concluded (using their structural
probit model) that subjects did appear to use different decision rules in the 6 ball and 7 ball designs. Evidently  not
being able to rely as easily on the representativeness heuristic lead a greater fraction of subjects to use
decision rules that more closely approximate a noisy version of Bayes rule in the 7 ball trials.

\begin{figure}[ht]
\caption{Data and classification thresholds in the Wisconsin experiments}
\label{fig:wisconsin_experiments}
\centering 
\begin{minipage}[t]{0.48\textwidth} 
    \centering
    \includegraphics[height=2.5in]{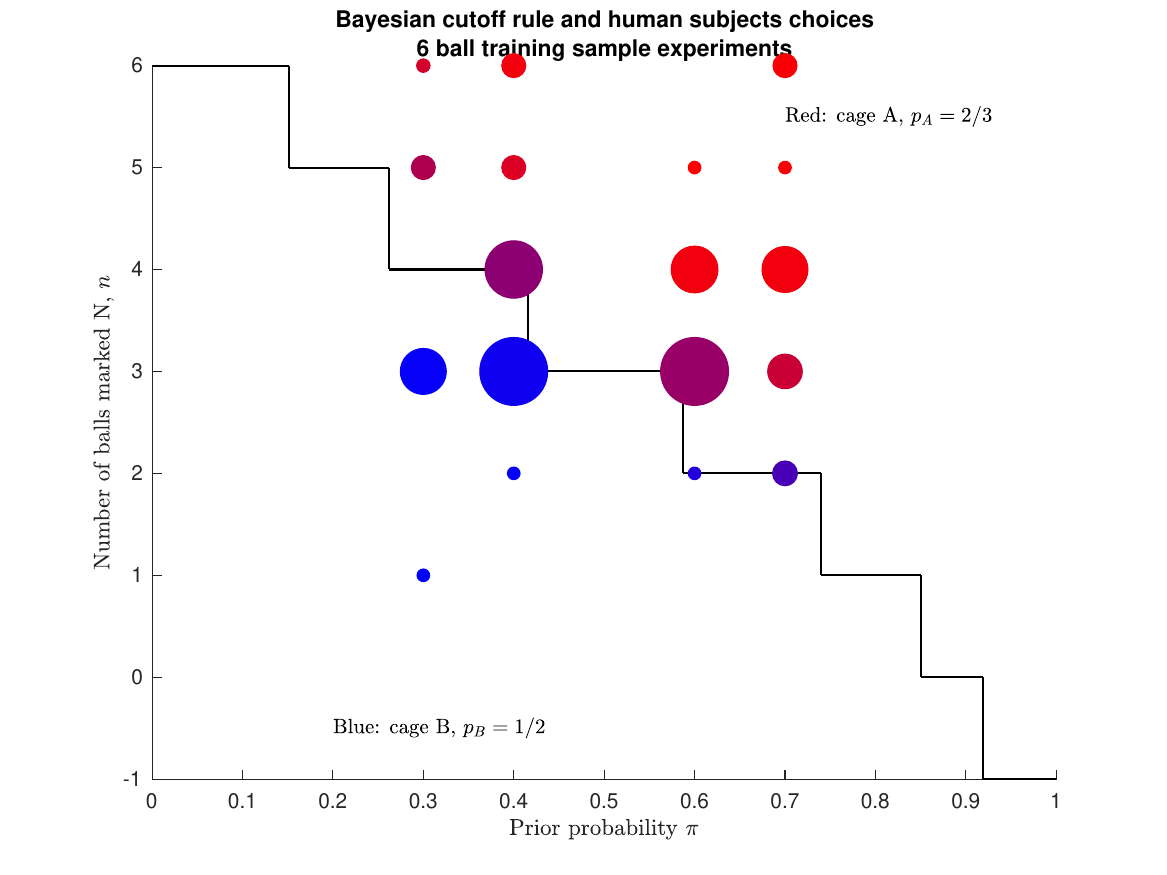}
\end{minipage}%
\hfill 
\begin{minipage}[t]{0.48\textwidth}
    \centering
    \includegraphics[height=2.5in]{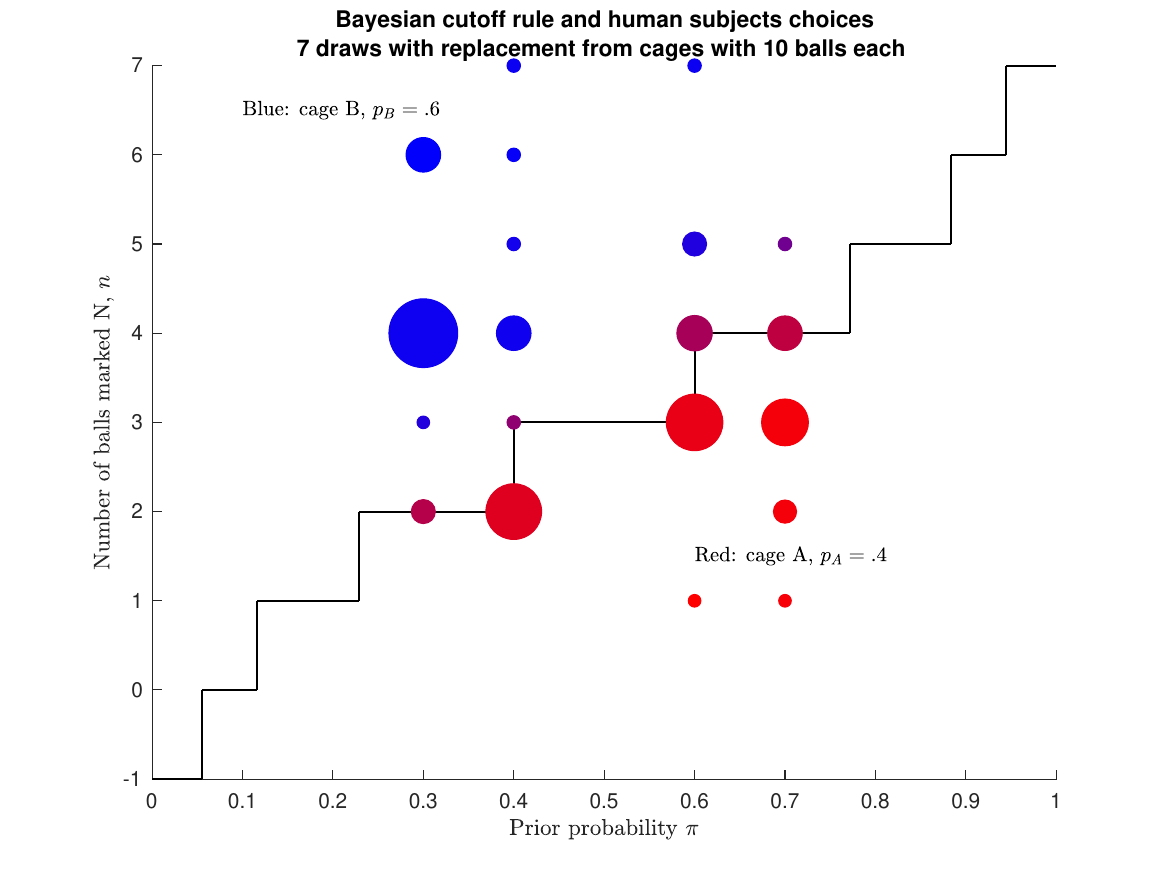}
\end{minipage}
\end{figure}

Figure~\ref{fig:wisconsin_experiments} summarizes the data on subject choices in the 6 and 7 ball trials.
The size of each circle is proportional to the number of observations and the color
of the circles represents the share of subjects choosing cage A or B. Thus, deep blue circles are those where
nearly all of the subjects chose cage B, deep red circles are those where nearly all the subjects chose cage A,
and the purple colored circles are those where there was a mix of subjects choosing cage A and B, where we  mixed
red and blue in proportion to the fraction of subjects choosing A and B.

We plot the cutoffs implied by  Bayes Rule as a function of the prior probability $\pi$ to make
it easier to assess whether subjects' responses are consistent with the optimal decision rule: at points on one side of the cutoff line cage A is optimal and on the other side cage B is the optimal choice.  We see how the change in experimental design changed the threshold from downward sloping in $\pi$ in the 6 ball design  to upward sloping in the 7 ball design. It is evident that choices are on
average consistent with Bayes Rule.  Subjects become ``confused'' (indicated by purple circles) for trial outcomes near the classification threshold or ``indifference curve.''  The circles further
away from it are colored deep red or blue, so these are the easy cases where subject choices largely accord with Bayes Rule. It seems there is less confusion in the 7 ball design but this may be an artifact of fewer observations laying on near the classification threshold compared to the 6 ball design.

Table~\ref{tab:sl_wisconsin_ml_estimates} presents the maximum likelihood estimates from the FM algorithm
for the 4 parameter specification of the structural logit model where $\eta=0$.\footnote{\footnotesize Estimation results
from the EC algorithm result in similar overall conclusions.}
We stopped the FM algorithm when it reached
$K=3$ types even though the likelihood ratio test indicated that $K=4$ types is a significantly better fit because
with only 79 subjects, there would be too few subjects assigned to some of the types in the 4 type specification, raising
the possibility of model overfitting and less reliable inferences. 

\begin{table}[htbp]
\caption{FM estimates of the structural logit model for 6 ball experiments, $LL=-654.25$}                                         
\begin{center} 
\label{tab:sl_wisconsin_ml_estimates}
\fontsize{11pt}{11}\selectfont
        \setlength{\tabcolsep}{2pt}
        \renewcommand{\arraystretch}{1.15}
                \begin{tabular}{|l|c|c|c|} 
\hline
Parameter (std error) & Type 1 ($\hat\lambda_1=.32$)  & Type 2 ($\hat\lambda_2=.24$) & Type 3 ($\hat\lambda_3=.44$)  \\
\hline
$\sigma$ (noise parameter) & .07         & .22   & .32 \\
                           &       (.08) & (.06) & (.05) \\
\hline
$\beta_0$ (bias/intercept) & .14        & .06  & -.11 \\
                           & (.18)       & (.11) & (.08) \\
\hline
$\beta_1$ ($\mbox{LLR}(n)$ coefficient) & 1.48 & 1.25 & 2.48 \\
                                     & (1.91)& (.36) & (.67) \\
\hline
$\beta_2$ ($\mbox{LPR}(\pi)$ coefficient) & 1.40 & 2.27 & 1.39 \\
                                     & (1.81)& (.70) & (.36) \\
\hline
$P$-value for $H_o: \beta_0=0,\beta_1=\beta_2$ & .74 & .02 & .01 \\
\hline
$P$-value for $H_o: \beta_0=0,\beta_1=\beta_2=1$ & .015 & .003 & .015 \\
\hline
\end{tabular}%
\end{center}
\end{table}

Of the three types, type 1 is the most ``Bayesian'' and also the least ``noisy''. The estimated noise parameter
$\sigma$ is less than a third of the value estimated for the other two types.  There is no significant overall
bias in the beliefs of any of the three types as evidenced by the fact that all three estimates of the 
bias term $\beta_0$ in subjective posterior beliefs in equation~(\ref{eq:subjectiveposterior})
is insignificantly different from zero. The next to the last row in the table presents the P values
of a Wald hypothesis test that $\beta_0=0$ and $\beta_1=\beta_2$, and we see that there is no evidence against
this for type 1 subjects but the hypothesis is rejected for type 2 and type 3 subjects. 
The type 2 subjects are those whose beliefs are consistent with the Representativeness
heuristic, i.e. they  put less weight on the prior information via $\mbox{LPR}$ relative to the sample information
via  $\mbox{LLR}$. The type 3 ``conservative subjects'' do the opposite.

The last row of the table shows that we can strongly reject the hypothesis that any of the three types are ``noisy Bayesians'' i.e.
who satisfy the hypothesis $H_o: \beta_0=0, \beta_1=1, \beta_2=1$,  though with possible noise affecting their choices, $\sigma > 0$. 
However  because the type 1 subjects satisfy the restriction that $\beta_0=0$ and $\beta_1=\beta_2$, their implied decision rule
is nearly optimal, as we show below. So at least in terms of their decision rule or {\it choice behavior\/} the type 1 subjects
are close to optimal  even if their subjective posterior beliefs may be a distorted version of true Bayesian beliefs.

\begin{figure}[ht]
\caption{Subjective vs Bayesian Posterior Beliefs, 6 ball experiments}
\label{fig:posterior_beliefs}
\centering 
\begin{minipage}[t]{0.32\textwidth} 
    \centering
    \includegraphics[width=\linewidth]{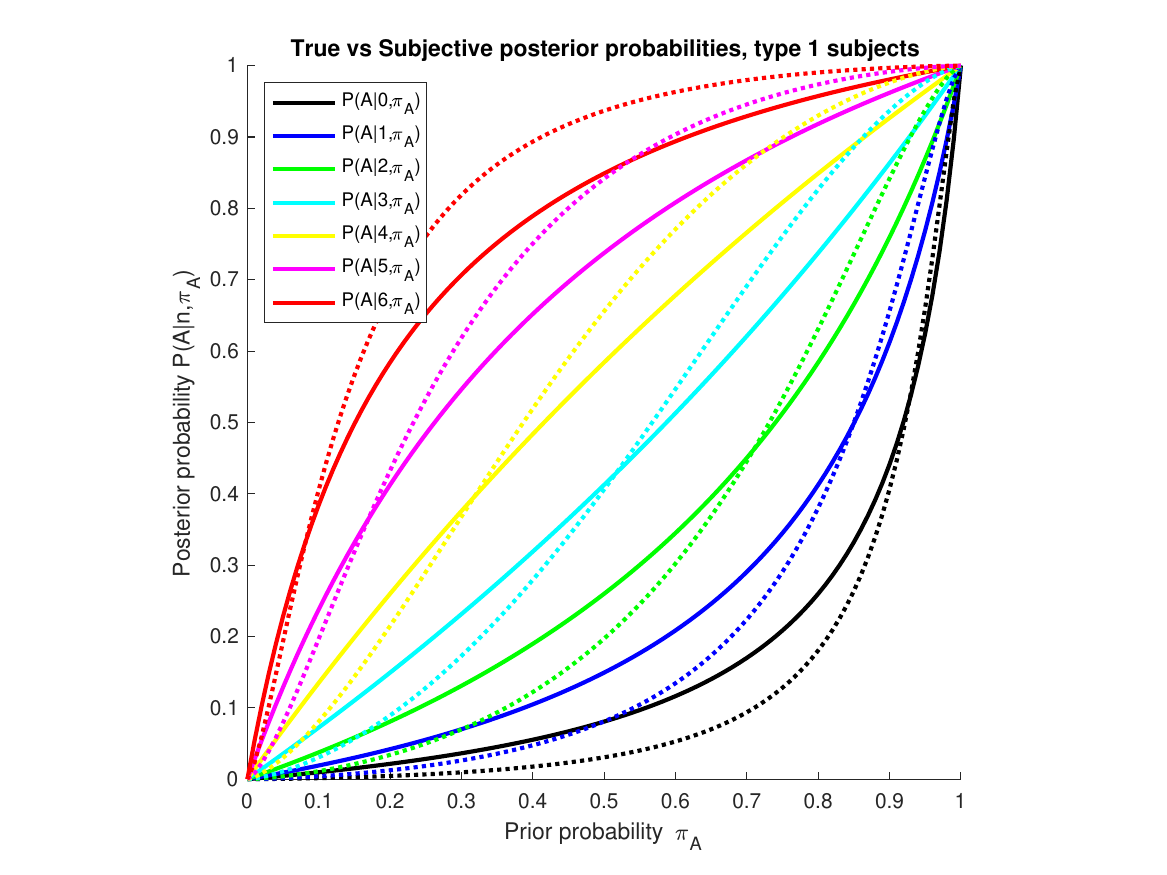}
\end{minipage}%
\hfill
\begin{minipage}[t]{0.32\textwidth}
    \centering
    \includegraphics[width=\linewidth]{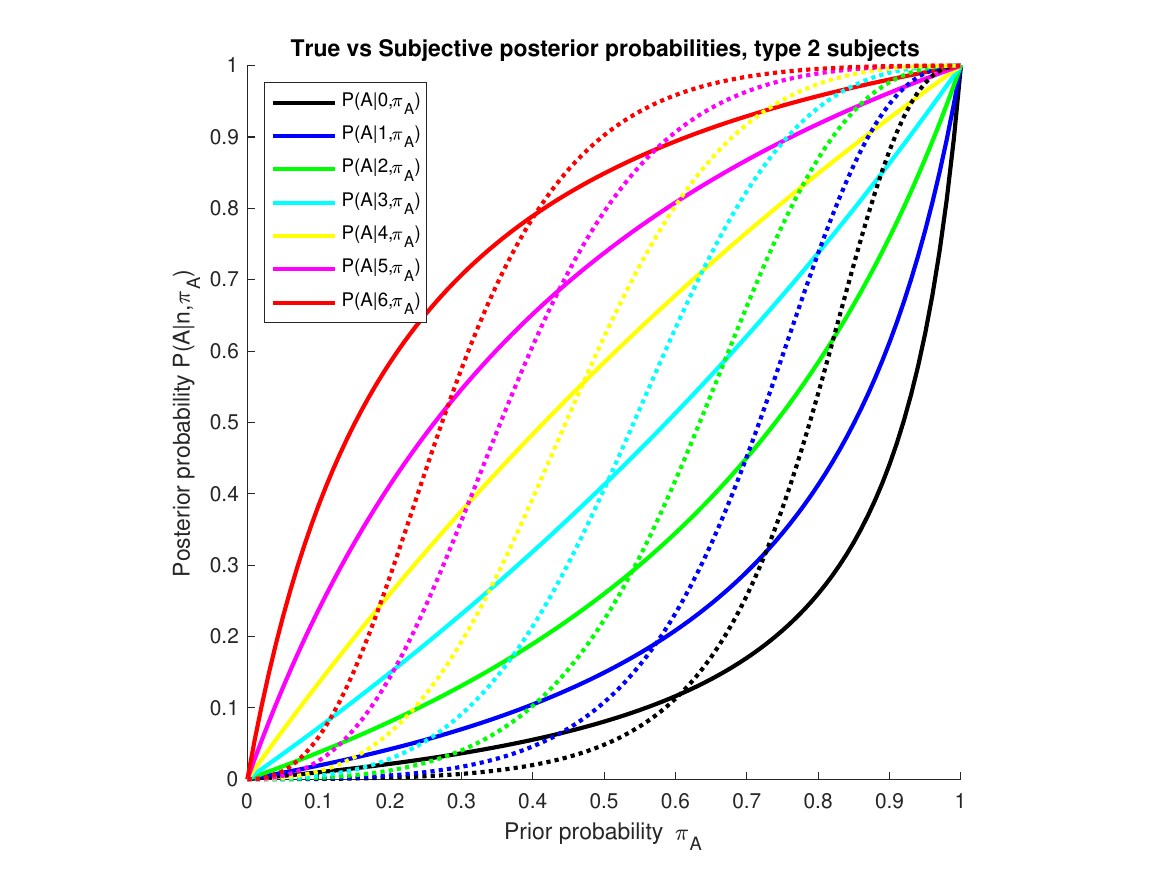}
\end{minipage}%
\hfill
\begin{minipage}[t]{0.32\textwidth}
    \centering
    \includegraphics[width=\linewidth]{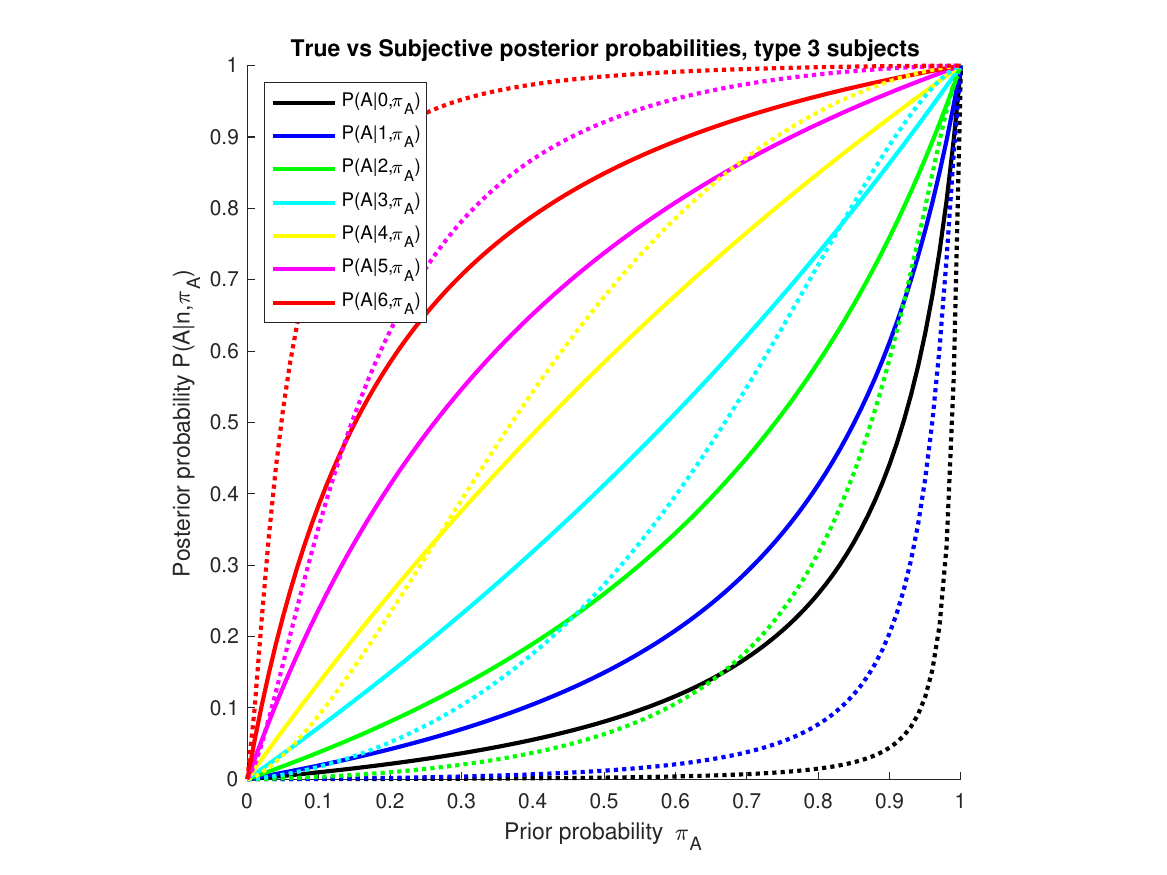}
\end{minipage}
\end{figure}

Figure~\ref{fig:posterior_beliefs} plots the estimated subjective posterior beliefs for the three types of subjects (dotted lines)
and compares them to the true Bayesian beliefs (solid lines) for all possible combinations $(\pi,d)$ where $\pi \in [0,1]$ and
$d \in \{0,1,\ldots,6\}$. 
It is visually apparent that the subjective posterior beliefs of the Type 1 subjects are closest
to Bayesian, whereas there are obvious distortions in the estimated posterior beliefs of the Type 2 and 3 
subjects. Type 2 subjects display a pattern of {\it overconfidence\/} overestimating the probability
of A relative to Bayes Rule for sufficiently high values of $\pi$ and under-estimating it for sufficiently low
values of $\pi$. Type 3 subjects display a different type of overconfidence. For high values of $d$
($d \in \{0,1,2,3\}$) these subjects' subjective beliefs systematically under-estimate the probability
of A relative to Bayes Rule for {\it all\/} values of $\pi$, but for high values of $d$ ($d \in \{5,6\}$) this reverses and they systematically
over-estimate the probability of A relative to Bayes Rule, for all $\pi \in [0,1]$. 

The distortions in subjective posterior beliefs imply suboptimal choices even
in the absence of ``noise'' in their ultimate decisions. If $\sigma$ were zero, the optimal decision rule
can be described by a cutoff rule $\overline\pi(d)$ that differs for each realized value of $d$.
These cutoffs occur at the intersections of the solid or
dashed curves and a horizontal line at a probability value of $1/2$. It is  optimal for the subject
who observes information $(\pi,d)$ to choose cage B if $\pi < \overline\pi(d)$ and cage A if $\pi \ge \overline\pi(d)$.
From this it is evident that the Type 1 subjects have the least distortion in the implied optimal (noise-free)
decision rules, whereas the Type 3 subjects have the most distorted decision rule. We show below that
this implies a significantly higher loss for type 3 subjects relative to type 1 subjects.

\begin{figure}[ht]
\caption{Fraction of Correct Responses}
\label{fig:fraction_of_optimal_responses}
\centering 
\includegraphics[height=3.0in]{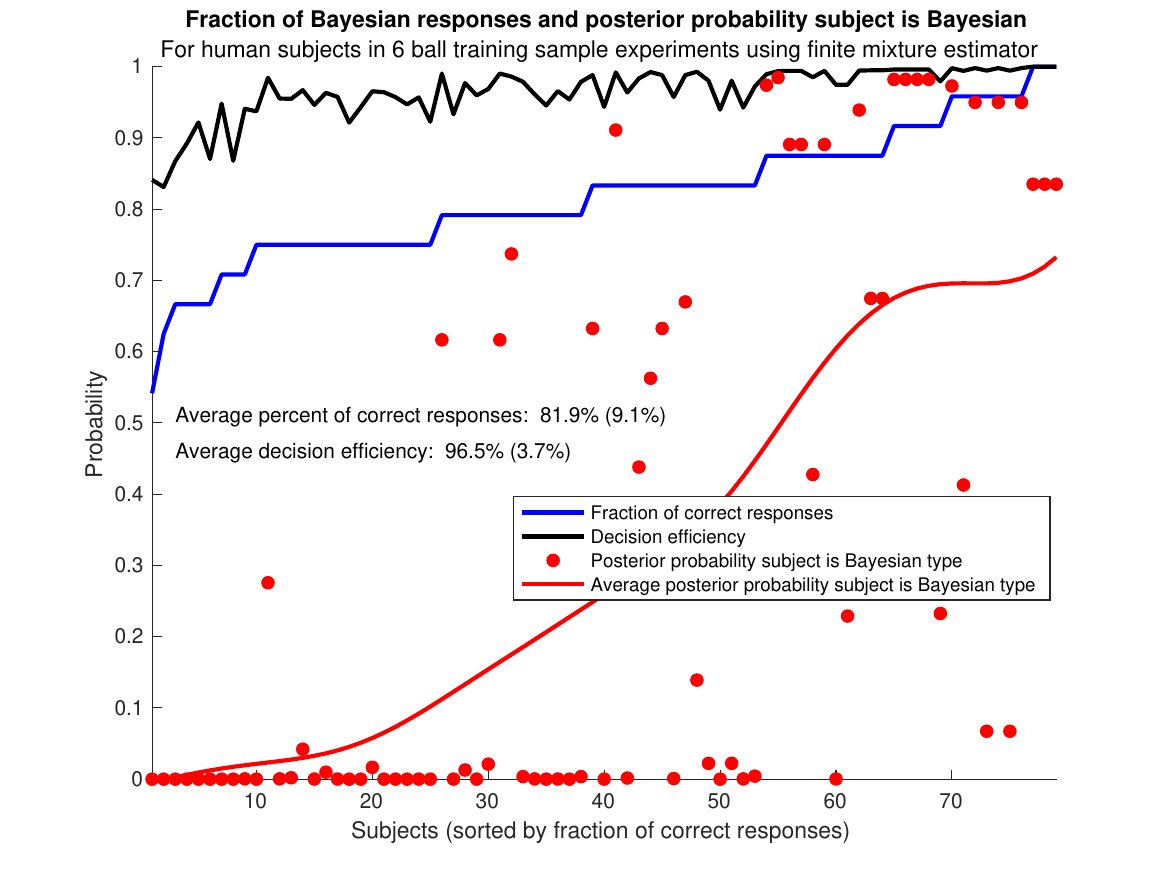}
\end{figure}

Before showing this, we present figure~\ref{fig:fraction_of_optimal_responses}. The blue curve in this figure 
plots the fraction of ``correct'' responses for each of the 79 subjects, i.e. the fraction
of their responses that coincide with the optimal choice of a Bayesian decision maker given
the same information $(\pi_t,d_t)$ in each trial $t$. We see that the fraction of correct responses ranges from a
low of 54\% to a high of 100\% with an average of over 81\% correct responses. The black and red curves
help relate this simpler ``percent correct'' metric of performance to Bayesian behavior. The black curve
plots subject-specific decision efficiencies, i.e. the ratio of the sum of each subject's expected win probabilities
in all trials to the corresponding total for a perfect Bayesian decision maker. We see that average efficiency is
quite high, 96.5\%, which exceeds the average of 81.9\%  correct responses of these subjects. This is due to the
fact that most of the ``mistakes'' that these subjects made were on the hard cases where the true posterior probability
was close to 1/2, and thus, mistakes for these cases are not very costly for these subjects.

The red curve in figure~\ref{fig:fraction_of_optimal_responses} plots the posterior probability (implied by the FM estimates) 
that each subject is a type 1 ``noisy Bayesian''
and the solid red curve is a local linear regression showing the local average fraction
of noisy Bayesians ordered by percent correctly answered.  We see a strong positive
relation between ``correctness'' and the probability of being a Type 1 subject, ranging from a low of 0\% for 
subjects who only answered 54\% of the trials correctly to a 
high of 70\% for subjects who answered 100\% correctly.\footnote{\footnotesize 
Of course, it is possible to get 100\%  due to unobservable random factors affecting
subjects' responses and this is why the posterior
probability of subjects who answered 100\% correctly is not 1. In fact, the posterior probability
of being a Type 1 subject can be rather low even if the subject answered 100\% correctly depending on the particular
sequence of trial values $\{(\pi_t,d_t)\}$ the subject is asked to classify. If most of these trials correspond
to the hard cases true posterior values close to 1/2, i.e. $\Pi(A|\pi_t,d_t) \simeq 1/2$, then it is harder to
distinguish whether the subject is a noisy Bayesian or some other type. This is reflected in a higher  posterior  probability that
the subject could have been a type 2 or 3 subject.}

\begin{figure}[ht]
\caption{Classification Thresholds and Expected Loss Functions, 6 ball experiments}
\label{fig:expected_loss_functions}
\centering 
\begin{minipage}[t]{0.48\textwidth} 
    \centering
    \includegraphics[height=2.5in]{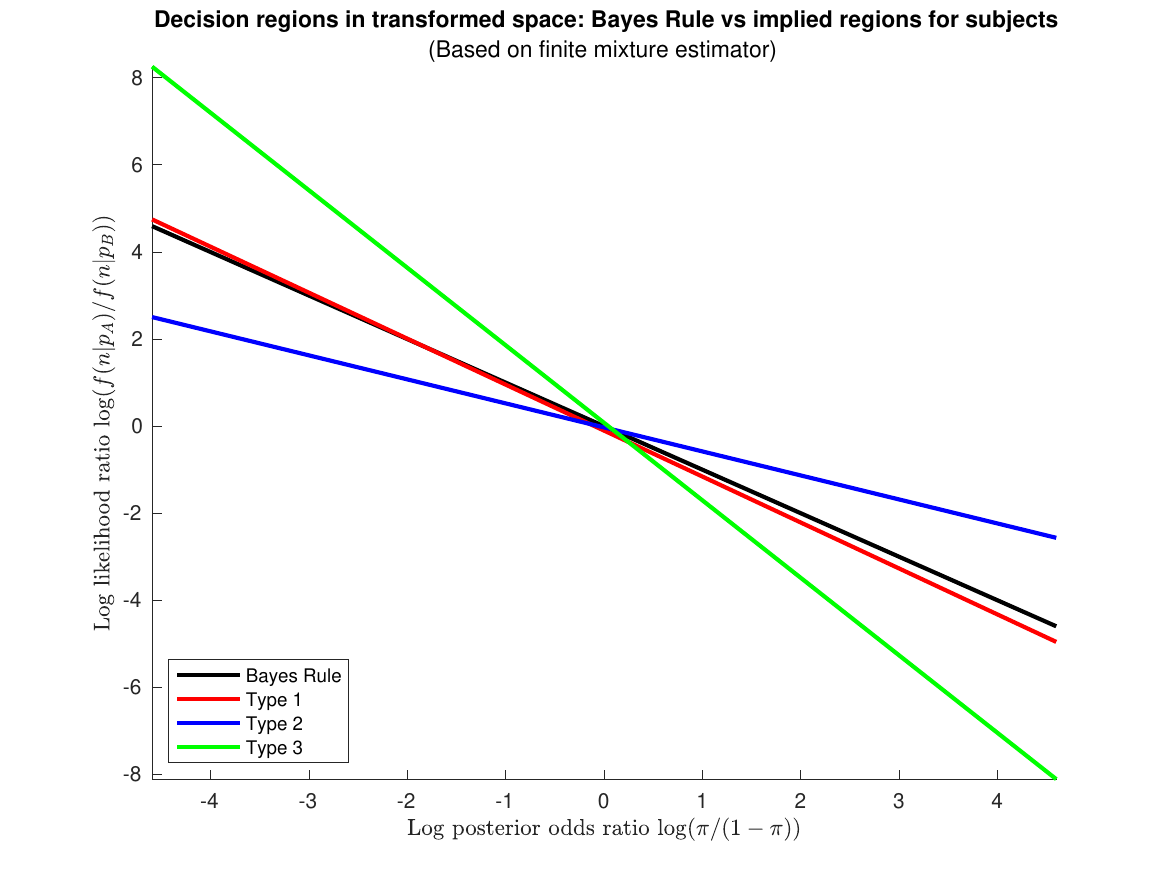}
\end{minipage}%
\hfill 
\begin{minipage}[t]{0.48\textwidth}
    \centering
    \includegraphics[height=2.5in]{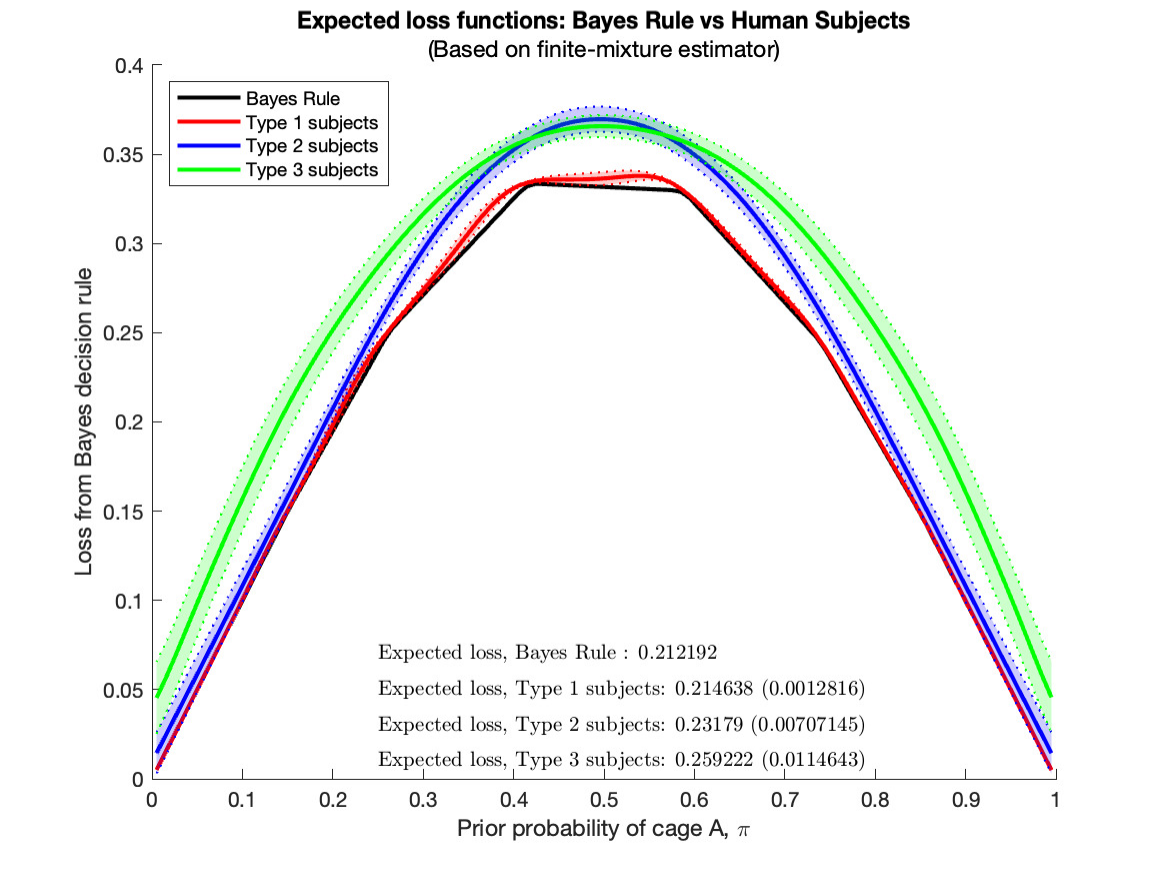}
\end{minipage}
\end{figure}

Figure~\ref{fig:expected_loss_functions} plots the classification thresholds (hyperplanes) for each of the three
types of subjects along with the prior-specific expected loss functions $L_P(\pi)$, see equation (\ref{eq:lossdef}).
Recall that along these lines the subject believes cages A and B are equally likely, so they each demarcate the 
decision regions each type of subject would make their classification decisions in the absence of noise (i.e. when $\sigma=0$).  Here it is
obvious that  the decision hyperplane of the Type 1 subjects essentially coincides with the Bayesian
classification threshold, but the hyperplane for the Type 2 subjects is significantly less steep
and the hyperplane for the Type 3 subjects is significantly more steep than the classification hyperplane
of a Bayesian. This allows us to readily assess the regions where subjects will make classification errors
relative to what an optimal Bayesian decision maker would choose. For example in the triangular region
to the left of $\mbox{LPR}(\pi)=0$ below the red and black lines and above the blue line, for values
of $(\mbox{LLR},\mbox{LPR})$ in this region a type 2 subject chooses cage A but a Bayesian (or type 1 subject)
chooses cage B. Similarly the triangular region below the green line and above the red and black lines
to the left of $\mbox{LPR}(\pi)=0$, represents misclassifications by Type 3 subjects: they 
choose cage B whereas a Bayesian (or type 1 subject) chooses cage A. 

The right hand panel of figure~(\ref{fig:expected_loss_functions}) plots the expected conditional loss functions $L_P(\pi)$
for the three types of subjects. The black curve plots the optimal Bayesian loss function $L_{\delta^*}(\pi)$ 
that necessarily minorizes the loss functions for the
human subjects by Lemma~\ref{lemma:bayes_optimality}. We see that the red expected loss curve for the type 1 subjects nearly coincides
with the black Bayesian loss function, indicating that the Type 1 subjects are nearly optimal Bayesian decision makers,
even if their subjective posterior beliefs are somewhat distorted relative to Bayes Rule.  However the green and blue curves
are significantly higher than the black and red expected loss curves, showing that the Type 2 and 3 subjects are using 
distinctly suboptimal decision rules. 

The most suboptimal decision makers are the type 3 subjects whose $L_P(\pi)$ loss
function majorizes the other three expected loss functions. Note that the other loss functions are close to zero
at values of $\pi$ equal to 0 or 1, whereas the green curve is significantly positive at those values similar to
what we found for the poorest performing types of California subjects. This is a reflection
of the high amount of ``decision noise'' $\sigma$ for the type 3 subjects that we noted in our discussion of the
estimation results in table~\ref{tab:sl_wisconsin_ml_estimates}. For example the model predicts
that a type 3 subject will choose cage A about 
5\% of the time even when $\pi=0$ when there is no chance that cage A was used to draw the sample, and conversely for
values of $\pi$ near 1.

We also calculated expected win probabilities using
the empirical distribution of $\pi$ for all three types. The EC algorithm predicts that the type 1 (less noisy
Bayesians) have a  decision efficiency score of 99.5\% (standard error 0.2\%), whereas type 2 (Conservative) subjects have
a decision efficiency of 96.2\% (standard error $0.9$\%), and type 3 (Representativeness) subjects have an efficiency
index of 94.5\% ($0.1$\%). While we can strongly reject the hypothesis that any of the three types is a fully
optimal decision maker, the overall efficiency of the Wisconsin subjects is quite high, 96.5\% (0.5\%) which 
is higher than the 93\% overall efficiency score for the California subjects.\footnote{\footnotesize Overall efficiency
of the Wisconsin subjects in the 7 ball design was 95.5\% (0.6\%).}

\begin{figure}[ht]
\caption{Predicted vs Actual CCPs in Training and Evaluation Samples}
\label{fig:predicted_vs_actual}
\centering 
\begin{minipage}[t]{0.48\textwidth} 
    \centering
    \includegraphics[height=2.5in]{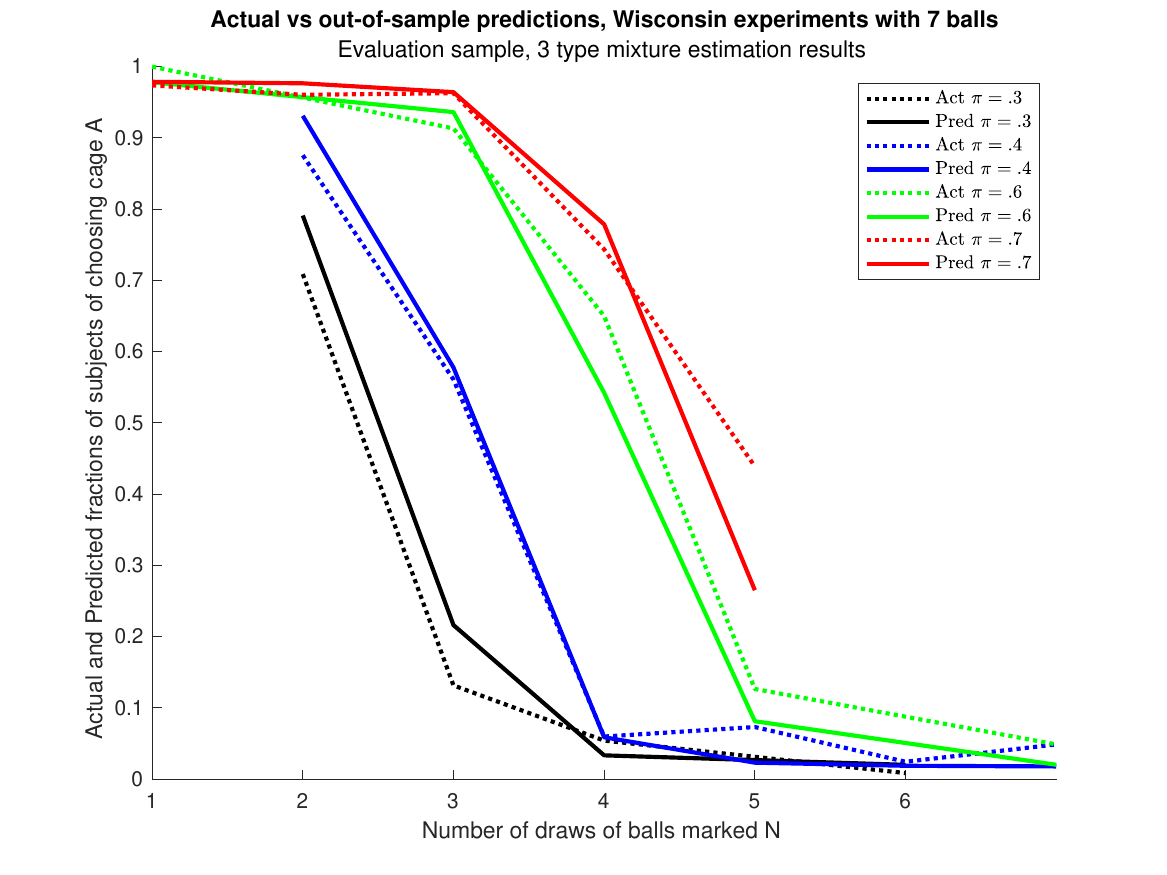}
\end{minipage}%
\hfill 
\begin{minipage}[t]{0.48\textwidth}
    \centering
    \includegraphics[height=2.5in]{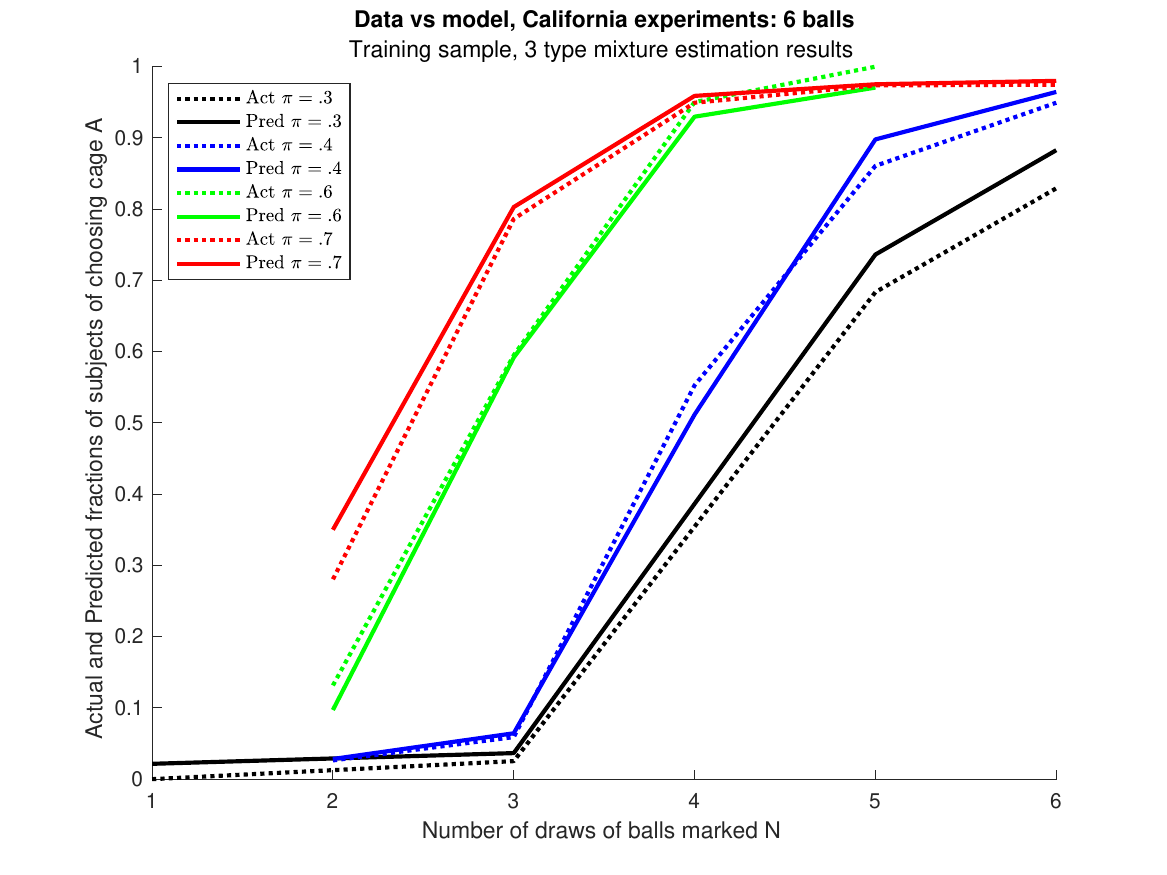}
\end{minipage}
\end{figure}

Finally, we consider how well the structural model can predict a ``policy change'' i.e. a change in the
experimental design from the 6 ball California design to the 7 ball Wisconsin design. We illustrate this in figure~\ref{fig:predicted_vs_actual}.
The left panel of the figure shows the in-sample fit, i.e. it compares the predicted CCPs  from the structural
model to the non-parametric estimates of the CCPs for subjects in the 6 ball ``training sample''. Note that we incorporated
the subject-specific unobserved heterogeneity in the model-predicted CCPs in this figure: that is, we  computed the 
structural logit CCPs by averaging the subject-specific CCPs where each subject was assigned their most likely
type from the output of the EC estimation algorithm.
We plot separate lines for the 4 separate priors used in these experiments, $\pi \in \{.3,.4,.6,.7\}$ where the dashed
lines plot the non-parametric CCPs as a function of $n$ and the solid lines are the predictions of the structural logit model.

The right hand panel illustrates the ability of the structural logit model to predict how subject behavior
changes in the $D=7$ ball evaluation sample.  The 4 curves are now {\it downward sloping in $d$\/} and of course
this is due to a change in binomial parameters from $(p_A,p_B)=(2/3,1/2)$ in the California
design to $(p_A,p_B)=(.4,.6)$ under the Wisconsin design. 
We see that the structural logit model provides fairly accurate predictions of the dramatic shift in subjects'
decision rules under the two experimental designs, suggesting that the structural logit model provides a good approximation
to subject behavior.

However there is a more subtle aspect in which subjects' decision rules might have changed  that constitutes
evidence against the ``structural stability'' of the structural logit model. That is, if the structural logit
model was a completely correct model of subject behavior, then if we were to estimate the parameters of the
structural logit model using the 7 ball evaluation sample, the structural parameters $\theta$ should not be statistically significantly
different  from the parameters estimated in the 6 ball training sample.  However 
we find that when we do use the 7 ball experiments as the estimation or
training sample,  there is a significant change in the structural parameter estimates: a smaller fraction, 23\%, 
of subjects  are classified as ``noisy Bayesians''. The noisy Bayesians also exhibit significantly
greater noise in their responses compared to the type 1 subjects identified
using the 6 ball training sample, with $\hat\sigma=.43$.   
Further, we find that type 1 and 3 subjects have relatively low inefficiency but type 2 subjects exhibit
high inefficiency due to the large values of $\sigma$ noted above. In the 7 ball data, we find that the inefficient, high noise
subjects constitute 23\% of the sample, whereas the other 77\% of subjects  have higher efficiency due to
lower estimated values of $\sigma$. 

Overall the fraction of subjects using ``efficient'' decision rules increased from 32\% 
in the 6 ball training sample to 77\% when we use the 7 ball experiments as the training sample.  
A likelihood ratio test strongly rejects the hypothesis that subjects were using the same decision rules
in both the 6 ball and 7 ball experimental designs: P-value $.002$.\footnote{\footnotesize When we estimated the
3 type structural model via the FM method using the 221 California subjects  we obtained roughly similar results as we obtained
for the Wisconsin subjects when we estimated their parameters for the same 6 ball California design. We found
that 21\% of the California subjects were ``noisy Bayesians'', 50\% overweight $\mbox{LLR}$ and hence show
behavior consistent with the representativeness heuristics, and 29\% overweight $\mbox{LPR}$ and hence
exhibit behavior consistent with conservatism. However the estimated $\sigma$ parameter for the noisy Bayesians 
was $\sigma=.74$ and this high level of noise made these subjects the most suboptimally
behaving subgroup among the California subjects.} However the average inefficiency of all Wisconsin student
subjects under the 7 ball design is 10.7\% (standard error, 1.4\%), which is significantly larger than
the 8.1\% average inefficiency of all subjects for the 6 ball design. Apparently the switch to the
7 ball design reduced the ability of subjects to rely on the representativeness heuristic. 
This may have induced more subjects to expend more mental effort resulting in more  efficient
decision rules. Despite this, the 23\% of the sample who continued to use the representativeness heuristic
also had significant noise affecting their responses, so the significantly higher inefficiency among this
group outweighed the improvement in efficiency among the other subjects, causing overall inefficiency to
increase in the 7 ball Wisconsin design.

\subsection{Reanalysis of Holt and Smith Experiments}
\label{section:holtsmith}

The identification problem for subjective beliefs using only binary choice data
discussed in section~\ref{section:identification} suggests the need for caution in drawing
conclusions about the fraction of subjects who have subjective posterior beliefs that are well approximated by Bayes Rule,
though we can be confident our inferences on overall inefficiency of human subjects since this
measure is based on the CCP which is non-parametrically identified. 

In this section we reanalyze experiments reported in \citet{HS2009} that directly elicited
subjective posterior beliefs. This was done via the {\it Becker-DeGroot-Marshak\/} (BDM) mechanism
which incentivizes rational subjects to truthfully report their
subjective posterior probabilities. They conducted two separate experiments: one at Holt's laboratory at the
University of Virginia involving 22 subjects, and a second one done via the Internet involving 30 subjects.
In both experiments the design parameters involved fixing $p_A=2/3$ and $p_B=1/3$ but varying the number of
draws $D$ and the priors across multiple trials for each subject.  $D$ took values from $\{0,1,2,3,4\}$ and $\pi$ varied
from $\{1/3,1/2,2/3\}$. Here we focus on the reanalysis of the first experiment with 22 subjects and, given space constraints, summarize the key findings from our analysis of their second web-based experiment in a footnote.

The BDM mechanism was implemented as follows: after seeing 
the prior $\pi$ and the result of the random drawing  $d$ from the selected cage/cup, subjects were 
asked to report a probability $p_r \in [0,1]$ that determines a payoff they would receive in a second stage 
gamble. This gamble, denoted by $\tilde G_R$, involves drawing a random probability $\tilde p \sim U(0,1)$ and paying the subject a
monetary reward of $R$ according to the following rule: if $\tilde p < p_r$  the subject receives
$R$ if the observed sample was drawn from cup A, otherwise if $\tilde p \ge p_r$ the  subject receives
$R$ with probability $\tilde p$. It is not hard to show that the subject's expected payoff from reporting $p_r$ in this
second stage BDM lottery is 
\begin{equation}
                 E\{\tilde G_R|p_r,d,\pi,p_A,p_B,D\}=R\left[\frac{1-p_r^2}{2} + p_r \Pi_s(A|d,\pi,p_A,p_B,D)\right],
\label{eq:bdmpayoff}
\end{equation}
where $\Pi_s(A|d,\pi,p_A,p_B,D)$ is the subjective posterior probability that the sample was drawn from cup A given
the information $(d,\pi,p_A,p_B,D)$. Further, it is easy to see that the report $p_r$ that maximizes expected payoff 
is $p_r^*=\Pi(A|d,\pi,p_A,p_B,D)$.

Note that the BDM mechanism incentivizes {\it truthful reporting\/} but not necessarily {\it correctness\/} of subject reports.
The latter depends on their capability as ``intuitive Bayesians'' and their ability to do internal mental calculations
(i.e. the extent to which they do ``fuzzy math''). A drawback of the BDM mechanism is that it can be confusing to
subjects and potentially harder for them to determine the optimal report $p_r$ than to determine the 
posterior probability of cup A. \citet{Holt2019} notes that ``The use of incentivized elicitation procedures is the norm in research
experiments, but there are some problems.'' One of the problems is that BDM relies heavily on the presumption
of rationality of the human subjects, including being able to calculate the expected payoff in equation (\ref{eq:bdmpayoff})
as a function of $p_r$ and then realize that the payoff maximizing report is $p^*_r=\Pi(A|d,\pi,p_A,p_B,D)$. If subjects
are not so good in probability and math, the second stage BDM mechanism might actually mislead or confuse them and would
then add extra ``decision noise'' into the experimental results. 

Indeed, \citet{Holt2019} acknowledges that the 
``BDM procedures may be difficult for subjects to comprehend.'' (p. 110). The instructions to subjects in the \citet{HS2009} experiments instructed them 
that their payoff is maximized by truthfully reporting their subjective posterior. To the extent subjects trusted and followed this advice the BDM mechanism
may not necessarily confuse subjects or add extra ``decision noise''.  

Subjects in Holt and Smith's experiments were not asked to
make an additional binary choice of which cup they believed the observed sample was more likely to have been drawn from.
It seems quite reasonable to assume that if subjects would have been asked to make such a choice (perhaps incentivized by
an additional payment for selecting the correct cup) that they would have chosen cup A if $\Pi(A|d,\pi,p_A,p_B,D) > 1/2$
and cup B otherwise.\footnote{Or randomly guess if $\Pi(A|d,\pi,p_A,p_B,D)=1/2$.} This implies that
$\sigma=0$ in our structural logit model specification in equation~(\ref{eq:dr}), and we use this to generate the binary
decision rule implied by subjects' beliefs in our analysis of implied inefficiency of subjects' decision rules below.
Any inefficiency in subjects' decision making is then due to  ``calculational noise'' $\nu$ and/or bias and incorrect
weights $\beta$ on the constant, $\mbox{LPR}$ and $\mbox{LLR}$  in the formula for subjective posterior
beliefs $\Pi_s(A|d,\pi,p_A,p_B,D,\beta,\nu)$ in equation~(\ref{eq:subjectiveposterior}). 

We estimated $\beta$ and the parameter $\eta$ under the assumption that $\nu \sim N(0,\eta^2)$ by maximum likelihood
using the log reported prior odds ratio regression specification in equation~(\ref{eq:slpor}).\footnote{\footnotesize A small
fraction of subjects reported posterior probabilities of 0 or 1 for which the log reported posterior odds ratio is undefined.
Rather than exclude these observations we estimated a truncated regression specification where we assume that a value
of $0$ is reported when the subjective posterior is lower than some lower threshold $\underline p$ and report a value
of $1$ when it exceeds an upper threshold $\overline p$. It is not hard to show that the maximum likelihood estimates
of these additional parameters are the min and max of the subset of reported subjective posterior values that are strictly in
the $(0,1)$ interval. We verified that all conclusions are robust to simply excluding the observations with reports of 0 or 1,
or recoding them to arbitrary values such as $.00001$ and $.99999$.}
We also estimated multi-type versions of these models using the EC algorithm and finite mixture approaches. We find
a strong improvement in the likelihood from going from 1 to 2 types, but we stopped at $K=2$ types because of the
relatively small number of subjects (22 and 24 in experiments 1 and 2, respectively). 

The results from the EC and finite mixture models are quite similar, both in the parameters and the estimated
fractions of each type. We omit the actual parameter values and describe the results more informally
and graphically below. We will discuss the results from the finite mixture method, though the results from the
EC algorithm are virtually identical.  For experiment 1 the two types can be described as 
1)``noisy Bayesians'' (45\% of the subjects) and 2) conservatives, i.e. those who put more weight on $\mbox{LPR}$ than $\mbox{LLR}$.
For the noisy Bayesian subjects, we cannot reject the hypothesis that $(\beta_0^*,\beta_1^*,\beta^*_2)=(0,1,1)$  (i.e. values that result
in Bayesian beliefs), though there is significant ``calculational noise'' as evidenced by the large and significant
estimate of $\hat\eta=0.91$ (std error $(0.14)$). We strongly reject this hypothesis for the type 2 ``conservative'' subjects 
and find a small bias against choosing cup A. However we note that the type 2 subjects are far less noisy than
the type 1's with an estimated value of $\hat\eta=0.40$, less than half the  value we estimated for type 1 subjects.

\begin{figure}[ht]
\caption{Predicted vs Actual Median Beliefs and Loss Functions: Holt Smith Experiment 1}
\label{fig:hs_exp1_pva_lf}
\centering 
\begin{minipage}[t]{0.48\textwidth} 
    \centering
    \includegraphics[height=2.5in]{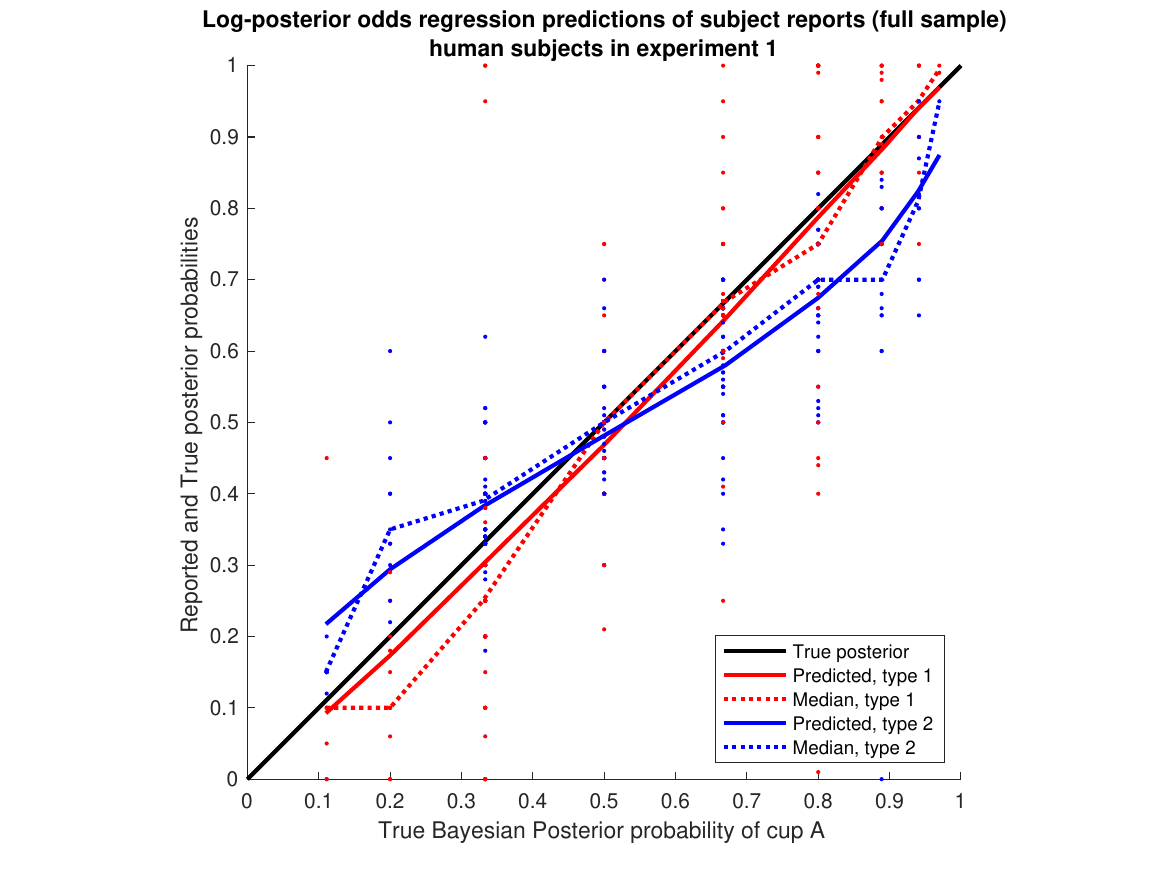}
\end{minipage}%
\hfill 
\begin{minipage}[t]{0.48\textwidth}
    \centering
    \includegraphics[height=2.5in]{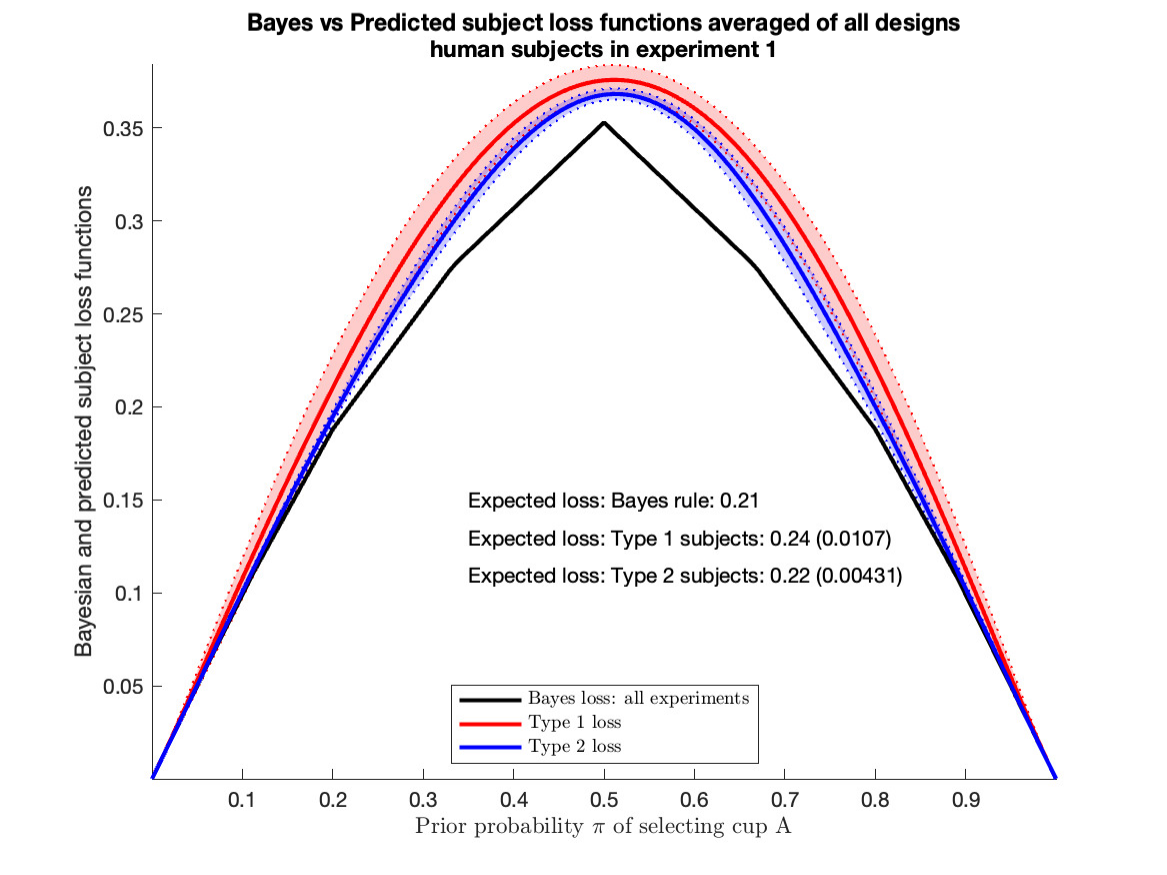}
\end{minipage}
\end{figure}

The left hand panel of figure~\ref{fig:hs_exp1_pva_lf} provides a scatterplot of the actual subject responses
where the x axis corresponds to the implied true Bayesian posterior probability for the $(\mbox{LPR},\mbox{LLR})$ pairs
for each subject and trial. For reference the black 45 degree line is the true Bayesian posterior probability and
the dotted lines are the median values of the subjects' responses. Using the EC algorithm we can classify each
subject as type 1 (noisy Bayesian) or 2 (conservative), and the dotted lines plot the median
of the subject responses for these two types. Finally the solid red and blue lines are the predicted
medians of subject responses from the estimated structural logit model. We see that the model fits the data well
and that the Bayesian subjects have median posterior beliefs that are quite close to the 45 degree line. However the
median beliefs of the type 2 conservative subjects increase less steeply than the true Bayesian posterior does, reflecting
``underconfidence'' i.e. when the true posterior exceeds $1/2$ the median beliefs of type 2 subjects is below the
Bayesian posterior, and conversely when the true posterior is less than $1/2$. As we noted in the discussion
in section~\ref{section:identification} underconfidence itself does not necessarily imply inefficiency of the
decision rule. Instead, inefficiency is more sensitive to the calculational noise, i.e. the value of $\eta$.

The right hand panel of figure~\ref{fig:hs_exp1_pva_lf} plots the loss functions implied by decision rules
that pick cup A when $\Pi_s(A|d,\pi,p_A,p_B,D,\nu) > 1/2$ and cup B otherwise. Due to the higher
value of $\eta$ for the type 1 (noisy Bayesian) subjects, their expected win probability is lower 
than the win probability of the type 2 conservative subjects (blue line). 
When we compute the average win probability for all subjects weighted by the empirical distribution
of $(\pi,D)$ values used in the design of experiment 1, we find an average decision efficiency score of 96\% with a standard
error of $0.7$\%. A Wald test strongly rejects the hypothesis that the human subjects are fully efficient decision makers.
However overall, the decision efficiency for these subjects is high and in line with results we find for subjects
in El-Gamal and Grether's California and Wisconsin experiments.

\begin{figure}[ht]
\caption{Stability of beliefs: 1,2 vs 3,4 trials from Holt Smith Experiment 1}
\label{fig:hs_exp1_cfp}
\centering 
\begin{minipage}[t]{0.48\textwidth} 
    \centering
    \includegraphics[height=2.8in]{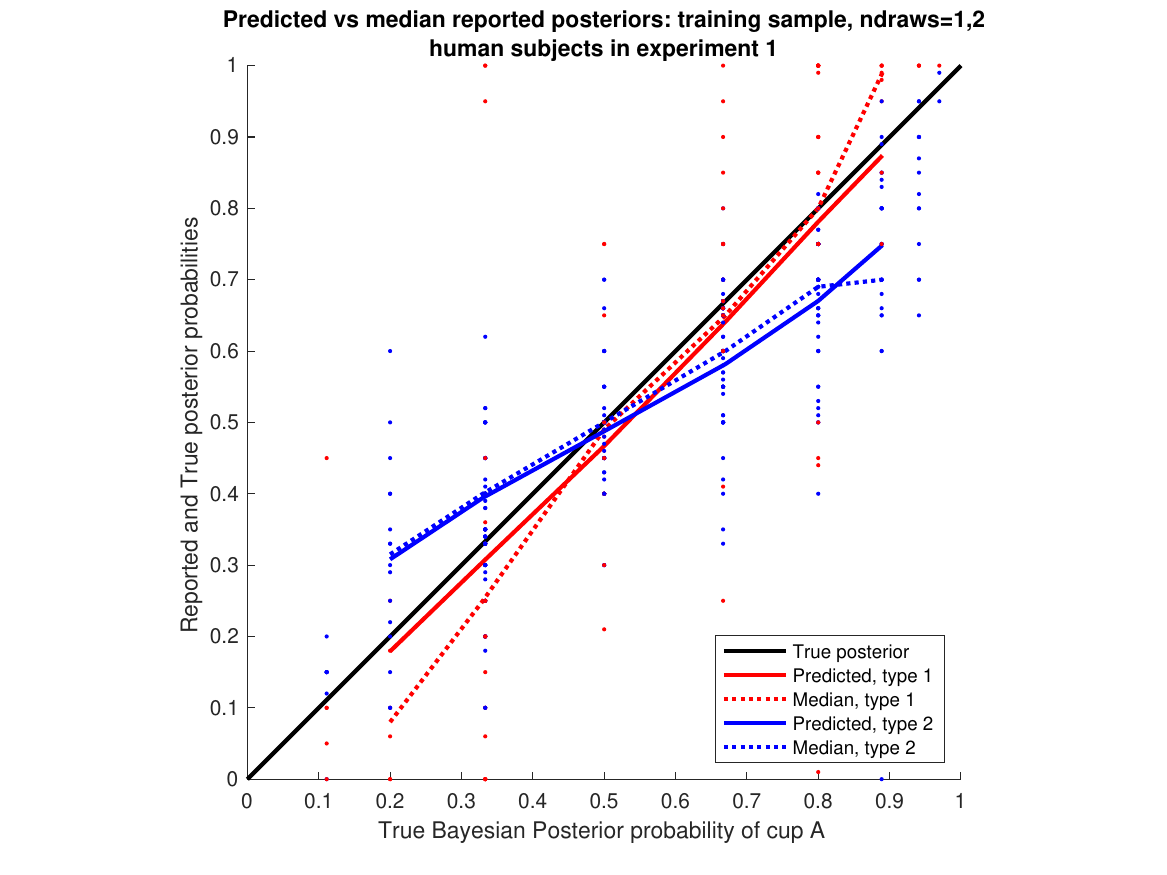}
\end{minipage}%
\hfill 
\begin{minipage}[t]{0.48\textwidth}
    \centering
    \includegraphics[height=2.8in]{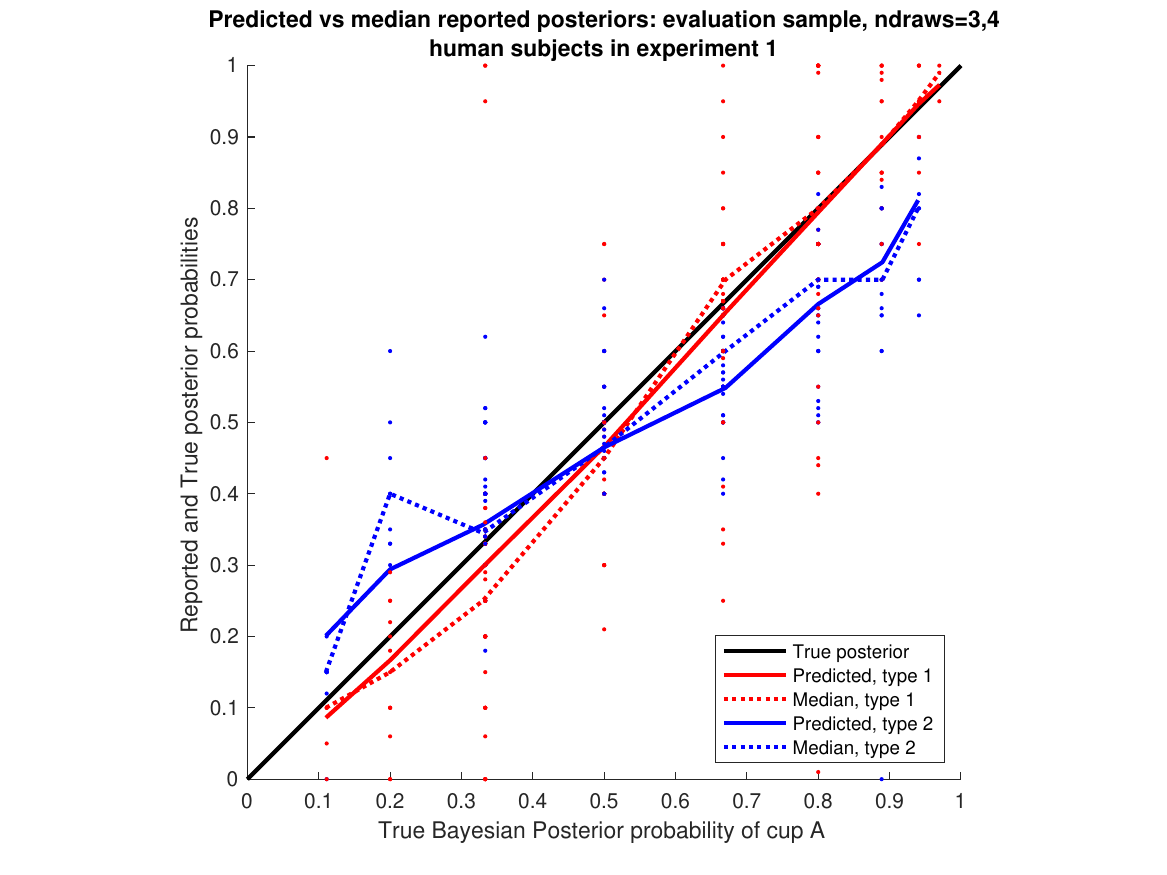}
\end{minipage}
\end{figure}

Since the design of experiment 1 has the total number of draws from the cups $D$ varying from trial to trial
over the values $D \in \{0,1,2,3,4\}$, this provides an opportunity to evaluate the ability of the structural
logit model to predict ``out of sample beliefs'' and test for structural stability. We estimated the model
for values of $D \in \{1,2\}$ only as the ``training sample'' and then compare how well the estimated model
predicts median reported beliefs for values of $D \in \{3,4\}$ that we refer to as the ``evaluation sample.''
The left hand panel of figure~\ref{fig:hs_exp1_cfp} plots the actual median and fitted median beliefs for type
1 and 2 subjects in the training sample and the right hand panel show both for the evaluation sample, where the
model predictions are based on parameters estimated in the training sample where $D \in \{1,2\}$ only.
Visually, the structural model does a good job in predicting median beliefs of the subjects in the
evaluation sample.

However we do observe that the fractions of subjects estimated to be type 1 or type 2 subjects changes
when we restrict the sample. For example in the training sample, we find only 27\% of the subjects are type 1
noisy Bayesians, whereas in the evaluation sample we estimate that 55\% of subjects are noisy Bayesians. Further
the amount of noise, as estimated by $\eta$ is higher in the training sample ($\hat\eta=1.22$) compared to the
evaluation sample $(\hat\eta=.74)$. Thus, we find evidence of ``structural instability'' and a likelihood ratio
test strongly rejects the hypothesis that subjects are using the same decision rule in both the training and
evaluation samples. Evidently, the training sample where $D \in \{1,2\}$ is a harder decision problem
than the evaluation sample where $D\in \{3,4\}$ and subjects change their decision rules to cope with this. 
The parameter for the standard deviation of the calculational errors is higher for the harder decision problem. Of course, this is also true for a Bayesian:
the loss function for a Bayesian when $D \{1,2\}$ majorizes the loss function for a Bayesian when $D\in \{3,4\}$.\footnote{\footnotesize
We also analyzed data from 30 subjects in Holt and Smith's experiment 2 which was conducted online. The overall conclusions are similar to those from our reanalysis of experiment 1, except that the EC algorithm no longer finds any noisy Bayesians: 62\% of subjects put excessive
weight on the prior and the remaining 38\% put too much weight on the data. The  level of calculational noise for  these subjects, 
$\eta$, is also significantly higher. 
The higher degree of noise in subjects' reports implies significantly higher loss and thus lower efficiency.
Average decision efficiency for all subjects in all trials in experiment 2 was 91\% (0.8\%), lower than the 
93\% efficiency of the subjects in El-Gamal and Grether's 6 ball California experiments and lower than 
the 96\% efficiency of human subjects in Holt and Smith's experiment 1.}

\subsection{Conclusions from the re-analysis of human subject data}
\label{section:humanconclusions}

Below we list the key takeaways from our reanalysis of human subject data from
\citet{EGG1995}, \citet{EGG1999} and \citet{HS2009}. 

\bigskip

\begin{blockquote}
\singlespacing
\begin{itemize}

\item[1.] We confirm their strong rejection of the hypothesis that {\it all subjects\/} behave as Bayesian decision makers (or even
noisy Bayesians). Using the EC algorithm and finite mixture methods we find substantial unobserved 
heterogeneity among different subjects as well as significant idiosyncratic random noise that affects their responses. 
\smallskip
\item[2.] However subsets of subjects in the El-Gamal and Grether experiments are classified as
noisy Bayesians, ranging from 32\% in the Wisconsin 6 ball experiments to 77\% in the Wisconsin 7 ball experiments.
In the California and Wisconsin  experiments subjects provide only binary responses so a ``revealed belief approach'' is required
to make these inferences and we showed that the identification of beliefs is fragile. In the experiments run by Holt and Smith posterior beliefs  were elicited via the BDM mechanism,
enabling us to directly identify their posterior beliefs. In experiment 1 we find that 45\% of subjects are classified as noisy Bayesians,
but in their web-based experiment 2, none of the subjects are.
\smallskip

\item[3.] Even though subject beliefs are difficult to identify in the El-Gamal and Grether experiments, their decision
rules are identified. We find high overall decision efficiency, ranging from 93\% of the optimal Bayesian win probability
in their California experiments, to 96.5\% for subjects in their Wisconsin experiments that used the 6 ball
California design. 
\smallskip
\item[4.] Holt and Smith did not ask subjects to make a binary choice of the more likely cup, but we can impute
this choice based on their elicited posterior probability of cup A. We find significant calculational noise
in the reported posterior probabilities, but the overall efficiency of a decision rule that selects cup A if the
subjective posterior probability exceeds 1/2 is still high: 96\% for the subjects in their experiment 1 and 
91\% in their web-based experiment 2.
\smallskip
\item[5.] The structural logit model provides a good approximation to the different decision rules
used by subjects and ``out-of-sample'' forecasts of how their
behavior changed in responses to changes in experimental design.
However our analysis confirms the main conclusion of \citet{EGG1999}
 that subjects appear to use different decision rules in different experimental designs. 
Thus, the structural logit model can only approximate the more complicated underlying cognitive processes governing  human subject responses.
\end{itemize}
\end{blockquote}


\section{Optimality and revealed beliefs of AI subjects}
\label{section:ai_subjects}
We will now analyze the performance of AI subjects collected from a series of experiments that replicated human studies, using various versions of ChatGPT: beginning with the earliest GPT-3.5, moving on to GPT-4, and culminating in the latest version, GPT-4o. These iterations demonstrate significant advancements in general-purpose AI capabilities.

We begin by introducing the design of the prompts used to conduct experiments with GPTs. We first replicate the Wisconsin experiments, where subjects are required to perform binary classification tasks, as outlined in \citet{EGG1999}. We also elicit the subjective posterior beliefs of AI subjects using prompts very similar to the experiments in \citet{HS2009}. Although it is less clear whether there are specific incentives for AI subjects to exert less effort or misreport, we retain the BDM mechanism in our prompts to ensure that our design aligns as closely as possible with the original human experiments.



We use the structural logit model developed in Section \ref{section:structural_model} to interpret the GPT data and test whether subjects are Bayesian. We do this
to provide a parallel analysis of human and GPT subjects, and due to the flexibility and superiority of the structural logit model relative to 
other behavioral models as we discussed in Section \ref{section:human_subjects}. We also allow for unobserved heterogeneity using both the EC and finite mixture methods. In our GPT experiments, we intentionally injected variability in their temperatures to proxy for unobserved heterogeneity in human subjects. The temperature parameter in a GPT  controls the scale of random ``noise'' in their responses and thus it is akin to heterogeneity in the extreme value $\sigma$ parameter in equation~\ref{eq:structural_logit}.\footnote{In GPTs, the temperature refers to a parameter that controls the randomness of the model's responses. Lower temperature results in more deterministic and focused responses, making the model more likely to choose the most probable next words. A higher temperature increases randomness, allowing for more varied responses. The default temperature for the three versions of GPTs we consider here is 0.7. } We show that the EC and finite mixture methods can effectively detect and control for the variability in temperature, which in the case of GPT subjects is actually an observed covariate rather than unobserved heterogeneity.

We use the accuracy and decision efficiency metrics defined in section 2 to compare the performance of GPT and human subjects. We find that GPTs are subject to biases that lead to suboptimal decisions. However, we
observe a rapid growth in overall ability across successive generations of GPT subjects, causing their performance to improve from sub-human levels in GPT-3.5 to human-like proficiency in GPT-4, and ultimately to superhuman and nearly perfect Bayesian classifications in the latest version, GPT-4o.

\subsection{Prompt Design for Experiments in ChatGPT}

Similar to \cite{GPTstudy2023}, we conduct our experiments by submitting inquiries through the public OpenAI application programming interface (API). Using APIs allows us to conduct massive experiments in a timely and cost-effective manner. 
We formulate the prompt by drawing an analogy between GPTs and human experiments. Each version of a GPT represents a distinct school where human experiments have been conducted. We consider the varying temperature settings for GPTs as analogous to different students or human subjects. Once we define the schools and subjects, we repeated submitting inquiries using our prompts through APIs, as each human subject participates in the trials in different experiments.



\ref{appendix:B} outlines the algorithm we developed to implement the experiments, which involves looping over different versions of GPTs (schools), temperature settings (subjects), experiments, and trials. We then parse the responses from ChatGPT and collect its choices of either Cage $A$ or $B$ in the Wisconsin experiment,\footnote{Occasionally, GPTs may stop prematurely before providing an answer regarding the choice between $A$ and $B$. In such cases, we resubmit the same inquiry until the GPT delivers a classification. We consider this process to be natural, as it mirrors our everyday use of ChatGPT—if it fails to provide a satisfactory answer due to an unexpected stop, we simply ask again. In our implementation, it takes a maximum of 5 iterations to resolve any missing answers in our experiment.} or the reported subjective probability in the case of the Holt and Smith experiment. \ref{appendix:C} provides further details about the prompts we use. 

One notable difference between the prompt in our study and that of \citet{GPTstudy2023} is that we allow GPTs to report their reasoning process, rather than solely providing the conclusion as done in \citet{GPTstudy2023}.\footnote{For example, the prompt for the investment experiment in \citet{GPTstudy2023} ends with "First please \textit{only} tell me the number of points for investing Asset A, then please \textit{only} tell me the number of points for investing Asset B." Including the word \textit{only} forces GPTs to report only the final conclusion without the reasoning process.} We deviate from them for two reasons. First, in the daily use of GPTs, people rarely impose such restrictions. Second, when answering our questions, GPTs typically answer them step by step. This chain of thought may potentially enhance GPT performance, as suggested by \citet{wei2022chain}. 

\subsection{Analysis of the Wisconsin Experiments in GPTs}

We begin by analyzing the binary classification results from the Wisconsin experiments conducted with the GPTs. Following  our re-analysis of human subjects, instead of pooling the data from the 6 and 7 ball design we focus on the 6 ball design and use the results from the 7 ball design as a validation set to test ``structural stability'' of the structural logit model. 

\begin{figure}[H]
 \centering
 \caption{Choices of GPT Subjects: 6-ball Experiments}%
 \label{fig:data_summary_gpt_wisconsin_day1}
 \begin{minipage}[t]{0.32\textwidth} 
    \centering
    \includegraphics[width=\linewidth, trim=15 1 15 .5, clip=true]{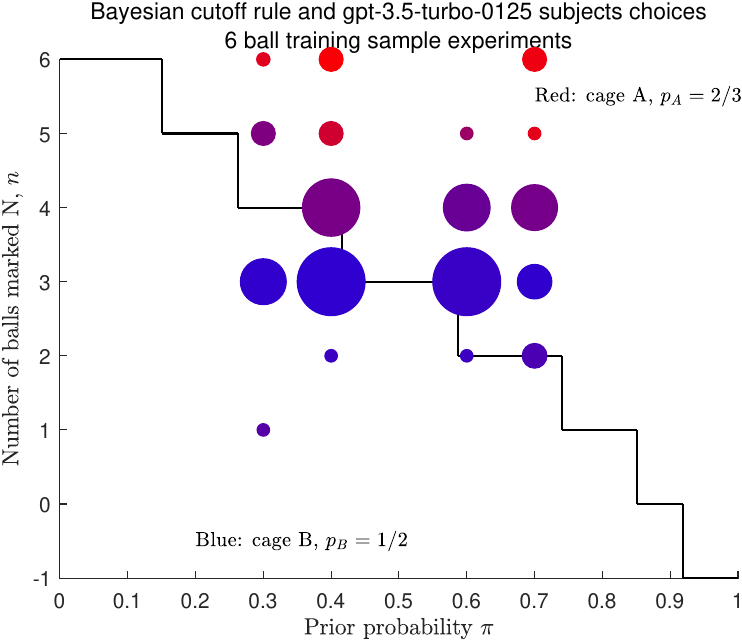}
    (a) GPT 3.5 
     \label{fig:data_summary_gpt_wisconsin_day1_3dot5} 
 \end{minipage}
 \hfill 
 \begin{minipage}[t]{0.32\textwidth}
    \centering
    \includegraphics[width=\linewidth, trim=15 1 15 .5, clip=true]{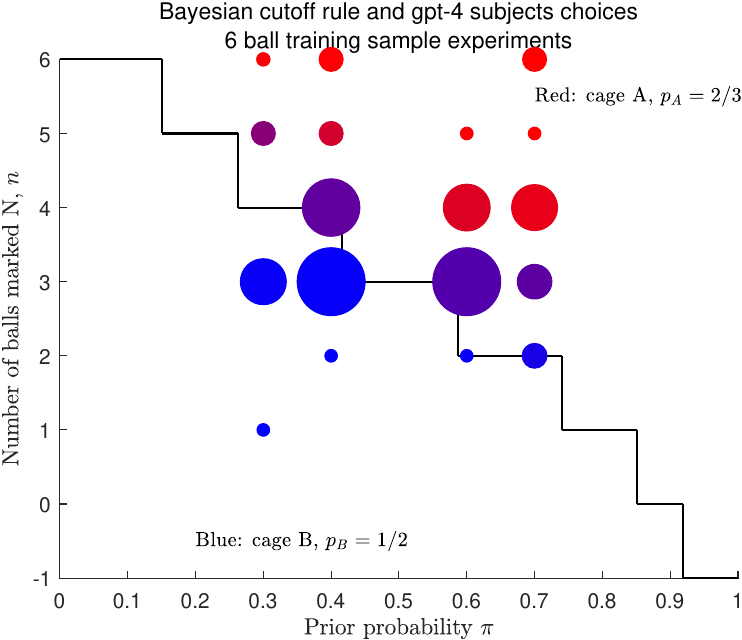}
    (b) GPT-4 
     \label{fig:data_summary_gpt_wisconsin_day1_4} 
 \end{minipage}
 \hfill 
 \begin{minipage}[t]{0.32\textwidth}
    \centering
    \includegraphics[width=\linewidth, trim=15 1 15 .5, clip=true]{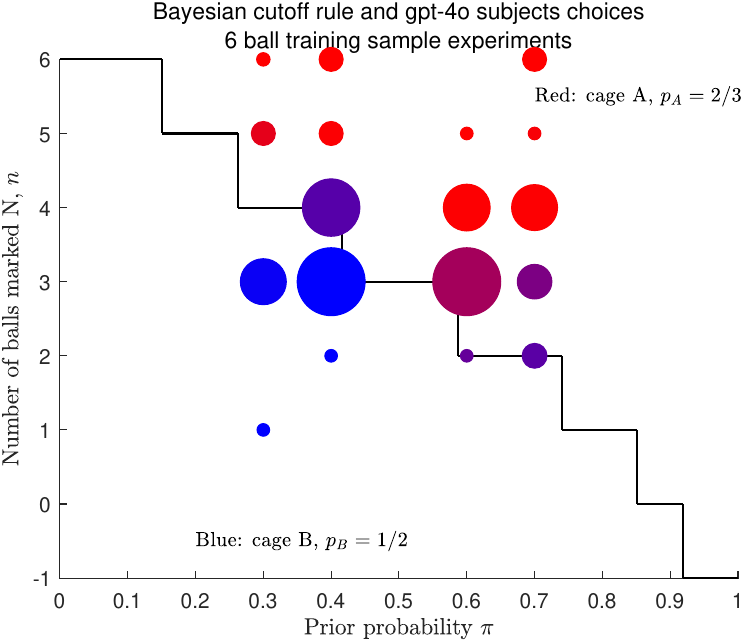}
    (c) GPT-4o 
     \label{fig:data_summary_gpt_wisconsin_day1_4o} 
 \end{minipage}
\end{figure}

Figure \ref{fig:data_summary_gpt_wisconsin_day1} summarizes the choice probabilities in the first day experiment. Each circle represents a trial, with prior probabilities, $\pi$, on the $x-$axis and number of $N$-draws, $n$, on the $y-$axis. The size of each circle indicates the number of choices made by subjects, while the color reflects their selections: deep blue indicates more subjects chose cage $B$, and deep red indicates more chose cage $A$. The black step function represents the optimal integer cutoffs calculated using Bayes' rule. Given a prior probability, it is optimal for a subject to choose cage $A$ if $n$ exceeds the cutoff, and to choose cage $B$ if it falls below.

For Bayesian decision-makers, we expect red circles above and blue circles below the Bayesian cutoffs. GPT-3.5 subjects make more errors than human subjects and advanced GPTs, as illustrated by large purple circles above the cutoff in Figure \ref{fig:data_summary_gpt_wisconsin_day1_3dot5}. For example, in trials with 4 $N$-draws and prior probabilities $0.6$ and $0.7$, all should choose $A$ (pure red), but many GPT-3.5 subjects choose $B$, mixing red with blue and creating the purple circles. In trials with 3 $N$-draws and a prior probability $0.6$, almost all GPT 3.5 subjects choose Cage $B$. In contrast, human subjects mix between Cages $A$ and $B$, aligning more closely with the Bayesian posterior probability of $1/2$.\footnote{The distribution of mistakes among trials are also different between GPT and human subjects. While a small fraction of human subjects make errors distributed across nearly all trials, GPT-3.5 tends to be uniformly accurate in some trials, with its errors concentrated in others.}


The accuracy of GPT-4 and GPT-4o is much improved. The two large purple circles disappear in Figure \ref{fig:data_summary_gpt_wisconsin_day1_4} and \ref{fig:data_summary_gpt_wisconsin_day1_4o}. However,
neither GPT-4 nor GPT-4o are perfect Bayesian decision makers. Mistakes still occur in certain trials, indicated by two purple circles for GPT-4 and one for GPT-4o, all positioned above but close to the cutoff. \footnote{We find a similar pattern in the second-day experiment that the accuracy improves with more advanced GPTs. We also note that, same as human subjects, GPTs make less mistakes in the 7-ball experiment than in the 6-ball experiment.}


We then estimate a multiple-type structural logit model using the FM method.\footnote{We also estimate the model using the EC algorithm. Although the EC algorithm typically identifies more types than the FM method, we prefer the FM method for two reasons. First, some types identified by the EC algorithm may be redundant due to subtle parameter differences or a small number of subjects in certain types. Second, starting from the estimates of EC algorithm, the FM method can always consolidate these into fewer types.} The estimates detect unobserved heterogeneity among GPT subjects, with different subjective posterior beliefs (determined by the belief parameters $\beta$'s) and level of idiosyncratic noise (captured by the extreme value scale parameter $\sigma$) associated with each type. 

The FM method identifies two types for GPT-3.5 subjects. Figure \ref{fig:beliefs_3dot5} plots the estimated subjective posterior beliefs against true posteriors for both types.\footnote{Subjective posterior beliefs, $\Pi_s(A|d,\pi,p_A,p_B,D,\nu)$ in equation \ref{eq:structural_logit}, depend on the trials, represented by $(d,\pi)$. Rather than plotting against the two-dimensional trial specifications, we summarize $(d,\pi)$ by calculating the corresponding Bayesian posterior probabilities, offering a one-dimensional summary. Recall that we assume $\nu=0$ in the Wisconsin experiment. } Both types deviate from the 45-degree line, showing non-Bayesian beliefs. However, when subjective posteriors are on the same side of $1/2$ as Bayesian posteriors, they can result in identical decisions to those of Bayesian decision makers. 58\% of subjects are type 1, aligning perfectly with Bayesian decisions when true posteriors are below $1/2$, but often differing when subjective posteriors bounce to zero as true posteriors cross $1/2$. Type 2 subjects have smaller wiggles in their subjective posteriors, but can still jump to different directions than the Bayesian posterior, leading to different decisions than Bayesian decision makers. 

Turning to the estimated $\sigma$ parameters, we find that type 1 subjects have smaller values than type 2 subjects. The difference in the estimated noise parameters reflects the known heterogeneity in GPT 3.5 subjects injected by our random assignment of temperatures. The mean temperature for type 1 subjects is 0.43, nearly half that of the noisier type 2 subjects. This supports the notion that a higher ``temperature'' corresponds to greater ``noise'' in subject responses and demonstrates the effectiveness of our algorithm in detecting unobserved heterogeneity.

\begin{figure}[H]
 \centering
 \caption{Estimated classification hyperplanes for GPT Subjects: 6-ball Experiments}%
 \label{fig:hyperplanes_wisconsin}
 \begin{minipage}[t]{0.32\textwidth} 
    \centering
    \includegraphics[width=\linewidth, trim= 15 1 15 .5, clip=true]{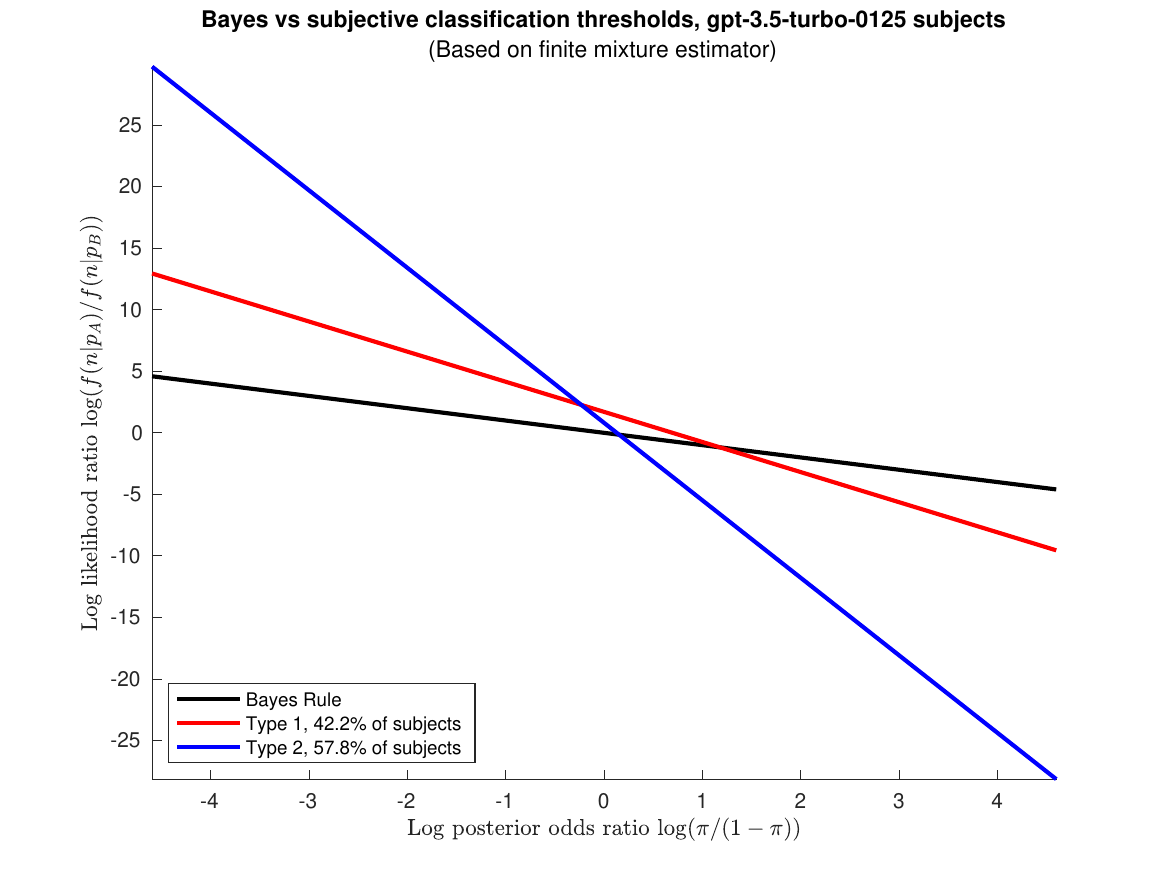}
    (a) GPT 3.5 
 \end{minipage}%
 \hfill
 \begin{minipage}[t]{0.32\textwidth}
    \centering
    \includegraphics[width=\linewidth, trim= 15 1 15 .5, clip=true] {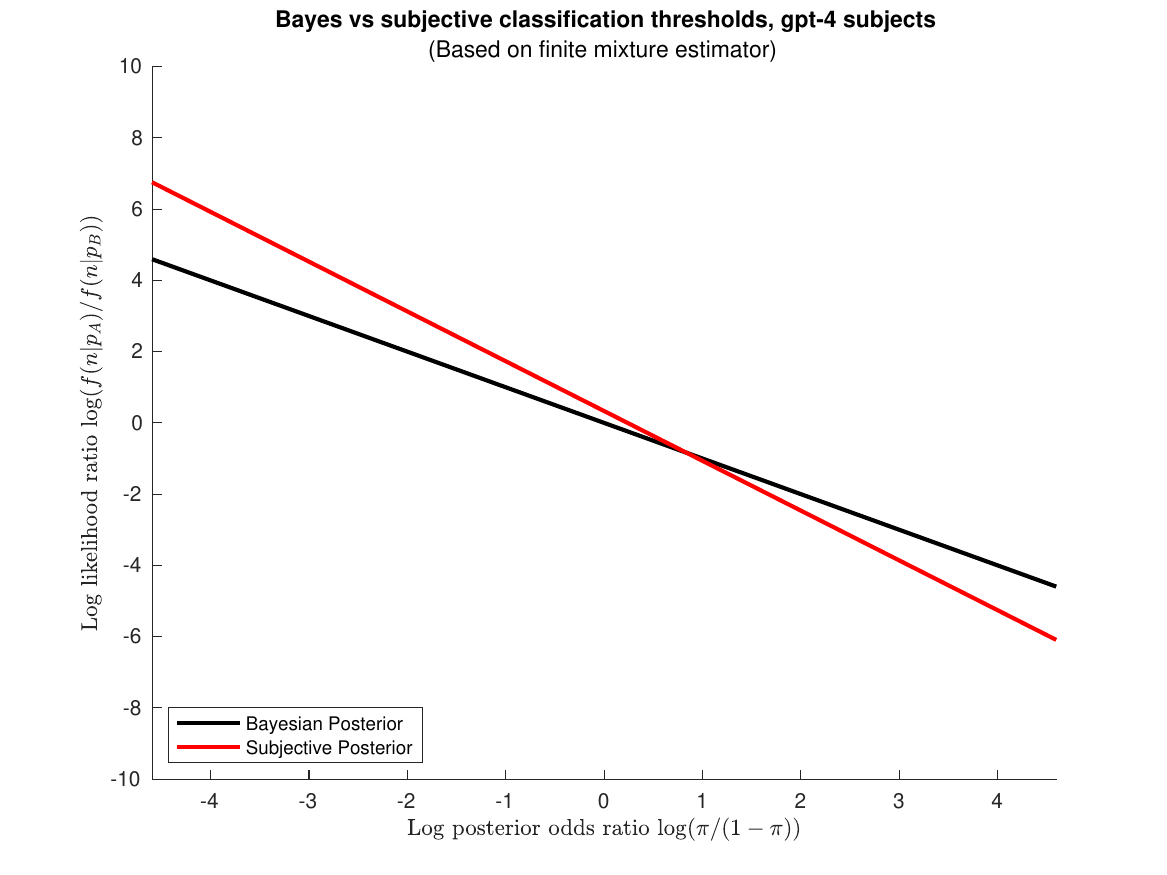}
    (b) GPT-4 
 \end{minipage}%
 \hfill
 \begin{minipage}[t]{0.32\textwidth}
    \centering
    \includegraphics[width=\linewidth, trim=15 1 15 .5, clip=true] {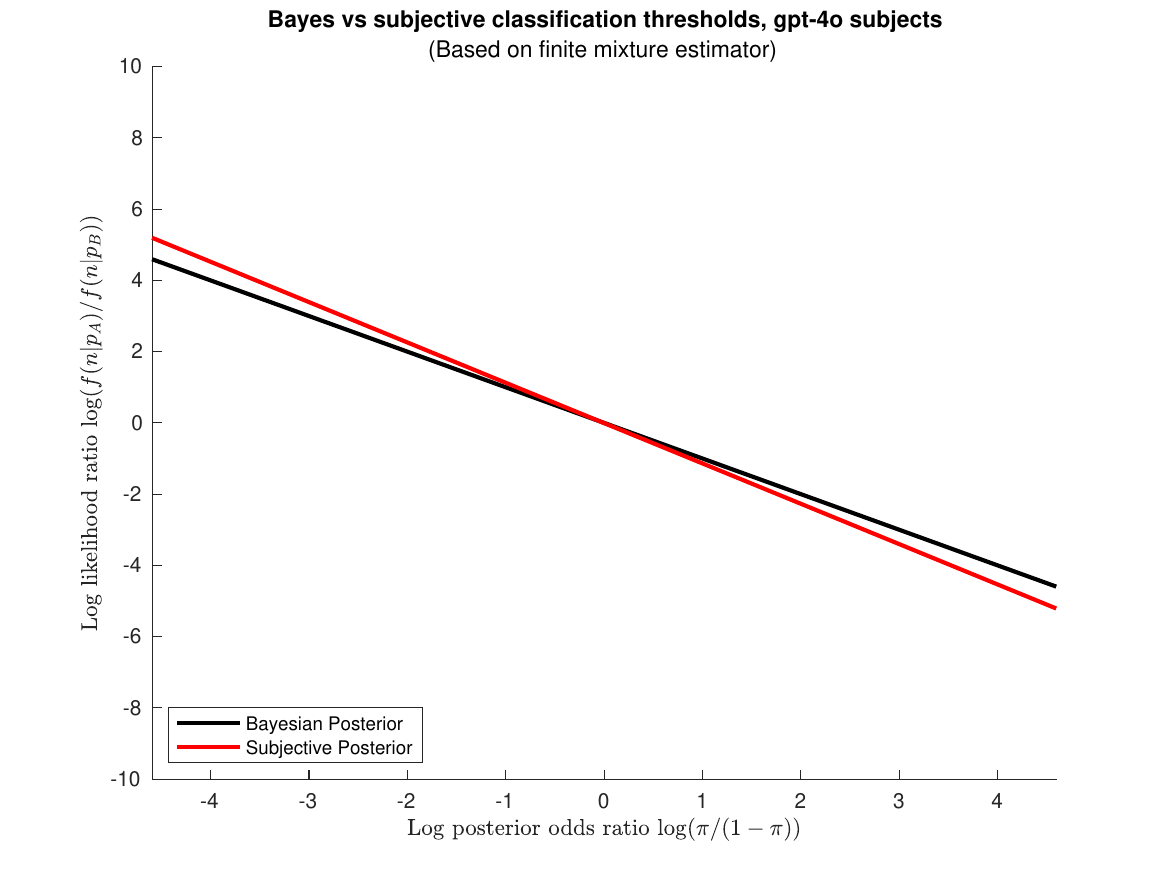}
    (c) GPT-4o 
 \end{minipage}
\end{figure}

The FM method only identifies a single type of GPT-4 and GPT 4o subjects, respectively. The estimated posterior beliefs in Figure \ref{fig:beliefs_4} and \ref{fig:beliefs_4o} are almost flat, but they do have a tiny positive slope, leading to the same CCPs as Bayesian decision makers.\footnote{We estimate the model from multiple starting values, all converging to the same single-type estimate. To further validate that GPT-4o acts as a single-type "noisy Bayesian" decision maker, with subjective posterior beliefs showing a slight positive slope against true posteriors, we estimate a noisy Bayesian model with $\beta$ parameters constrained to $[0, -1, -1]$, estimating only $\sigma$. The likelihood ratio test does not reject the noisy Bayesian model.} The seemingly flat beliefs conceal the differences in subjective beliefs between GPT-4 and GPT-4o. Figure \ref{fig:hyperplanes_wisconsin} shows the implied classification hyperplanes, whose slopes equal the negative ratio of coefficients for LLR ($\widehat{\beta}{1}$) and LPR ($\widehat{\beta}{2}$). The Bayesian classification hyperplanes, which have a slope of $-1$, are in black. Although the $\beta$ values are small, the ratios are close to -1, so GPT-4o subjects assign nearly equal weights to LLR and LPR and their subjective classification hyperplane is closer to the Bayesian hyperplane than GPT-4, which places more weight on LPR than on LLR.

We next turn to evaluate the model fit and ``structural stability'' of the multiple-type structural logit model when applying to data generated by GPTs. Figure \ref{fig:wisconsin_gpt3dot5_ccps} compares the model-predicted CCPs for GPT-3.5 with the data from the estimation sample (on the left) and the second-day experiment as a validation sample (on the right). Although the in-sample fit is good,\footnote{There are two exceptions. It underestimates the probability of choosing Cage $A$ with $N=2$ at a low prior probability of $\pi=0.3$, and it significantly overestimates the probability with $N=5$ at a high prior probability of $\pi=0.6$.} the out-of-sample predictions, unlike those for human data, are quite poor. This is because the choice probabilities in GPT 3.5 lack monotonicity with respect to both LPRs and LLRs, which can be a very weak requirement for rationality. The structural logit model, however, imposes such monotonicity and therefore, is unable to capture the complicated nature of mistakes taken by GPT 3.5. We can extend the simple structural logit model to capture such non-monotonicity by incorporating higher order terms of LLRs and LPRs. However, due to limited observations, more flexible specifications may lead to overfitting. Therefore, we choose to retain our simple structural logit models.



Likelihood ratio tests reject the hypothesis of structural stability of the model with GPT-4 and GPT-4o subjects. However, the rejection is weaker for more advanced GPTs. This suggests that while the structural logit model doesn't perfectly capture GPT behavior, its accuracy improves as subjects become more rational.\footnote{We conjecture that for GPT-o1, as it converges to nearly perfect Bayesian behavior, we will see structural stability and the structural logit model (the subcase of a near noiseless Bayesian) will almost perfectly describe its behavior.}

\begin{figure}[H]
 \centering
 \caption{Percent correctly answered for GPT Subjects: 6-ball Experiments}%
 \label{fig:pca_wisconsin}
 \begin{minipage}[t]{0.32\textwidth} 
    \centering
    \includegraphics[width=\linewidth, trim= 15 1 15 .5, clip=true]{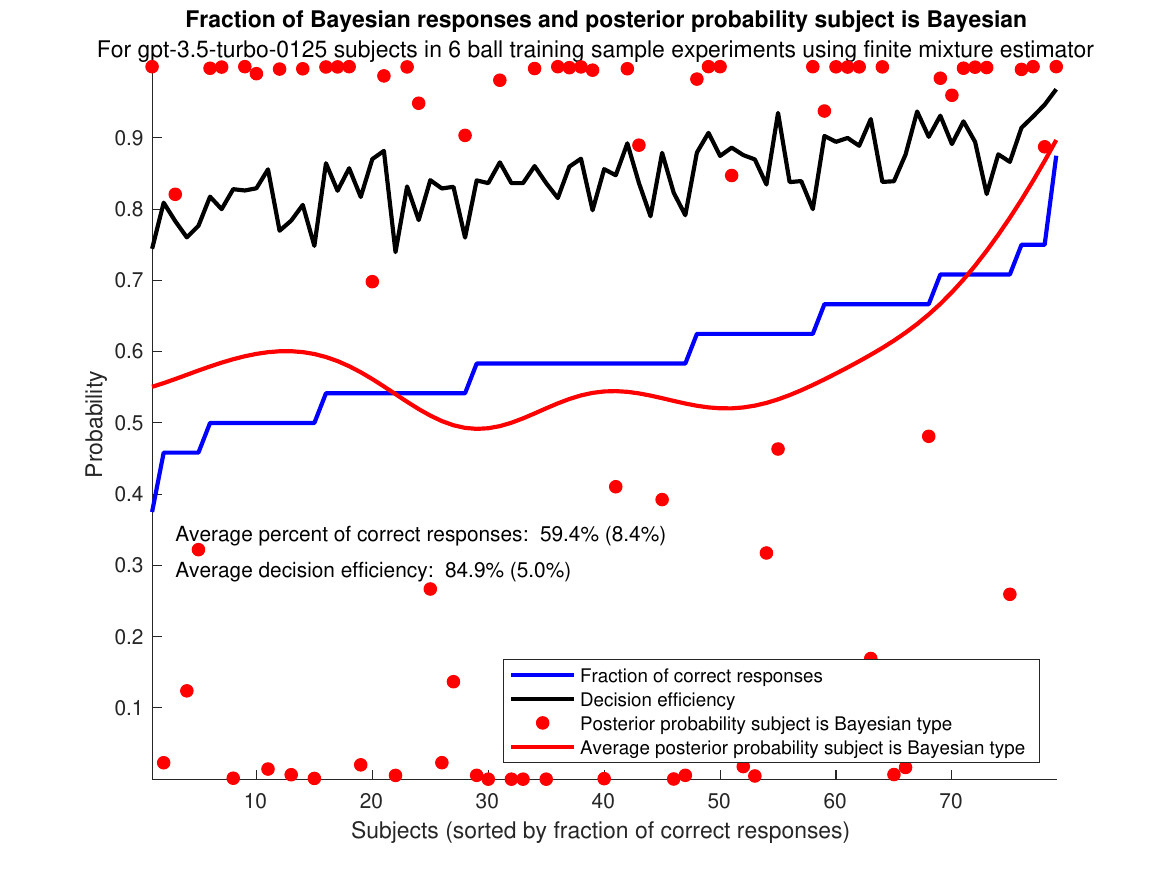}
    (a) GPT 3.5 
     \label{fig:pca_3dot5} 
 \end{minipage}%
 \hfill
 \begin{minipage}[t]{0.32\textwidth}
    \centering
    \includegraphics[width=\linewidth, trim= 15 1 15 .5, clip=true] {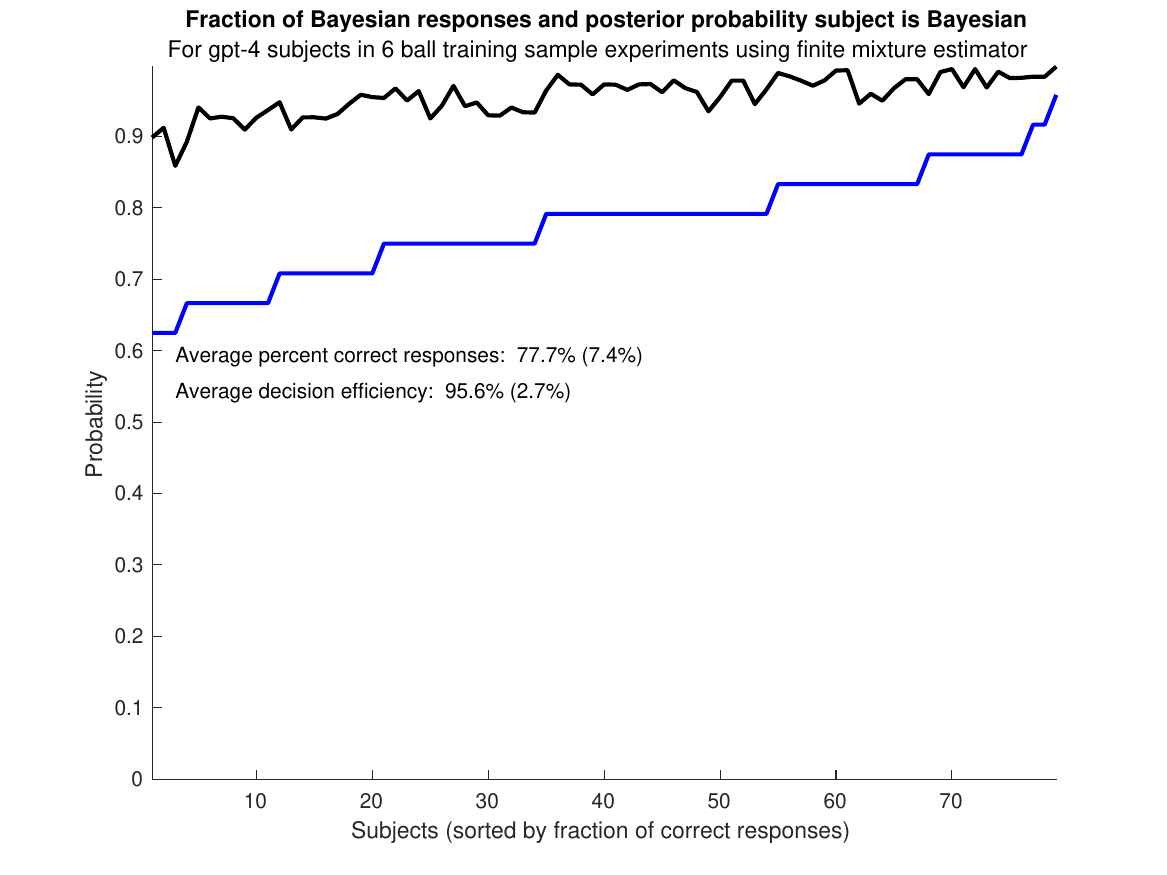}
    (b) GPT-4 
     \label{fig:pca_4} 
 \end{minipage}%
 \hfill
 \begin{minipage}[t]{0.32\textwidth}
    \centering
    \includegraphics[width=\linewidth, trim=15 1 15 .5, clip=true] {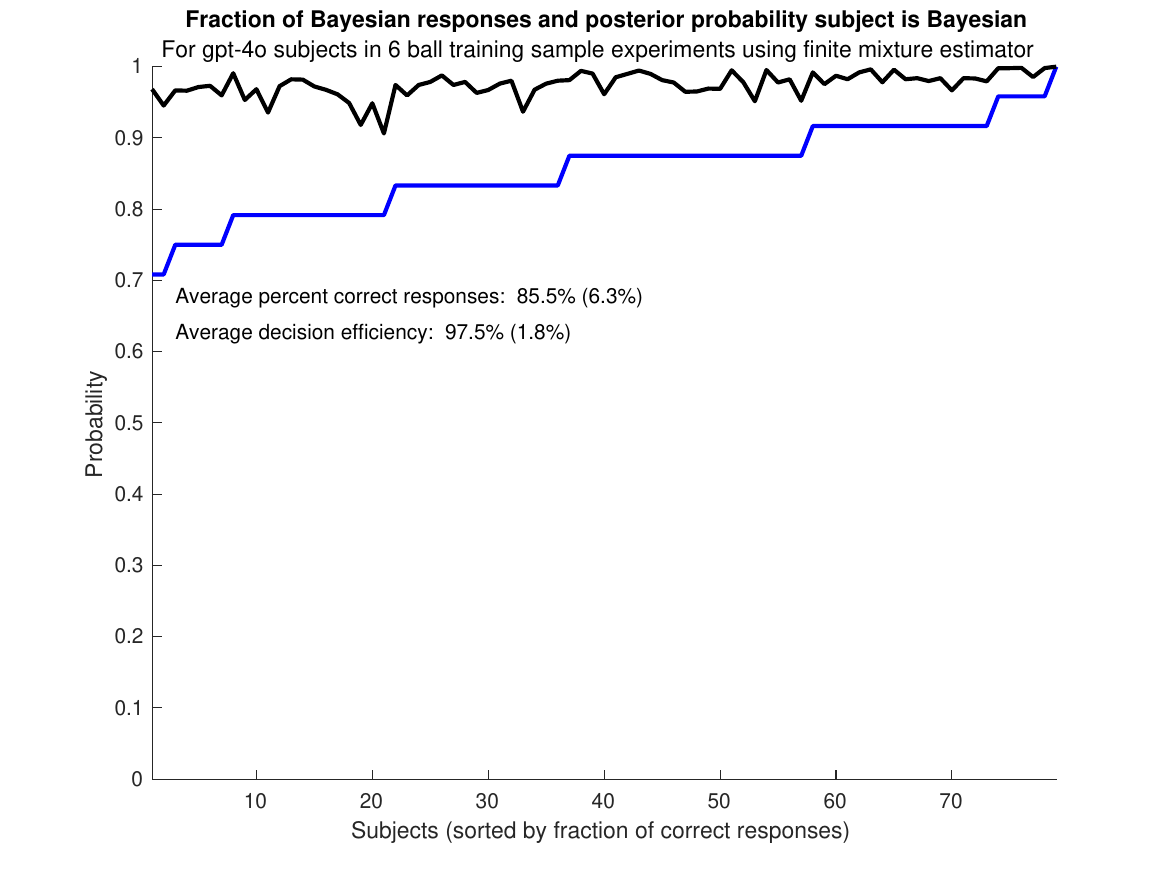}
    (c) GPT-4o 
     \label{fig:pca_4o} 
 \end{minipage}
\end{figure}

Next we compare the performance of the three generations of GPTs  by calculating accuracy and decision efficiency as in Section \ref{section:statistical_decision_theory}. The blue step function in Figure \ref{fig:pca_3dot5} shows that accuracy varies from below 40\% to 90\% among GPT 3.5 subjects, with a very low average of just 59.4\%. The accuracy curve moves up for GPT-4 and rises further for GPT-4o, highlighting the improvement in accuracy across successive generations of GPTs.

The rapid increase in rationality is further supported by the measure of decision efficiency, plotted in Figure \ref{fig:pca_wisconsin} by black curves, which shift upwards from \ref{fig:pca_3dot5} to \ref{fig:pca_4o}. Decision efficiency generally increases as subjects achieve higher accuracy, but there are noticeable fluctuations at a local level. They arise because a subject with a slightly higher correct response rate may struggle on ``easy'' cases—where the true posteriors are more extreme and easier to differentiate—while randomly guessing correctly in ``hard'' cases, where the true posterior is around 1/2. As a result, a subject with higher accuracy may show lower decision efficiency. GPT-4 and GPT-4o have much smaller wiggles than GPT-3.5 reflecting the lower level of estimated noise, $\sigma$, affecting their responses.

When multiple types are identified, as is the case with GPT 3.5, Figure \ref{fig:pca_wisconsin} selects the most Bayesian type and calculates the posterior probability that a subject belongs to this type, represented by the red dots. The red Bayesian-type curve plots the average posterior probability of the most Bayesian type, estimated using a local linear regression of these red dots. The fluctuation in the red dots and the non-monotonicity of the Bayesian-type curve show that higher accuracy in a subject doesn't necessarily make it more likely to be a Bayesian type. 

Table \ref{table:6ballsummary} summarizes the overall performance of humans and various GPTs in the Wisconsin 6 ball design experiments, averaging across all estimated types. In terms of both efficiency and accuracy the GPT-3.5 subjects perform significantly worse than humans. Although humans are comparable to GPT-4, they are less efficient than GPT-4o. This underscores the rapid transition in GPT performance  from ``subhuman'' to ``superhuman'' with just a few upgrades over a relatively short period. 

The bottom row in Table \ref{table:6ballsummary} reports the number of subject types detected by the FM method. We find that there are fewer types identified for the later versions of GPT, suggesting that these subjects are converging  a single type that is very close to a Bayesian decision maker. We find more heterogeneity among human subjects than for GPTs, whose behavior is more uniform, despite our intentional introduction of heterogeneity via temperature. 
 


\begin{table}[H]
\centering
\caption{Performance of GPTs and humans in the 6-Ball Experiments}
\label{table:6ballsummary}
\begin{tabular}{lcccc}
\toprule
                            & GPT 3.5       & GPT-4         & GPT-4o            & Humans \\
\midrule
Efficiency            & 84.9          & 96.0          & 97.5              & 96.5 \\
                            & (0.6\%)       & (0.3\%)       & (0.2\%)           & (0.5\%) \\
Accuracy         & 59.4          & 77.7          & 85.5              & 81.9 \\
                            & (8.4\%)       & (7.4\%)       & (6.3\%)           & (9.1\%) \\
No. of Types             & 2             & 1             & 1                 & 3 \\
\bottomrule
\end{tabular}
\end{table}

\subsection{Analysis of the Holt and Smith Experiments in GPTs}

Now we turn to an analysis of elicited posteriors using the Holt and Smith design with GPT subjects. These data allow us to directly observe subjective posterior beliefs which are otherwise challenging to identify when subjects report only binary choices.

We controlled for subject heterogeneity in the GPT subjects using the FM method.\footnote{We obtain very similar results using the EC method.} As with the analysis of human data, we restrict it to at most two types due to the limited number of subjects involved in the experiment.

We replicate both the experiments run at the University of Virginia (experiment 1) as well as online (experiment 2). In contrast to our reanalysis of human data, we find that the estimated subjective posterior beliefs are very similar for GPT subjects in the two experiments. This is not surprising, as humans may be more easily distracted in online than in offline experiments, as suggested by the larger variance in the error terms. Such differences do not appear in experiments with GPTs. Therefore, we focus on presenting the results from the experiment 1 where the average human decision efficiency was higher, 96\% compared to 91\% for human subjects in experiment 2.

\begin{figure}[H]
 \centering
 \caption{True vs Estimated Subjective Posterior Probabilities}%
 \label{fig:hs_model_fit_ccps}
 \begin{minipage}[t]{0.32\textwidth} 
    \centering
    \includegraphics[width=\linewidth]{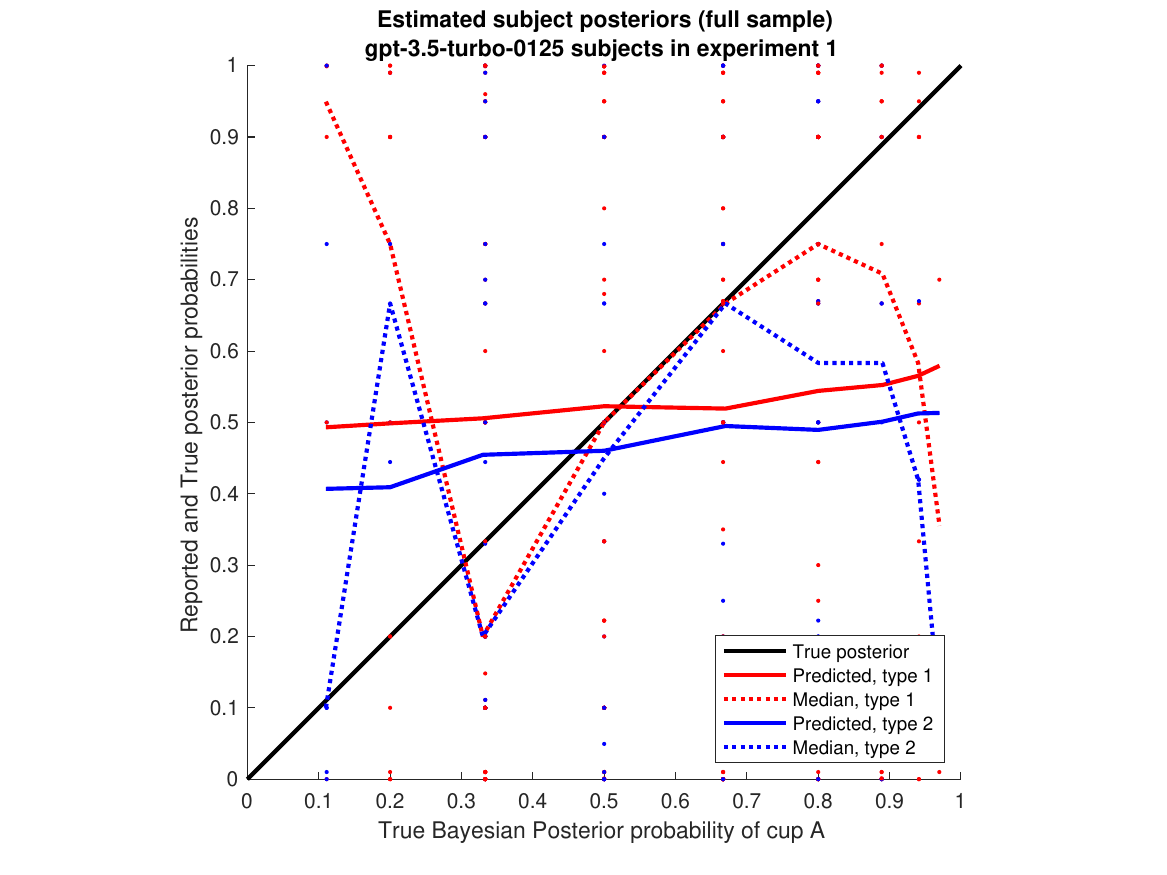}
    (a) GPT 3.5 
     \label{fig:hs_model_fit_ccp_3dot5} 
 \end{minipage}%
 \hfill
 \begin{minipage}[t]{0.32\textwidth}
    \centering
    \includegraphics[width=\linewidth]{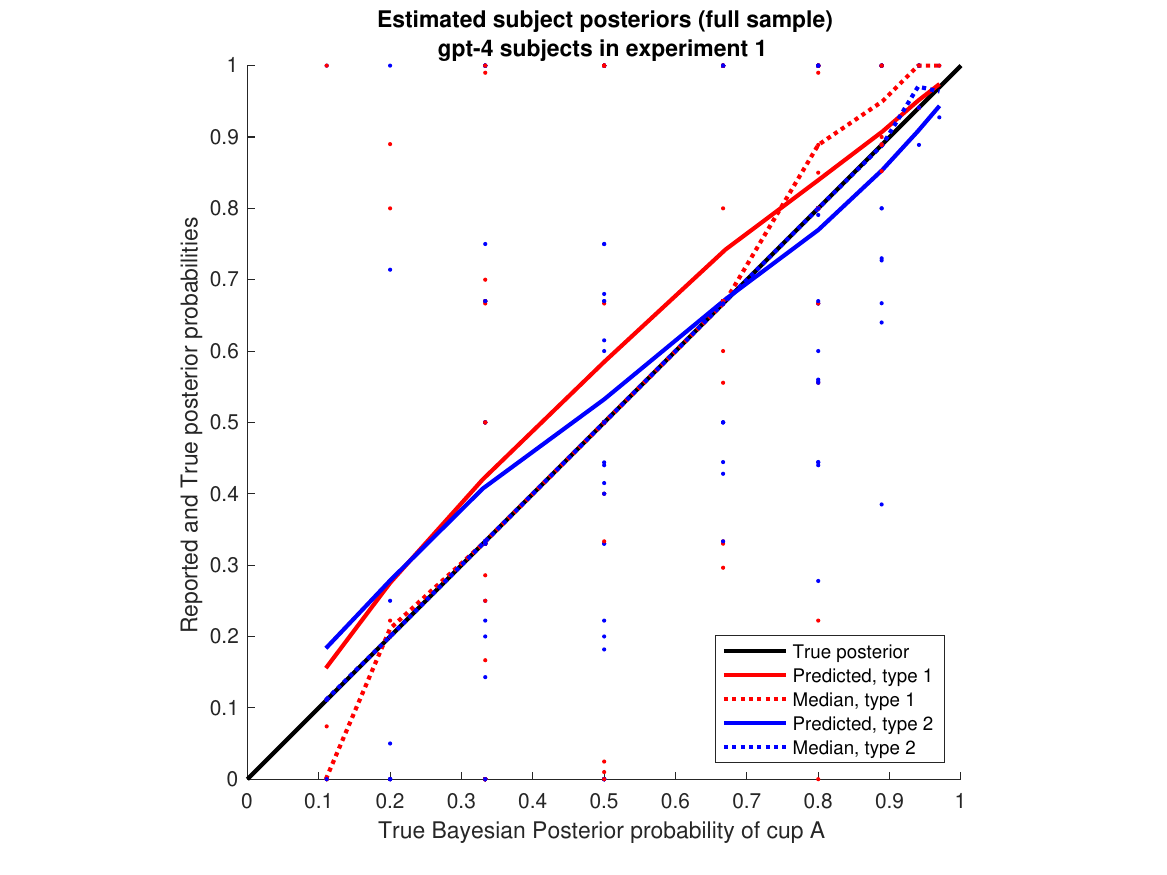}
    (b) GPT-4 
     \label{fig:hs_model_fit_ccp_4} 
 \end{minipage}%
 \hfill
 \begin{minipage}[t]{0.32\textwidth}
    \centering
    \includegraphics[width=\linewidth]{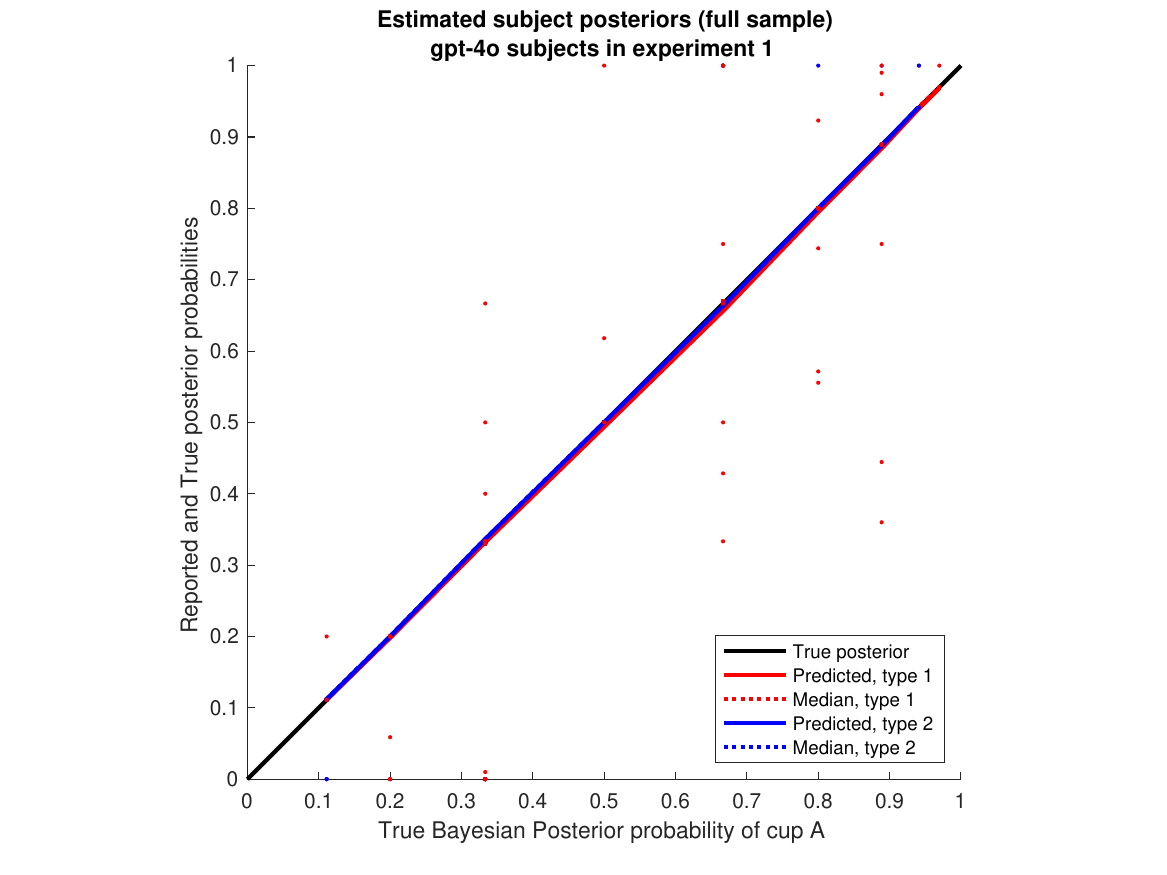}
    (c) GPT-4o 
     \label{fig:hs_model_fit_ccp_4o} 
 \end{minipage}
\end{figure}

Figure \ref{fig:hs_model_fit_ccps} plots, for the 3 versions of GPT subjects, their reported posterior probabilities against the corresponding Bayesian posterior probabilities, each represented as a dot.\footnote{In one trial, a subject from GPT 3.5 reported a posterior probability of $2$, which is excluded from Figure \ref{fig:hs_model_fit_ccp_3dot5}. However, we do not exclude this observation from our sample.} The EC algorithm  identifies two types of subjects for all three GPTs, denoted by red dots (type 1) and blue dots (type 2), respectively. 

The dotted lines illustrate the median reported posterior probabilities derived from the data. It is evident that each successive generation of
GPT subjects report posterior beliefs that increasingly align with the
true Bayesian posterior, as evidenced by their proximity to the 45-degree Bayesian line. GPT-3.5 frequently makes numerical mistakes in calculating its posterior, with both the frequency and severity of these inaccuracies escalating when true posterior probabilities are more extreme, particularly in the ranges below 0.2 and above 0.8, where discrepancies are especially pronounced. For GPT-4, the median subjective posterior probabilities of type 1 subjects closely align with the Bayesian 45-degree line, except that they may slightly overestimate the posterior probabilities when the true posteriors are close to 1. Type 2 subjects also approximate Bayesian decision-making in most trials, except for those with true posterior probabilities near 0, where subjects tend to underestimate probabilities, and near 1, where they tend to overestimate. The median subjective posteriors for both types of subjects in GPT-4o are almost identical to the Bayesian decision makers. 

The solid red and blue lines in Figure \ref{fig:hs_model_fit_ccps} represent the model-predicted median of subject responses, constructed from the estimated $\beta$'s. Consistent with the findings from the Wisconsin experiments, we observe a markedly improved model fit as subjects approach Bayesian decision-making. GPT-3.5 exhibits numerous errors in its responses, which makes it challenging for the simple structural logit model to capture the full range of atypical behaviors. 

The structural logit model fits much better in GPT-4 and GPT-4o. However, both subject types identified in GPT-4 consistently overestimate posterior probabilities when the true posteriors are below 0.7, while underestimating them otherwise. Both types assign less weight to LPR, however, the type 1 subjects are closer to Bayesian, as evidenced by a smaller difference between the coefficients on LPR and LLR. 

The model-predicted and data posterior probabilities both coincide with the Bayesian 45-degree line, illustrating an almost perfect model fit. Both types of GPT-4o subjects put almost equal weights on LPR and LLR and have negligible bias, so their posterior beliefs are close to Bayesian. The two types are different mainly in their estimated noise parameters. 

Beyond the parameter $\beta$'s in the subjective beliefs, the estimated noise parameter $\widehat{\eta}$ reveals another aspect of improvement, namely, the level of calculation noise in their subjective beliefs reduces in each successive generation. For GPT 3.5, the degree of noise is large for both types, with $\widehat{\eta}=2.8$ for type 1 subjects and $\widehat{\eta}=2.1$ for type 2 subjects. In contrast, the level of noise is significantly smaller for GPT-4, with $\widehat{\eta}=0.8$ for type 1 subjects and $\widehat{\eta}=1.6$ for type 2 subjects. By the time we get to GPT-4o 
there is almost no estimated noise for type 2 subjects,  $\widehat{\eta}=0.0025$, which means that effectively they are perfect, noiseless Bayesians. In contrast, type 1 subjects have a slightly larger noise of $\widehat{\eta}=0.4$, making them noisy Bayesian decision makers.

\begin{figure}[H]
 \centering
 \caption{Loss Functions by Types}%
 \label{fig:hs_loss_by_types}
 \begin{minipage}[t]{0.32\textwidth} 
    \centering
    \includegraphics[width=\linewidth, trim= 15 1 15 .5, clip=true]{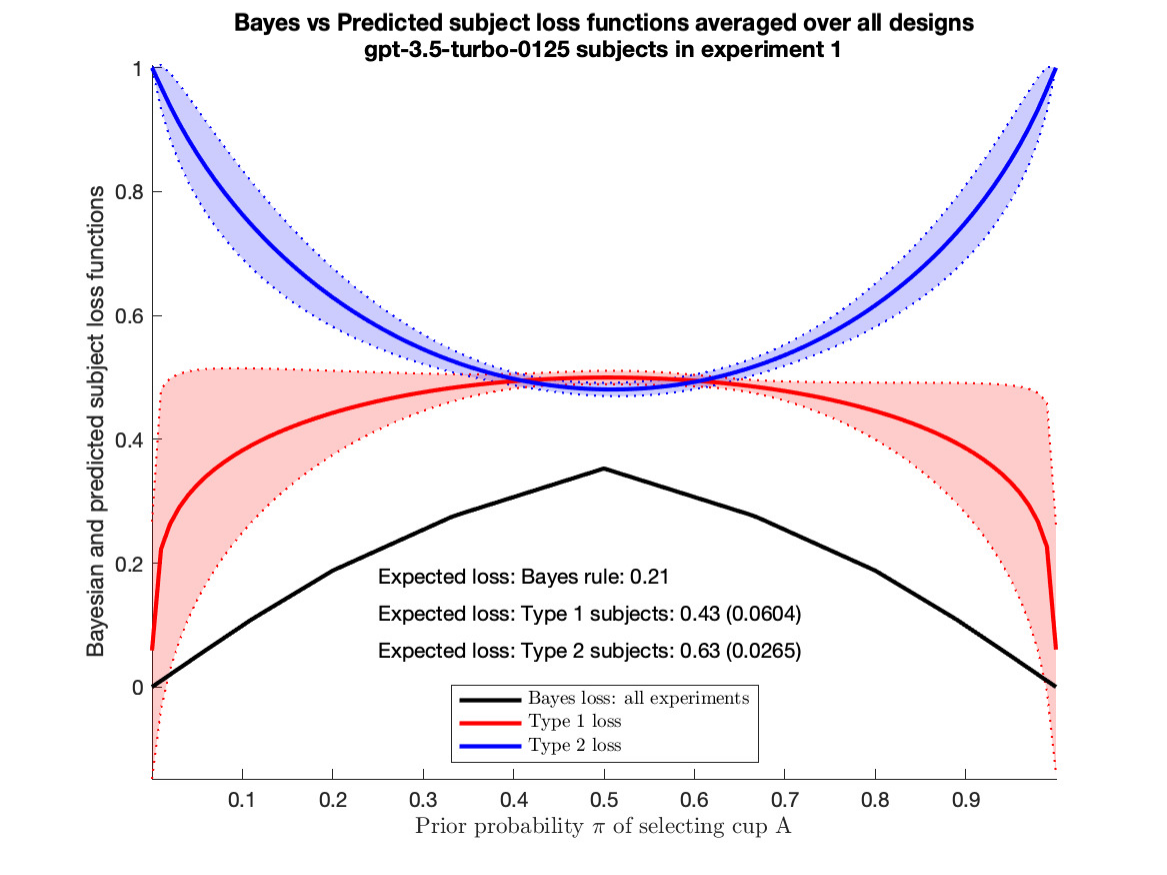}
    (a) GPT 3.5 
    \label{fig:hs_loss_by_type_3dot5} 
 \end{minipage}%
 \hfill 
 \begin{minipage}[t]{0.32\textwidth}
    \centering
    \includegraphics[width=\linewidth, trim= 15 1 15 .5, clip=true]{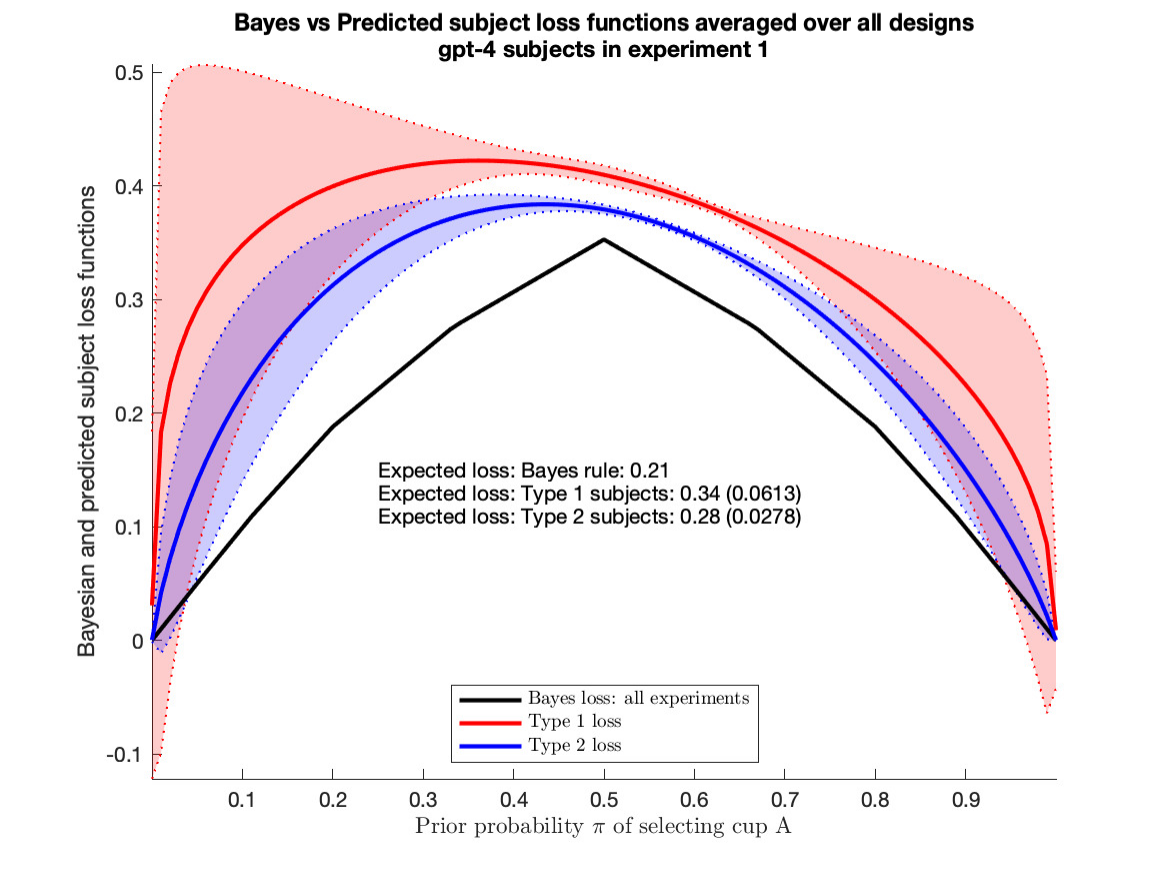}
    (b) GPT-4 
     \label{fig:hs_loss_by_type_4} 
 \end{minipage}%
 \hfill 
 \begin{minipage}[t]{0.32\textwidth}
    \centering
    \includegraphics[width=\linewidth, trim= 15 1 15 .5, clip=true]{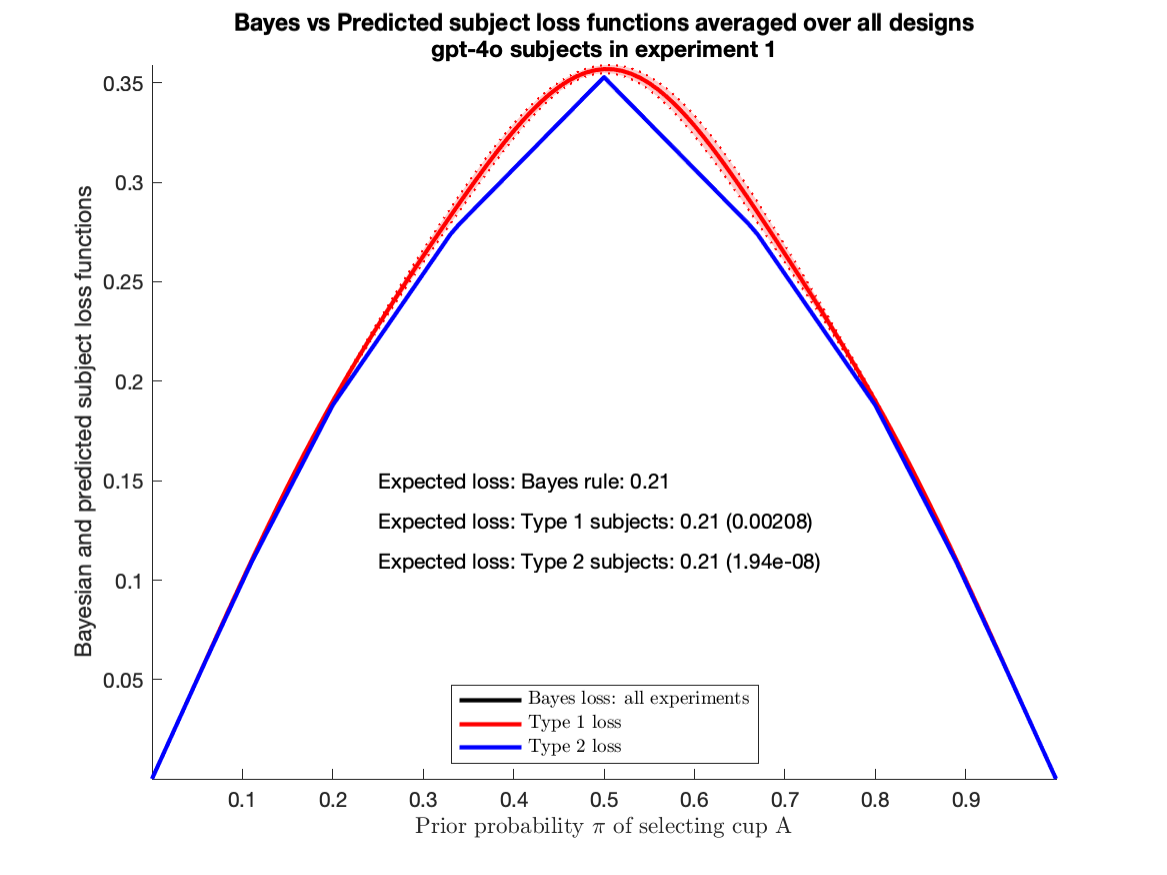}
    (c) GPT-4o 
     \label{fig:hs_loss_by_type_4o} 
 \end{minipage}
\end{figure}

To facilitate a comparison of decision efficiency across different GPTs and between GPTs and human subjects, we calculate the expected loss functions for various prior probabilities of selecting cage A and plot them in Figure \ref{fig:hs_loss_by_types}. Both the proximity of subjective beliefs to Bayesian beliefs and smaller noise levels contribute to a smaller expected loss. Because the subjective beliefs are closer to Bayesian beliefs and there is less noise affecting the beliefs of the more advanced GPT 
versions, the expected loss functions are closer to the efficiency benchmark as shown in moving from Figure \ref{fig:hs_loss_by_type_3dot5} to Figure \ref{fig:hs_loss_by_type_4o}.

Finally, table~\ref{table:hssummary} summarizes the performance of human and 
GPT subjects in Holt and Smith's experiment 1. The table reports our decision efficiency and accuracy scores as well as a new measure of ``accuracy'', namely, the $R^2$ of a regression of the reported posterior on a constant and the true posterior. Values closer to 1 for each of these measures can be interpreted as ``more Bayesian''.  We see a steady progression in efficiency and accuracy of the GPT subjects from GPT 3.5 to GPT 4o. The human subjects are superior to GPT 3.5 and 4 in terms of efficiency, but GPT-4o shows superhuman performance, especially in terms of decision efficiency and accuracy as measured by $R^2$, reflecting the fact that reported posteriors by GPT subjects are significantly closer to the true Bayesian posterior on average than the noisier human subjects. 

The EC algorithm finds 2 types of subjects for each version of GPT and human subjects, and in each case the key difference is the level of estimated noise (captured by the standard deviation parameter $\eta$ representing ``calculational errors'' in the structural logit model). Except for GPT 3.5, both types of subjects are ``noisy Bayesians'' but one of the types makes significantly larger calculational errors than the other. Among the GPT-4o subjects, 45\% are classified as essentially perfect noiseless Bayesians 
with 100\% decision efficiency. For human subjects, 45\% are Bayesians also, but due to the greater level of noise in their response their decision efficiency is lower, 95\%.

For GPT 4 and GPT 4o this noise correlates directly with the ``temperature'' parameter we set to generate heterogeneity among the GPT subjects. For example for GPT-4o, the noisier subjects had average temperature of .75 (std error .30), twice the average temperature of the less noisy subjects identified by the EC algorithm. Thus, as we would expect  higher temperature implies noisier responses by GPT subjects and this noise leads to lower efficiency and accuracy. We can obtain even better performance for the GPT-4o subjects by reducing the temperature parameter.

\begin{table}[H]
\centering
\caption{Performance of humans and GPT in the Holt-Smith Experiments}
\label{table:hssummary}
\begin{tabular}{lcccc}
\toprule
                            & GPT 3.5       & GPT-4         & GPT-4o            & Humans \\
\midrule
Efficiency             &  75.0              & 93.0     & 99.4 & 96.0 \\
                       & (1.0\%)       & (0.4\%)       & (0.3\%)  & (0.7\%)   \\
Accuracy              &   58.1        & 84.0          &   98.2   &  87.4      \\
                      & (0.1)       & (0.1)        &    (0.02)       &  (0.1) \\
$R^2$                 & 0.7          & 41.8              & 88.0  &  63.5 \\
 
No. of Types             & 2             & 2             & 2                 & 2 \\
\bottomrule
\end{tabular}
\end{table}

\subsection{Context Effects}

Echoing David Grether's criticism that humans are more influenced by context effects in "real world" situations, we test whether GPT subjects exhibit similar biases. We randomly sample 100 trials in the Wisconsin 6-ball experiment for each GPT model, where temperatures were near the median, and replace \textit{bingo cages and balls} in the prompt as \textit{ponds and fishes}, where humans may have more experience and GPTs might learn from that. Table \ref{tab:context_effects} compares the accuracy of the original bingo experiments with the reframing pond experiments. Context effects are rejected by the t-test for all three GPT models we considered, as none are even close to the conventional significance levels.



\begin{table}[H]
\centering
\caption{Accuracy Comparison between Bingo and Pond Experiments}
\label{tab:context_effects}
    \begin{tabular}{lcccccc}
        \toprule
        Model & Bingo & Pond & t-stat & p-value & N\\
        \midrule
        GPT-3.5           & 0.63 & 0.69 & -0.84 & 0.40 & 100 \\
        GPT-4             & 0.85 & 0.83 & 0.45  & 0.66 & 100 \\
        GPT-4o            & 0.94 & 0.92 & 0.63  & 0.53 & 100 \\
        \bottomrule
    \end{tabular}
\end{table}

\subsection{Conclusions from the analysis of GPT subjects}

Below we list the key conclusions from our analysis of the El-Gamal and Grether Wisconsin experiments and the Holt and Smith experiments using GPT subjects:

\begin{blockquote}
\begin{itemize}
\singlespacing
\item We find a rapid improvement in the accuracy and decision efficiency in successive versions of GPT, from sub-human performance for GPT 3.5, to approximately human for GPT-4, and super-human performance for GPT-4o. 
\smallskip
\item The estimates of the multiple-type structural logit model reveal three key aspects of this improvement in performance over successive generations of GPT: (i) decreasing bias, (ii) increasingly equal weight on the data and the prior,  and (iii) lower levels of idiosyncratic calculational and decision noise. 
    \smallskip
\item We find less unobserved heterogeneity for GPTs than humans in the Wisconsin experiments, especially for GPT-4 and 4o, where we find only one type of subject,  noisy Bayesians. In the Holt and Smith experiment,  40\% of subjects are classified as perfectly Bayesian, while the remaining 60\% are best described as noisy Bayesians.
\smallskip
\item As in the case of human subjects, a likelihood ratio test of ``structural stability''  rejects the hypothesis that the structural logit model is the correct model of the choices of GPT subjects. However, the P-values of these rejections increase with each successive version of  GPT, indicating that the structural logit model provides an increasingly good approximation to the behavior of successive generations of GPT.      
\end{itemize}
\end{blockquote}

\section{Analysis of Errors from AI Subject Response Text}
\label{section:textual}

Most econometric models treat humans as black boxes and apart from some research from neuroscience, we do not know exactly how humans process information. However, from the textual responses of GPT, we have the unique advantage of observing the reasoning of GPT subjects, opening the door to analyze where GPTs make mistakes. There are two challenges in such an analysis: the errors made by GPTs can be highly diverse, and the textual responses are not well-structured. 

We overcome the first challenge by exploiting the simple structure of the binary decision problem, which allows us to classify errors into nine binary error flags under four broad categories. To obtain a distribution of GPT errors across categories, we then develop a GPT grader to efficiently process large-scale unstructured responses and determine the value of error flags within each category. Key inputs for the GPT grader include a reference answer using Bayes' rule, detailed grading rubrics, original experiment prompts and responses. We present our grading prompt in \ref{appendix:grading_prompts}.

We use a more advanced version, GPT o3-mini, to grade three less advanced models considered, ensuring the grader has superior performance in the binary classification tasks and general intelligence. We also manually reviewed 50 randomly sampled textual responses for each model to verify the GPT grader's performance. Overall, our independent cross-check confirmed the accuracy of GPT o3-mini's grading and classification of errors in the responses from ChatGPT 3.5, 4 and 4o.\footnote{In \ref{appendix:grader_validation}, we report the error rates for the same set of 50 samples graded by GPT o3-mini and a human grader. Additionally, we include the grading results using the most advanced version of GPT o1, which, while offering slightly superior performance, is significantly more costly.}

\subsection{Error Taxonomy}

We focus on the textual responses from the 6-ball Wisconsin experiment. Table \ref{tab:textual_error_distribution} presents the four broad categories and the error flags under each category.\footnote{To demonstrate the error flags, we provide excerpts from textual responses as example answers classified under each error flag in \ref{appendix_error_flag_example}. } The first type of error under Panel A examines whether GPTs understand the context and correctly interpret experimental parameters, including cage composition (i.e., the number of $N$ balls in each cage), sample size ($D$), and the observed number of $N$ balls ($d$). 

The second type evaluates whether GPT subjects are ``conceptually'' Bayesian by checking if subjects take the prior information into consideration and if they use the sample information as inputs to their decision. Failure to use this information can explain choices that are consistent with conservatism and representativeness. We emphasize the \textit{consideration} of both prior and likelihood in the second category, leaving the examination of \textit{numerically correct} posterior calculation to the third category under Panel C. 

In Panel C, we check whether the prior probability ($\pi_A$),\footnote{To calculate the prior probability, subjects must understand the process of rolling a 10-sided die and divide the specified range of results for Cage A by 10. While this is straightforward for humans, we occasionally find that GPT subjects use an incorrect numerator or denominator when calculating the prior probability for Cage A. See \ref{appendix_error_flag_example} for an example. }  the likelihood probabilities ($f(d|p_A,D)$ and $f(d|p_B,D)$) and the posterior ($\frac{\pi_A f(d|p_A,D)}{\pi_A f(d|p_A,D)+(1-\pi_A) f(d|p_B,D)}$) or posterior odds ($\frac{\pi_A f(d|p_A,D)}{(1-\pi_A) f(d|p_B,D)}$) are correctly calculated. 

There are a few points worth discussing. First, we explicitly instruct the GPT grader to allow for the omission of binomial coefficients when calculating the likelihood, as they will cancel out and do not affect the posterior calculation. Second, we find that some GPT subjects, instead of calculating the posterior probability $\Pi(A|d,\pi,p_A,p_B,D)$ as in equation \ref{eq:bayes_rule}, make decisions by calculating and comparing the product of prior and likelihood, $\pi \times f(d|p_A,D)$ and $(1-\pi) \times f(d|p_B,D)$. This approach is consistent with Bayes' rule, and in such cases, the subject passes error flag 8. Third, similar to the identification challenge discussed in \ref{section:identification}, it is sufficient for subjects to make correct decisions if the posterior is on the same side of $1/2$ as the Bayesian posterior. However, since GPT subjects usually report the posterior probabilities, we apply a stricter grading rubric by marking the subject as making an error in flag 8 if their calculated posterior (or posterior odds) was incorrect regardless of whether it leads to the same decision as the true posterior. 

The final category under Panel D examines whether the final decision (Cage $A$ or $B$) aligns with the previous reasoning, which substantiates the decision noise $\varepsilon$ in the structural logit model. We instruct the GPT grader to read the overall reasoning of the textual responses and then predict the expected outcome based on the reasoning flow. Almost all GPT-4 and 4o subjects answer the problem by calculating the posterior or posterior odds. In such cases, we define error flag nine, final decision contradicting the previous reasoning, as choosing cage $A$ when the posterior of Cage $A$ is below $\frac{1}{2}$, or when $\pi \times f(d|p_A,D)$ is smaller than $(1-\pi) \times f(d|p_B,D)$. In other cases, we ask the GPT grader to explicitly predict the outcome based on the reasoning just before the final answer and compare whether this prediction is consistent with the subjects' report.  

\subsection{Grading Results}

We apply the grading prompt to evaluate 500 randomly selected text responses for each GPT model, regardless of whether the final decision aligns with Bayes' rule. The output of the grading algorithm assigns a value to each error flag, with 1 indicating that the student's response contains a mistake. Table \ref{tab:textual_error_distribution} presents the error rates for nine types of errors. We also report the fraction of responses inconsistent with the Bayes' rule at the bottom of the Table.\footnote{The sum of error rates across all categories doesn't necessarily match the fraction of incorrect responses for two reasons. First, errors are not mutually exclusive. If a subject ignores the prior (error flag 4 = YES), they will also have errors in calculating the prior (error flag 6 = YES) and subsequently the posterior (error flag 8 = YES). Second, since the final decision's consistency with Bayes' rule only requires comparing the posterior to $\frac{1}{2}$, an error in calculating the posterior does not necessarily lead to a mistake in the final decision.} 

\begin{table}[H]
  \centering
  \caption{Error Distribution from GPT Responses}
  \begin{adjustbox}{width=\textwidth}
    \begin{tabular}{rlccc}
    \toprule
    \toprule
          & Error Flag (Yes/No) & GPT 3.5 (\%)  & GPT 4 (\%) & GPT 4o (\%) \\
    \midrule
    \multicolumn{2}{l}{Panel A. Data read-in errors} \\
     \midrule
    1     &  Error reading the compositions of the two cages  & 0.4     & 0.4     & 0.0 \\
    2     &  Error reading the number of balls drawn from  the two cages  & 0.0     & 0.0     & 0.0 \\
    3     &  Error reading the outcome of the draws  & 0.0     & 0.0     & 0.0 \\
     \midrule
    \multicolumn{2}{l}{Panel B. Errors in the application of Bayes Rule} \\
     \midrule
    4     &  Ignoring the prior  & 82.4     & 1.4     & 0.6 \\
    5     &  Ignoring the likelihood  & 0.2    & 0.0     & 0.0 \\
     \midrule
    \multicolumn{2}{l}{Panel C. Errors in computing the posterior probability} \\
     \midrule
    6     &  Error calculating prior probability  & 83.4     & 2.2     & 0.6 \\
    7     &  Error calculating the likelihood  & 70.0     & 67.8     & 13.6 \\
    8     &  Error calculating the posterior (or posterior odds)  & 98.4     & 72.0     & 21.8 \\
     \midrule
    \multicolumn{2}{l}{Panel D. Errors in the final decision} \\
     \midrule
    9     &  Final decision contradicting the previous reasoning & 4.4     & 3.0     & 4.6\\
     \midrule\\
  \multicolumn{2}{l}{ Fraction of responses inconsistent with Bayes' rule} & 35.2 & 14.2 & 9.2 \\  
 \bottomrule
    \end{tabular}%
  \label{tab:textual_error_distribution}%
  \end{adjustbox}
\end{table}%

Table \ref{tab:textual_error_distribution} reinforces our conclusion about the rapid improvement from GPT-3.5 to GPT-4 and GPT-4o, as GPT-4 and GPT-4o make significantly fewer errors in almost all types. This not only reflects higher accuracy, but also indicates fewer mistakes in calculating posterior probabilities or other steps even when final decisions align with Bayes' rule.
Panel A shows that GPT subjects rarely make mistakes in reading experimental parameters, suggesting a good understanding of the experimental setup. 
We observed a remarkable transformation from non-Bayesian to conceptual Bayesian reasoning from GPT-3.5 to GPT-4. More than 80\% of GPT-3.5 decisions rely only on likelihood, lacking a Bayesian rationality even conceptually. This aligns with our findings in Section \ref{section:ai_subjects}. Our manual review of the text responses shows that they often use representativeness heuristics, making decisions by matching observed patterns in the sample with the composition of the two cages, or simply comparing the likelihood of each cage being the sample's source. Interestingly, GPT-3.5 almost never ignores information from the sample, suggesting it is less prone to conservative bias. GPT-4 and GPT-4o almost always demonstrate at least "conceptual Bayesian" reasoning, either by explicitly writing down Bayes' formula or informally considering both prior and likelihood information.  
Panel C highlights another major achievement in transitioning from "conceptual Bayesian" to "perfect Bayesian" when moving from GPT-4 to GPT-4o. In GPT-4o, 78\% of responses correctly calculate the posterior, compared to just 28\% for GPT-4. GPT-3.5 struggles to calculate both the likelihood and prior probability, often due to overlooking prior information.
Panel D indicates that all models experience decision noise. Surprisingly, although GPT-4o generally outperforms GPT-4, its final decisions are more likely to be inconsistent with the calculated posterior, whether comparing it to $\frac{1}{2}$ or comparing the numerator and denominator of the posterior odds.  To understand it, we delve deeper into the sample graded by the human grader, who also confirms this observation.\footnote{The human grader reports an error rate of 14\% for GPT-4o and 2\% for GPT-4 regarding the decision inconsistency error flag. See \ref{appendix_grader_validation}.} First, we find that GPT-4o is more likely to report the posterior as fractional numbers than GPT-4,\footnote{Out of the 50 samples, 86\% are reported as fractional numbers in GPT-4o, with the remainder as rounded decimals. This percentage decreases to 66\% in GPT-4. }   probably due to its ability to more accurately calculate the posterior, as shown in Panel C. However, the fractional numbers reported by GPT-4o, though precise, are more complex with more digits\footnote{See \ref{appendix_error_flag_example} for an example where the posterior for Cage A is calculated and reported as  $\frac{2612736}{4200459}$, which equals 0.62 as a decimal.  The GPT-4o model should have chosen Cage A, but instead, it chose Cage B.}, making comparisons for the final decision more challenging.\footnote{Out of 17 responses where GPT-4o reported posteriors as fractional numbers, 7 resulted in mistakes during the final decision. In contrast, none of the 7 cases with fractional posteriors from GPT-4 had such issues.  We note that only 1 decision was inconsistent with the calculated posterior out of 43 cases for GPT-4 and 33 cases for GPT-4o when using rounded decimals.} Therefore, although GPT 4o can calculate the posterior more accurately, it may shoot itself in the foot when making final decisions by comparing these  numbers. 
\\ 

\subsection{Conclusion from Analysis of Textual Responses}

We summarize our main finding from the analysis of textual responses here.
\begin{enumerate}
    \item Analyzing the textual responses confirms the rapid improvement from GPT 3.5, to GPT-4 and 4o as more advanced GPTs make less mistakes in almost all error categories.
    \item We observed a remarkable transformation from non-Bayesian to conceptual Bayesian reasoning from GPT-3.5 to GPT-4.
    \item Transitioning from GPT-4 to GPT-4o represents another major leap from ``conceptual Bayesian" to ``perfect Bayesian," achieving the accurate calculation of a Bayesian posterior.
\end{enumerate}

\section{Conclusion}
\label{section:conclusion}
This paper examines whether humans or ChatGPTs more closely resemble a Bayesian decision maker in a simple binary classification task. Although the task is straightforward, Bayesian rationality is fundamental in decision theory and broadly applicable to many important contexts.

We introduce a parsimonious structural logit model with unobserved heterogeneity to infer subjects' subjective beliefs. Our model demonstrates greater predictive power for human behaviors than those used in previous studies. Additionally, we develop a novel decision efficiency measure, allowing us to compare different groups with varying task difficulties, addressing limitations of the popular accuracy measure based on Bayes' rule consistency.

We estimate the structural logit model using data from human experiments by \citet{EGG1995}, \citet{EGG1999}, and \citet{HS2009}, alongside our replication using various ChatGPTs. Our study differs from previous literature in two key ways. First, we account for significant heterogeneity across GPT versions, showing that conclusions vary due to rapid improvements in Bayesian rationality. We document the evolution from sub-human performance in early GPT-3.5 to near-human performance in GPT-4, and superior performance in GPT-4o. Second, we highlight considerable heterogeneity among human subjects, as revealed by our model. The most efficient humans can closely resemble Bayesian decision makers, comparable to the most advanced GPTs. We also find less unobserved heterogeneity in GPT subjects, particularly in more advanced versions.

As a first step in unraveling the decision mechanisms of GPTs, we leverage their ability to display reasoning in textual responses to analyze where they make mistakes. By exploring the simple structure of binary choice models, we categorize errors and provide clear guidance for GPT-4o graders to interpret these responses. Examining the extensive textual data reveals a remarkable shift from non-Bayesian to conceptual Bayesian reasoning from GPT-3.5 to GPT-4, with GPT-4o making another exceptional transition to accurately calculating posterior probabilities.


\setcitestyle{maxcitenames=2} 

\singlespacing

\newpage
\appendix
\renewcommand{\thesection}{Appendix \Alph{section}}
\renewcommand{\thefigure}{\Alph{section}\arabic{figure}}

\setcounter{figure}{0}
\section{Proof of Lemma~\ref{lemma:two}}
\label{appendix:A}
This appendix provides the proof of identification of the
structural logit model when $\sigma > 0$, Lemma~\ref{lemma:two} of section~\ref{section:identification}.
We start by assuming that the agent's decision rule $P(A|\pi,n)$ is identified for all possible values
of $\pi \in [0,1]$ and all possible experiments involving draws from one of two bingo cages where the number
of draws from these cages, and hence the possible values of $n$, can be arbitrarily large. 
In fact because we assume that the experimenter can run an arbitrary number of experiments and
observe the subject outcomes, the experimenter has the freedom to design experiments with
arbitrary values for the log-likelihood ratio $\mbox{LLR}$ and the log-prior odds ratio $\mbox{LPR}$,
so we assume these quantities can take any value between $-\infty$ and $+\infty$ via an appropriately
designed set of experiments. Further, since the subject's decision rule depends on $(\pi,n)$ via
the quantities $(\mbox{LPR}(\pi),\mbox{LLR}(n))$ and the experimenter has the freedom to design experiments
that allow $\mbox{LPR}$ and $\mbox{LLR}$ to take on any values in $R^2$, we can treat the subject's
decision rule as a known function of two continuous arguments, $(\mbox{LPR},\mbox{LLR})$, which constitute
``sufficient statistics'' for the subject's posterior beliefs, and hence for the subject's decision rule.
In short, we can assume that the subject's decision rule is a known conditional probability
of the form  $P(A|\mbox{LPR},\mbox{LLR})$  mapping $R^2$ into $[0,1]$. 

Now, let the subject's true decision rule be given by the function
\begin{equation}
   P(A|\mathrm{LPR},\mathrm{LLR},\sigma^*,\beta^*) = \frac{ 1 }{ 1 + \exp\{[2\Pi_s(A|\mathrm{LPR},\mathrm{LLR},\beta^*)-1]/\sigma^*\} },
\label{eq:a1}
\end{equation}
and by assumption, this function is identified, i.e. its value is known for any pair $(\mbox{LPR},\mbox{LLR}) \in R^2$.
Now suppose there is some other structural logit that is observationally equivalent to the true one.
That is, suppose there is a function $P(A|\mbox{LPR},\mbox{LLR},\sigma,\beta)$ such that
\begin{equation}
P(A|\mbox{LPR},\mbox{LLR},\sigma^*,\beta^*) = P(A|\mbox{LPR},\mbox{LLR},\sigma,\beta), \quad \forall (\mbox{LPR},\mbox{LLR}) \in R^2.
\label{eq:a2}
\end{equation}
We will now that that if this is true, then  it must be the case that $\sigma=\sigma^*$, and $\beta=\beta^*$. That is,
the parameters of the structural logit model are identified.

We can show this under a {\it full support condition\/} which is a version of an ``identification at infinity''
type of argument. The full support condition implies that the experimenter can conduct sufficient experimentation
on subjects that the log-prior ratio $\mbox{LPR}$ takes on any value on the real line. It follows that by taking
the limit as $\mbox{LPR} \to \infty$ we have $\Pi_s(A|\mbox{LPR},\mbox{LLR}) \to 0$. Since equation (\ref{eq:a2}) holds
for all $(\mbox{LPR},\mbox{LLR}) \in R^2$ and is a continuous function of these variables, it follows that the equality must
hold in the limit so we have
\begin{equation}
      \frac{1}{ 1 + \exp\{1/\sigma^*\} } = \frac{1}{ 1 + \exp\{1/\sigma\} },
\end{equation}
which implies that $\sigma^*=\sigma$, so this parameter is identified. Using this result we can immediately conclude
from equations (\ref{eq:a1}) and (\ref{eq:a2}) that 
\begin{equation}
\Pi_s(A|\mbox{LPR},\mbox{LLR},\beta^*) = \Pi_s(A|\mbox{LPR},\mbox{LLR},\beta).
\end{equation}
Since $\Pi_s(A|\mbox{LPR},\mbox{LLR},\beta)=1/(1+\exp\{\beta_0+\beta_1\mbox{LPR}+\beta_2\mbox{LLR}\}$, it follows that we have
\begin{equation}
       \beta^*_0+\beta^*_1\mbox{LPR}+\beta^*_2\mbox{LLR} = \beta_0+\beta_1\mbox{LPR}+\beta_2\mbox{LLR}, \quad
\forall (\mbox{LPR},\mbox{LLR}) \in R^2.
\label{eq:a3}
\end{equation}      
Let $\mbox{LPR}=\mbox{LLR}=0$ (which is possible due to the full support assumption). If follows from equation (\ref{eq:a3}) that 
$\beta^*_0=\beta_0$. Next, set $\mbox{LLR}=0$ and $\mbox{LPR}=1$, and it follows that $\beta^*_1=\beta_1$.
Finally, set $\mbox{LLR}=1$ and $\mbox{LPR}=0$, and it follows that $\beta^*_2=\beta_2$. We conclude that the parameters
$(\sigma^*,\beta^*_0,\beta^*_1,\beta^*_2)$ of the structural logit model are identified.

\newpage

\setcounter{figure}{0}
\section{Algorithm for Data Collection from LLMs}

\label{appendix:B}
\begin{algorithm}[H]
\caption{Data Collection and Processing}
\begin{algorithmic}[1]
    \State \textbf{Initialize} data collection specifications based on the experiment specified by \texttt{run\_name}, generate textual prompts and model settings, and save to disk
    \State \texttt{\textbf{\textquotesingle send\textquotesingle}} prompts and model settings to OpenAI via API
    \While{any request is \textbf{not} completed}
        \For{every request}
            \State \texttt{\textbf{\textquotesingle retrieve\textquotesingle}} the request's status
            \If{the request has been completed \textbf{and not} retrieved}
                \State Retrieve the responses and save to disk
            \ElsIf{the request has failed}
                \State \texttt{\textbf{\textquotesingle resend\_failed\textquotesingle}} request(s)
            \ElsIf
                \State \textbf{continue}
            \EndIf
        \EndFor
        \State \texttt{\textbf{\textquotesingle finalize\textquotesingle}} the responses, which includes:
        \State \hspace{1em} Align responses with the original prompts
        \State \hspace{1em} Parse the responses into the final answers like Cage A or B or a numerical value
        \State \hspace{1em} Append metadata and informational columns
        \State \hspace{1em} Save different formats of the collected data to disk
        \State \hspace{1em} Check for invalid responses that cannot be parsed into a final answer
        \If{any response is invalid}
            \State \texttt{\textbf{\textquotesingle resend\_invalid\textquotesingle}} prompts and model settings to OpenAI via API
        \EndIf
    \EndWhile
\end{algorithmic}
\end{algorithm}

\newpage

\setcounter{figure}{0}
\newpage
\section{Prompts used to collect data from LLMs}
\label{appendix:C} 

In this section, we provide our prompt to replicate using LLMs the two experiments at the University of Wisconsin-Madison and the two experiments reported in \citet{HS2009}.

There are no \texttt{developer} / \texttt{system} messages, only one \texttt{user} message for each chat completion request for each trial. The exact numerical values will reflect the specifications of the actual trial, the followings are examples.

\subsection*{C.1 Wisconsin}
For the experiment that allows for \texttt{reasoning}, the example \texttt{user} message is:
\begin{lstlisting}
You are participating in a decision-making experiment, where you can earn money based on the number of correct decisions you make.

There are two identical bingo cages, Cage A and Cage B, each containing 6 balls. Cage A contains 4 balls labeled "N" and 2 balls labeled "G", while Cage B contains 3 balls labeled "N" and 3 balls labeled "G".

A 10-sided die is used to determine which of the two cages will be used to generate draws. If a random roll of the die shows 1 through 3, I will use Cage A; if it shows 4 through 10, I will use Cage B. You will not know the outcome of the roll of the die or which cage I use.

Once a cage is chosen at random based on the roll of the die, it is used to generate draws with replacement.

I have drawn a total of 6 balls with replacement. The result is 3 "N" balls and 3 "G" balls.
After observing this outcome, which cage do you think generated the observations? Your decision is correct if the balls were drawn from that cage.

YOU ARE WELCOME TO ALSO DESCRIBE YOUR REASONING, BROKEN INTO SEPARATE STEPS, TO EXPLAIN HOW YOU ARRIVED AT YOUR FINAL ANSWER.
Please state your answer in the following format at the end.
"Final answer: Cage A." or "Final answer: Cage B.".
\end{lstlisting}

For the experiment that prohibits \texttt{reasoning}, the last section of the \texttt{user} message is substituted with:
\begin{lstlisting}
PLEASE JUST REPORT YOU FINAL ANSWER AND DO NOT PROVIDE ANY REASONING AS TO HOW YOU ARRIVED AT YOUR FINAL ANSWER.
Please state your answer in the following format.
"Final answer: Cage A." or "Final answer: Cage B.".
\end{lstlisting}

\subsection*{C.2 Holt and Smith}
For the experiment that allows for \texttt{reasoning}, the example \texttt{user} message is:
\begin{lstlisting}
This is an experiment in the economics of decision making. Various agencies have provided funds for the experiment. Your earnings will depend partly on your decisions and partly on chance. If you are careful and make good decisions, you may earn a considerable amount of money, which will be paid to you, privately, in cash, at the end of the experiment. In addition to the money that you earn during the experiment, you will also receive $6. This payment is to compensate you for showing up today.

This experiment involves two stages. In stage 1 we will show you some information including the result of a drawing of 1 ball from one of two possible cages, each containing different numbers of light and dark balls. Then at the start of stage 2 you will report a number P between 0 and 1. After your report, we will draw a random number U that is equally likely to be any number between 0 and 1. Your payoff from this experiment will either be $1000 or $0 depending on your report P and the random number U.

Let's describe the two stages in more detail now. In stage 1 we will show you 1 ball that are drawn at random from one of two possible urns labelled A and B.

Urn A contains 2 light balls and 1 dark ball.
Urn B contains 1 light ball and 2 dark balls.

We select the urn, A or B, from which we draw the sample of 1 ball by the outcome of throwing a 6 sided die.
We do not show you the outcome of this throw of the die but we do tell you the rule we use to select urn A or B.

If the outcome of the die throw is 1 to 3 we select urn A.
If the outcome of the die throw is 4 to 6, we use urn B to draw the random sample of 1 ball.

Once you see the outcome of the sample of 1 ball, stage 1 is over and stage 2 begins.

At the start of stage 2 we ask you to report a number P between 0 and 1. Your payoff from this experiment depend on another random number, which we call U, which we draw after you report the number P. We draw the random number U in a way that every possible number between 0 and 1 has an equal chance of being selected.

Here is how you will be paid from participating in this experiment. There are two possible cases:

Case 1. If the number U is less than or equal to P then you will receive $1000 if the sample of 1 ball we showed you in stage 1 was from urn A and $0 otherwise.
Case 2. If the number U is between the number P you report and 1, you will receive $1000 with probability equal to the realized value of U, but with probability 1-U you will get $0.

OK, this is the setup. Let's now start begin this experiment, starting with stage 1.

We have tossed the die (the outcome we don't show to you) and selected one of these urns according to the rule given above (i.e. urn A if the die throw was 1 to 3, and urn B otherwise). We have drawn 1 ball from the selected urn and the outcome is D, i.e., Dark.

Now, we are at stage 2 where we are asking you, given the information from stage 1 to report a number P between 0 and 1 that in conjunction with the random number U will determine if you get either $1000 or $0 according to the rule given in cases 1 and 2 above.

Please report a number P between 0 and 1 that maximizes your probability of winning $1000 in this experiment.

YOU ARE WELCOME TO ALSO DESCRIBE YOUR REASONING, BROKEN INTO SEPARATE STEPS, TO EXPLAIN HOW YOU ARRIVED AT YOUR FINAL ANSWER P.
Please state your answer in the following format at the end.
Final answer: [your P value here].
\end{lstlisting}

For the experiment that prohibits \texttt{reasoning}, the last section of the \texttt{user} message is substituted with:
\begin{lstlisting}
PLEASE JUST REPORT P AND DO NOT PROVIDE ANY REASONING AS TO HOW YOU ARRIVED AT THE VALUE P.
Please state your answer in the following format.
Final answer: [your P value here].
\end{lstlisting}
\label{appendix_prompts}
\newpage

\setcounter{figure}{0}
\section{Grading Prompts}
\label{appendix:grading_prompts} 

We provide a grading prompt that evaluates GPT-generated textual responses and identifies errors. First, we introduce the grading task and the role of the grader to GPT-01. Next, we present the experiment design prompt along with the original responses from GPT students. The key input for our grading prompt is a meticulously crafted step-by-step solution to the binary choice problem, which includes calculating the prior probability, determining the likelihood, applying Bayes’ rule to find the posterior probability, and making a final decision based on the posterior. This solution is crucial as a benchmark for the GPT grader to understand the binary decision task and identify errors in the student responses. We then provide detailed grading rubrics that allow the GPT grader to classify different error types. Finally, the prompt specifies the format for outputting error flags. We develop Python code to loop over different student responses, parameters of trials, and the corresponding benchmark solutions. \\
\\
\textbf{General Instruction about the Role of the Grader}
\begin{lstlisting}
You are an expert and detailed grader for this task:
\end{lstlisting}
\textbf{Introduce the Experiment Prompt }
\begin{lstlisting}
You are participating in a decision-making experiment, where you can earn money based on the number of correct decisions you make.

There are two identical bingo cages, Cage A and Cage B, each containing 10 balls. Cage A contains 4 balls labeled "N" and 6 balls labeled "G", while Cage B contains 6 balls labeled "N" and 4 balls labeled "G".

A 10-sided die is used to determine which of the two cages will be used to generate draws. If a random roll of the die shows 1 through 6, I will use Cage A; if it shows 7 through 10, I will use Cage B. You will not know the outcome of the roll of the die or which cage I use.

Once a cage is chosen at random based on the roll of the die, it is used to generate draws with replacement.

I have drawn a total of 7 balls with replacement. The result is 3 "N" balls and 4 "G" balls.
After observing this outcome, which cage do you think generated the observations? Your decision is correct if the balls were drawn from that cage.

YOU ARE WELCOME TO ALSO DESCRIBE YOUR REASONING, BROKEN INTO SEPARATE STEPS, TO EXPLAIN HOW YOU ARRIVED AT YOUR FINAL ANSWER.
Please state your answer in the following format at the end.
"Final answer: Cage A." or "Final answer: Cage B.".
\end{lstlisting}
\textbf{Student textual responses}
\begin{lstlisting}
Based on the outcome of 3 "N" balls and 4 "G" balls, here is my reasoning broken down into steps:

Step 1: Calculate the probability of getting 3 "N" balls and 4 "G" balls from each cage.

- Probability of getting 3 "N" balls and 4 "G" balls from Cage A: (4/10)^3 * (6/10)^4 = 0.00324
- Probability of getting 3 "N" balls and 4 "G" balls from Cage B: (6/10)^3 * (4/10)^4 = 0.00307

Step 2: Compare the probabilities of the two cages.

- The probability of getting the observed outcome is slightly higher for Cage A than Cage B.

Step 3: Consider the initial probabilities of each cage being chosen.

- Cage A has a 6/10 chance of being chosen, while Cage B has a 4/10 chance.

Step 4: Combine the probabilities of the observed outcome and the initial probabilities of the cages being chosen.

- Probability of getting the observed outcome from Cage A: 0.00324 * 0.6 = 0.00194
- Probability of getting the observed outcome from Cage B: 0.00307 * 0.4 = 0.00123

Step 5: Compare the combined probabilities.

- The combined probability of getting the observed outcome is higher for Cage A than Cage B.

Final answer: Cage A.
\end{lstlisting}
\textbf{Step-by-step Solution Benchmark }
\begin{lstlisting}
Step 1: Prior from die roll.
  p(A) = 3/5 = 0.6000;  p(B) = 2/5 = 0.4000
  Prior ratio p(A)/p(B) = 3/2 = 1.500000
Step 2: Cage compositions.
  Cage A: p(N|A) = 2/5 = 0.4000, p(G|A) = 3/5 = 0.6000
  Cage B: p(N|B) = 3/5 = 0.6000, p(G|B) = 2/5 = 0.4000
Step 3: Observed outcome & binomial likelihood.
  Observed: 3 'N', 4 'G' (total 7).
  L(A) = 4536/15625 = 0.290304, L(B) = 3024/15625 |approx| 0.193536
  Likelihood ratio L(A)/L(B) = 3/2 = 1.500000
Step 4: Posterior components & probabilities (fraction & decimal).
  p(A)*L(A) = 13608/78125 |approx| 0.174182
  p(B)*L(B) = 6048/78125 |approx| 0.077414
  Post(A)   = 9/13 = 0.692308
  Post(B)   = 4/13 = 0.307692
  Posterior ratio Post(A)/Post(B) = 9/4 = 2.250000
Step 5: Decision.
  Final answer: Cage A.
\end{lstlisting}
\textbf{Grading Rubrics}
\begin{lstlisting}
Evaluate the student's answer based on the following criteria:

Part I. Did They Make a Mistake When Reading the Data?

Instructions: For (1), (2), and (3), we want to see if the student understands the basic experimental setup and correctly incorporates the trial parameters into their reasoning. Focus on whether they read the relevant information accurately, not on how they use it later. For example, if a student correctly identifies the number of N balls in cages A and B but makes a comparison error later, you should still answer YES if the criterion is reading the number of N balls correctly.

(1) Cage Composition: Do they explicitly mention or implicitly acknowledge the number of N and G balls in each cage? Answer Yes or No.

(2) Draw Count: Do they explicitly mention or implicitly acknowledge the total number of balls in the sample? Answer Yes or No.

(3) Observed Data: Do they explicitly mention or implicitly acknowledge the number of N draws from the sample? Answer Yes or No.

Part II. Are they conceptually Bayesian?

Instructions: A Bayesian decision maker should consider both prior information (the announced probability of using a cage) and posterior information (the likelihood that the sample was drawn from a cage). Criteria (4) and (5) assess whether the student incorporates both prior and posterior information in their reasoning. We are not looking for explicit numerical calculations, but both types of information should be part of their reasoning process.

(4) Ignoring Prior: Do they make a decision using only the likelihood or observed data, ignoring the prior conceptually? Answer Yes or No.

(5) Ignoring Likelihood: Do they make a decision using only the announced probability of using cage A, ignoring the sample information conceptually? Answer Yes or No.

Part III. Can they correctly calculate the Bayesian posterior probability?

Instructions: To answer correctly, the student should apply Bayes' rule and calculate the posterior probability accurately. This involves three steps:

Prior Probability: Calculate the prior probability that the sample is drawn from cage A and B, based on the announced probability in the experiment.
Likelihood: Calculate the likelihood that a sample is drawn from each cage, using the number of N draws in the sample and the cage composition.
Posterior Probability: Either calculate the posterior probability or compute the product of likelihood and prior for each cage.
Note: Values are equal if they round to the same number at two decimal places. For example, 0.333 and 0.33 should be treated as the same. If the student doesn't attempt the calculation or leaves it incomplete, answer No. If interrupted, also answer No.

(6) Prior Computed: Do they calculate the prior probability for each cage correctly? Answer Yes or No.

(7) Likelihood Computed: Do they calculate the likelihood correctly? Answer Yes or No. Note, that omitting the binomial coefficient is acceptable, as it is a constant for both cages.

(8) Posterior Computed: Do they apply Bayes' rule and calculate the posterior probability correctly? Alternatively, answer Yes if they correctly compute and compare the product of likelihood and prior probability. Answer Yes or No.

Part IV. Do they make a final decision that is consistent with their reasoning process? Instructions: The student should reach a conclusion based on their reasoning process, and the final answer should align with that conclusion. You should examine the student's reasoning and predict what they should report (e.g., cage A or cage B), then compare it to their actual report. Provide YES if they are consistent.

Here are two examples of inconsistency.

A student calculates the posterior probability of cage A as 2/3. Since 2/3 is greater than 1/2, they should report cage A but instead report cage B.
A student finds the product of likelihood and prior to be 2/3 for cage A and 1/3 for cage B. Since 2/3 is greater than 1/3, they should choose cage A but report cage B.
(9) Inconsistency: Based on the student's reasoning process and decision rule, determine what the final decision should be. Is the decision they actually make consistent with their reasoning? Answer Yes or No.
Note, that we are looking for cases where the student has reduced the answer to decimals or simple fractions, and decides against what they should have decided.
We do want to penalize fraction-to-decimal conversion errors if they clearly lead the student to the wrong final choice.
If the student never attempts a numeric calculation or fails to provide a final decision at all, also answer No.
\end{lstlisting}
\textbf{Output Instructions}
\begin{lstlisting}
Provide detailed reasoning first and then at the end output a JSON. Your answers must use: Yes = 1, No = 0.

    {{
      "1_correct_cage_composition": 1/0,
      "2_correct_draw_count_and_with_replacement": 1/0,
      "3_correct_use_of_observed_data": 1/0,
      "4_conceptual_ignore_prior_likelihood_only": 1/0,
      "5_conceptual_ignore_likelihood_prior_only": 1/0,
      "6_compute_prior_exact": 1/0,
      "7_compute_likelihood_exact": 1/0,
      "8_compute_posterior_exact": 1/0,
      "9_decision_consistency": 1/0
    }}
\end{lstlisting}
\label{appendix_grading_prompts}
\newpage

\setcounter{figure}{0}
\section{Examples of Textual Responses under Error Flags}

In this section, we excerpt from representative examples of \textit{students}' original textual responses, classified by the GPT grader under each category, along with the \textit{grader}'s responses explaining the reasoning behind its classifications. We don't present examples for Error Flags 2, 3 and 5 because they are rarely, if any, detected.\\
\\
\textbf{1. Cage Composition}
\begin{tcolorbox}[breakable, title={Student}]
Based on the given information, here is my reasoning broken into separate steps:

Step 1. Since there are more N balls in Cage A than in Cage B, and the outcome after 7 draws is 4 N balls and 3 G balls, it's more likely that the draws came from Cage A because there are more N balls available in that cage.

Step 2. The probability of getting a N ball from Cage A is 4/10 = 0.4, and the probability of getting a G ball from Cage A is $6/10 = 0.6$.

Step 3. The probability of getting a N ball from Cage B is 6/10 = 0.6, and the probability of getting a G ball from Cage B is $4/10 = 0.4$.

Step 4. Considering the outcomes of 4 N balls and 3 G balls after 7 draws, it is more probable that the draws came from Cage A because the probability of drawing a N ball from Cage A (0.4) is higher than drawing a N ball from Cage B (0.6).

Step 5. Therefore, my final answer is:
Final answer: Cage A.
\end{tcolorbox}

\begin{tcolorbox}[breakable, title={Grader}]
The student states there are more N balls in Cage A than in Cage B. This is incorrect because Cage A actually has 4 N balls and Cage B has 6 N balls. Thus, they misread the cage composition.
\end{tcolorbox}
\raggedright

\textbf{4. Ignoring Prior}
\begin{tcolorbox}[breakable, title={Student}]

Step 1: Calculate the probability of getting 3 "N" balls and 3 "G" balls from each cage.

- For Cage A:
The probability of drawing an "N" ball from Cage A is 4/6 = 2/3.
The probability of drawing a "G" ball from Cage A is 2/6 = 1/3.
The probability of getting 3 "N" balls and 3 "G" balls from Cage A is $(2/3)^3 * (1/3)^3 = 8/729$.

- For Cage B:
The probability of drawing an "N" ball from Cage B is 3/6 = 1/2.
The probability of drawing a "G" ball from Cage B is also 1/2.
The probability of getting 3 "N" balls and 3 "G" balls from Cage B is $(1/2)^3 * (1/2)^3 = 1/64$.

Step 2: Compare the probabilities of the outcomes from each cage.

- The probability of getting 3 "N" balls and 3 "G" balls is higher for Cage A (8/729) compared to Cage B (1/64).

Step 3: Make a decision based on the higher probability outcome.

Final answer: Cage A.

\end{tcolorbox}

\begin{tcolorbox}[breakable, title={Grader}]
The student never mentioned or used the die roll prior probabilities (3/10 for Cage A and 7/10 for Cage B) and made the decision solely based on the likelihood calculation. Thus they did ignore prior information.
\end{tcolorbox}
\raggedright

\textbf{6. Prior Computed}
\begin{tcolorbox}[breakable, title={Student}]
Step 1: Calculate the probability of drawing 6 "N" balls from each cage: For Cage A: $(4/6)^6 = 0.012$, For Cage B: $(3/6)^6 = 0.015$

Step 2: Compare the probabilities:
- The probability of drawing 6 "N" balls from Cage B is slightly higher than from Cage A.

Step 3: Consider the initial composition of the cages:
- Cage A has more "N" balls compared to Cage B, which could make it more likely to draw 6 "N" balls from that cage.

Final answer: Cage A.
\end{tcolorbox}

\begin{tcolorbox}[breakable, title={Grader}]
The student does not calculate or even consider the prior probabilities (the announced cage probabilities).
\end{tcolorbox}
\raggedright

\textbf{7. Likelihood Computed}
\begin{tcolorbox}[breakable, title={Student}]
Step 1: Calculate the probability of getting 2 "N" balls and 5 "G" balls from each cage.

For Cage A:
P(2 "N" balls and 5 "G" balls from Cage A) = P(2 "N" balls from Cage A) * P(5 "G" balls from Cage A) = $(4/10)^2 * (6/10)^5 \approx 0.0288$

For Cage B:
P(2 "N" balls and 5 "G" balls from Cage B) = P(2 "N" balls from Cage B) * P(5 "G" balls from Cage B) = $(6/10)^2 * (4/10)^5 \approx 0.013824$

Step 2: Compare the probabilities calculated above.

Since the probability of getting 2 "N" balls and 5 "G" balls from Cage A is higher than from Cage B, it is more likely that the observations were generated from Cage A.

Final answer: Cage A.
\end{tcolorbox}

\begin{tcolorbox}[breakable, title={Grader}]
Step 3: Observed outcome and binomial likelihood.
Observed: 2 'N', 5 'G' (total 7).
L(A) = $20412/78125 \approx 0.261274$
L(B) = $6048/78125 \approx 0.077414$
Likelihood ratio L(A)/L(B) = $27/8 = 3.375000$\\

The student attempts to compute the likelihoods, but they did not include the proper terms (or at least did not compute using the correct values as the proper binomial coefficient is omitted, and the numerical values do not match the correct likelihood calculation). Moreover, their computed numbers do not lead to the correct likelihood ratio.
\end{tcolorbox}
\raggedright
\textbf{8. Posterior Computed}
\begin{tcolorbox}[breakable, title={Grader}]
To determine which cage is more likely to have generated the observed outcome, we can use Bayes' theorem. The steps for reasoning through this problem are as follows:

Step 1: Determine Prior Probabilities\\
- **Probability of using Cage A**: Since the die shows 1 through 4 for Cage A, \( P(A) = \frac{4}{10} = 0.4 \).\\
- **Probability of using Cage B**: Since the die shows 5 through 10 for Cage B, \( P(B) = \frac{6}{10} = 0.6 \).\\

Step 2: Determine Likelihoods
- **Likelihood of drawing 5 "N" and 1 "G" from Cage A**:
  - Probability of drawing "N" from Cage A: \( \frac{4}{6} = \frac{2}{3} \).\\
  - Probability of drawing "G" from Cage A: \( \frac{2}{6} = \frac{1}{3} \).\\
  - Probability of the observed outcome from Cage A: 
    \[
    \left(\frac{2}{3}\right)^5 \times \left(\frac{1}{3}\right)^1 = \frac{32}{243}.
    \]

- **Likelihood of drawing 5 "N" and 1 "G" from Cage B**:
  - Probability of drawing "N" from Cage B: \( \frac{3}{6} = \frac{1}{2} \).\\
  - Probability of drawing "G" from Cage B: \( \frac{3}{6} = \frac{1}{2} \).\\
  - Probability of the observed outcome from Cage B:
    \[
    \left(\frac{1}{2}\right)^5 \times \left(\frac{1}{2}\right)^1 = \frac{1}{64}.
    \]

Step 3: Apply Bayes' Theorem
- We want to compare the posterior probabilities \( P(A \mid \text{observed outcome}) \) and \( P(B \mid \text{observed outcome}) \).\\

Using Bayes' theorem:
\[
P(A \mid \text{observed outcome}) \propto P(\text{observed outcome} \mid A) \times P(A) = \frac{32}{243} \times 0.4
\]
\[
P(B \mid \text{observed outcome}) \propto P(\text{observed outcome} \mid B) \times P(B) = \frac{1}{64} \times 0.6
\]

Calculate these products:
- For Cage A: \( \frac{32}{243} \times 0.4 = \frac{12.8}{243} \).
- For Cage B: \( \frac{1}{64} \times 0.6 = \frac{0.6}{64} \).\\

Step 4: Compare Posterior Probabilities
To compare the likelihoods, it is easier to compare the ratios:
- The ratio of the likelihood for Cage A to Cage B is:
  \[
  \frac{\frac{12.8}{243}}{\frac{0.6}{64}} = \frac{12.8 \times 64}{243 \times 0.6} = \frac{819.2}{145.8} \approx 5.62
  \]

Since the ratio is greater than 1, Cage A is more likely to have produced the observed outcome.\\

Conclusion
Given that the posterior probability is higher for Cage A, it is more likely that Cage A generated the observations.

**Final answer: Cage A.**
\end{tcolorbox}

\begin{tcolorbox}[breakable, title={Grader}]
The student applies Bayes rule by multiplying the (miscomputed) likelihood by the correct prior. Although they combine the two pieces of information, the incorrect likelihood for Cage A (and thus the resulting ratio of products) leads to a numerical error in the posterior computation. Therefore, the computation is not exact.  
\end{tcolorbox}
\raggedright
\textbf{9. Decision Inconsistency}
\begin{tcolorbox}[breakable, title={Student}]
To determine which cage is more likely to have generated the result of 3 "N" balls and 3 "G" balls, let's analyze the problem step by step.\\
Step 1: Determine the Probability of Choosing Each Cage
- The probability of choosing Cage A is determined by the roll of the die showing 1 through 7. So, \(P(A) = \frac{7}{10} = 0.7\).
- The probability of choosing Cage B is determined by the roll of the die showing 8 through 10. So, \(P(B) = \frac{3}{10} = 0.3\).\\
Step 2: Calculate the Probability of the Observed Result for Each Cage

**Cage A:**
- Probability of drawing "N" from Cage A = \(\frac{4}{6} = \frac{2}{3}\).
- Probability of drawing "G" from Cage A = \(\frac{2}{6} = \frac{1}{3}\).
- To get exactly 3 "N" and 3 "G" in 6 draws, we use the binomial probability formula:
  \[
  P(3N, 3G | A) = \binom{6}{3} \left(\frac{2}{3}\right)^3 \left(\frac{1}{3}\right)^3
  \]
  \[
  = 20 \times \left(\frac{8}{27}\right) \times \left(\frac{1}{27}\right)
  \]
  \[
  = 20 \times \frac{8}{729}
  \]
  \[
  = \frac{160}{729}
  \]

**Cage B:**
- Probability of drawing "N" from Cage B = \(\frac{3}{6} = \frac{1}{2}\).
- Probability of drawing "G" from Cage B = \(\frac{3}{6} = \frac{1}{2}\).
- To get exactly 3 "N" and 3 "G" in 6 draws:
  \[
  P(3N, 3G | B) = \binom{6}{3} \left(\frac{1}{2}\right)^3 \left(\frac{1}{2}\right)^3
  \]
  \[
  = 20 \times \left(\frac{1}{8}\right) \times \left(\frac{1}{8}\right)
  \]
  \[
  = 20 \times \frac{1}{64}
  \]
  \[
  = \frac{20}{64} = \frac{5}{16}
  \]

Step 3: Use Bayes' Theorem to Update Beliefs

We need to calculate \(P(A | \text{3N, 3G})\) and \(P(B | \text{3N, 3G})\). According to Bayes' Theorem:
\[
P(A | \text{3N, 3G}) = \frac{P(\text{3N, 3G} | A) \times P(A)}{P(\text{3N, 3G})}
\]
\[
P(B | \text{3N, 3G}) = \frac{P(\text{3N, 3G} | B) \times P(B)}{P(\text{3N, 3G})}
\]

The denominator \(P(\text{3N, 3G})\) is the same for both, so we can compare the numerators directly:

\[
P(A | \text{3N, 3G}) \propto \frac{160}{729} \times 0.7
\]
\[
P(B | \text{3N, 3G}) \propto \frac{5}{16} \times 0.3
\]

Calculating these:
- \(P(A | \text{3N, 3G}) \propto \frac{160}{729} \times 0.7 = \frac{112}{729}\)
- \(P(B | \text{3N, 3G}) \propto \frac{5}{16} \times 0.3 = \frac{1.5}{16} = \frac{1.5 \times 729}{16 \times 729} = \frac{1.5 \times 729}{11664}\)

Comparing these values, \(P(B | \text{3N, 3G})\) is greater.\\
Conclusion

The observed result is more likely with Cage B than with Cage A, given the probabilities calculated. Therefore, the draws are more likely to have come from Cage B.

Final answer: Cage B.
\end{tcolorbox}

\begin{tcolorbox}[breakable, title={Grader}]
Step 1: Prior from die roll.\\
  p(A) = 7/10 = 0.7000;  p(B) = 3/10 = 0.3000\\
  Prior ratio p(A)/p(B) = 7/3 $\approx$ 2.333333\\
Step 2: Cage compositions.\\
  Cage A: p(N|A) = 2/3$\approx $ 0.6667, p(G|A) = 1/3 $ \approx$ 0.3333\\
  Cage B: p(N|B) = $1/2 = 0.5000$, p(G|B) = $1/2 = 0.5000$\\
Step 3: Observed outcome and binomial likelihood.
  Observed: 3 'N', 3 'G' (total 6).\\
  L(A) = $160/729 \approx 0.219479$, L(B) = $5/16 = 0.312500$\\
  Likelihood ratio L(A)/L(B) = $512/729 \approx 0.702332$\\
Step 4: Posterior components and probabilities (fraction and decimal).\\
  p(A)*L(A) = $112/729 \approx 0.153635$, p(B)*L(B) = $3/32 \approx 0.093750$\\
  Post(A)   = $3584/5771 \approx 0.621036$, 
  Post(B)   = $2187/5771 \approx 0.378964$\\
  Posterior ratio Post(A)/Post(B) = $3584/2187 \approx 1.638775$\\
Step 5: Decision.\\
  Final answer: Cage A.\\
  
   Because the correctly computed product for Cage A ($\approx 0.1537$) is higher than that for Cage B ($\approx 0.09375$), the decision should have been Cage A. The student's final answer Cage B is inconsistent with the calculations.
\end{tcolorbox}
\raggedright

\label{appendix_error_flag_example}
\newpage

\setcounter{figure}{0}
\section{Additional Figures}

\label{appendix:D}

\begin{figure}[H]
 \centering
 \caption{Estimated subjective posterior beliefs of GPT Subjects: 6-ball Experiments}
 \label{fig:beliefs_gpt_wisconsin}
 \subfloat[GPT 3.5]{
     \includegraphics[width=2in, trim= 65 1 65 .5, clip=true]{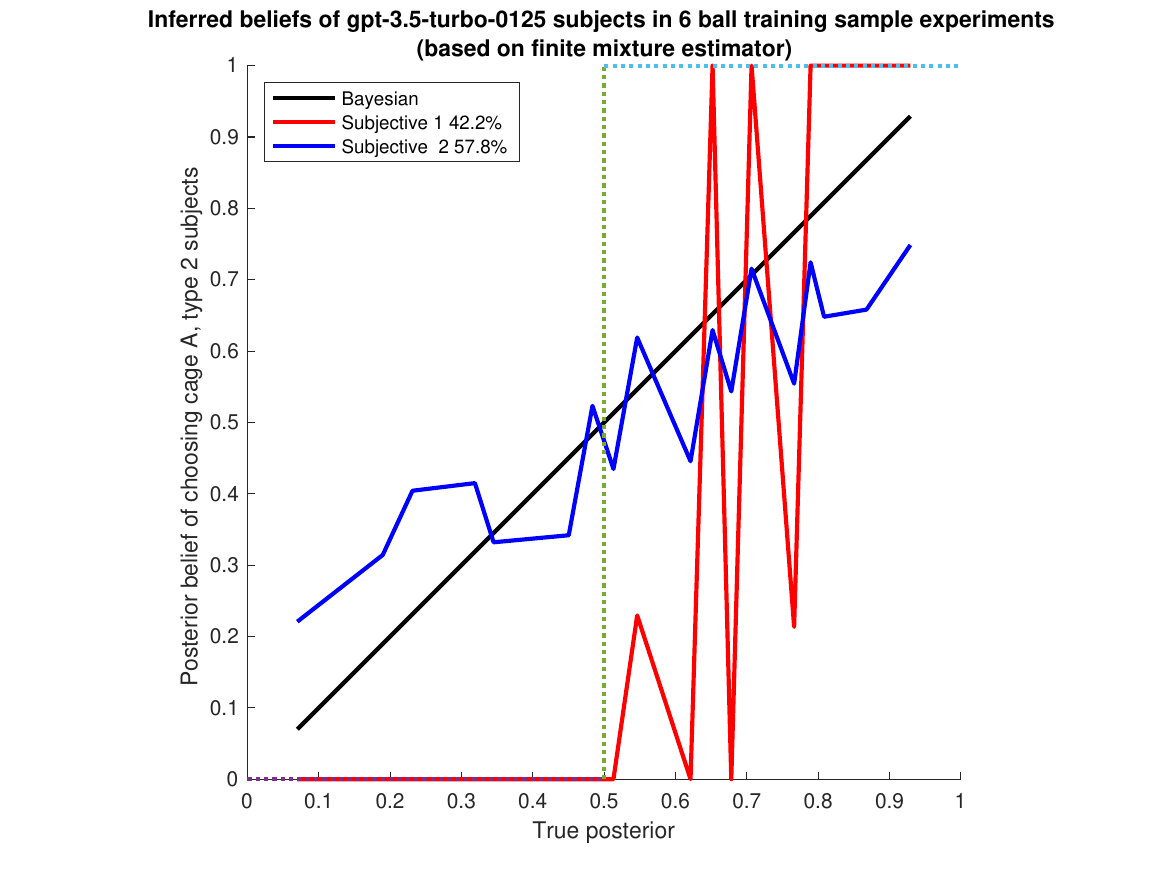}
      \label{fig:beliefs_3dot5}}
 \subfloat[GPT 4]{
      \includegraphics[width=2in, trim= 65 1 65 .5, clip=true]{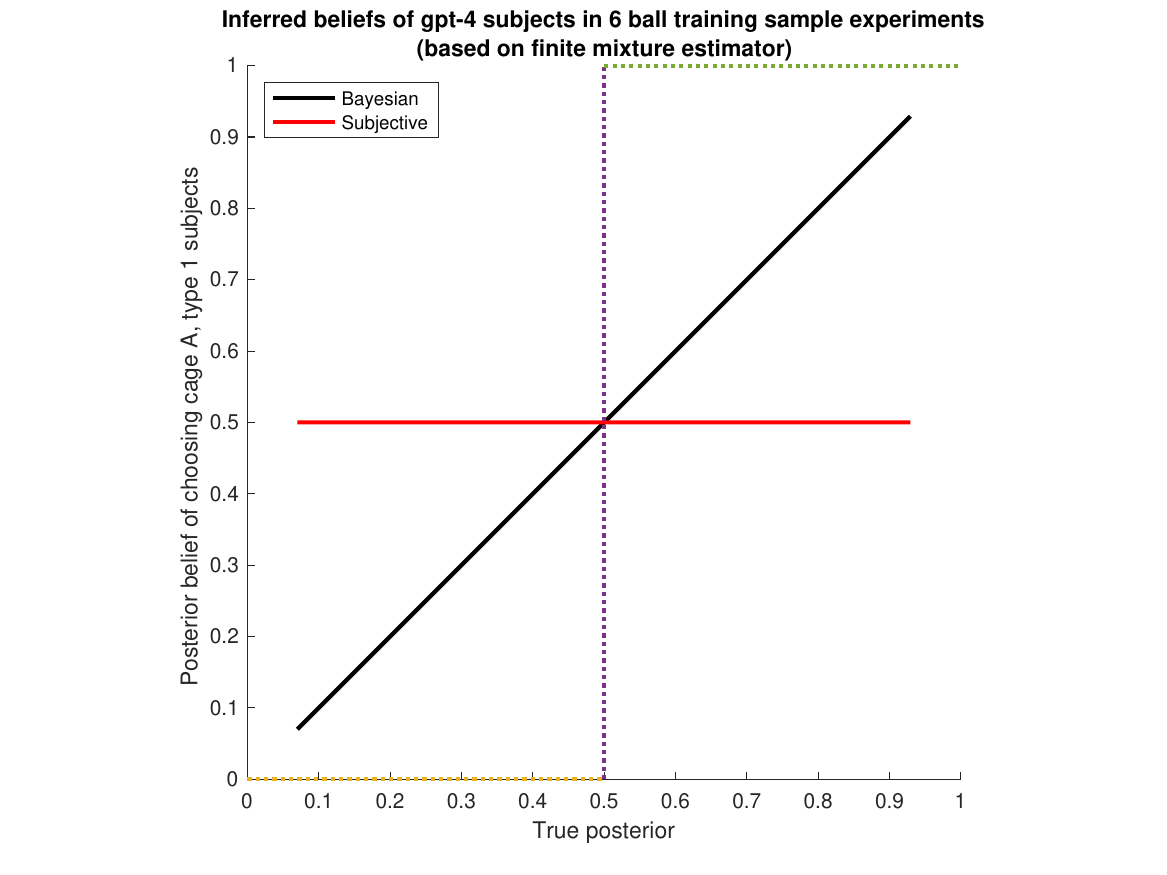}
      \label{fig:beliefs_4}}
 \subfloat[GPT 4o]{
     \includegraphics[width=2in, trim=65 1 65 .5, clip=true]{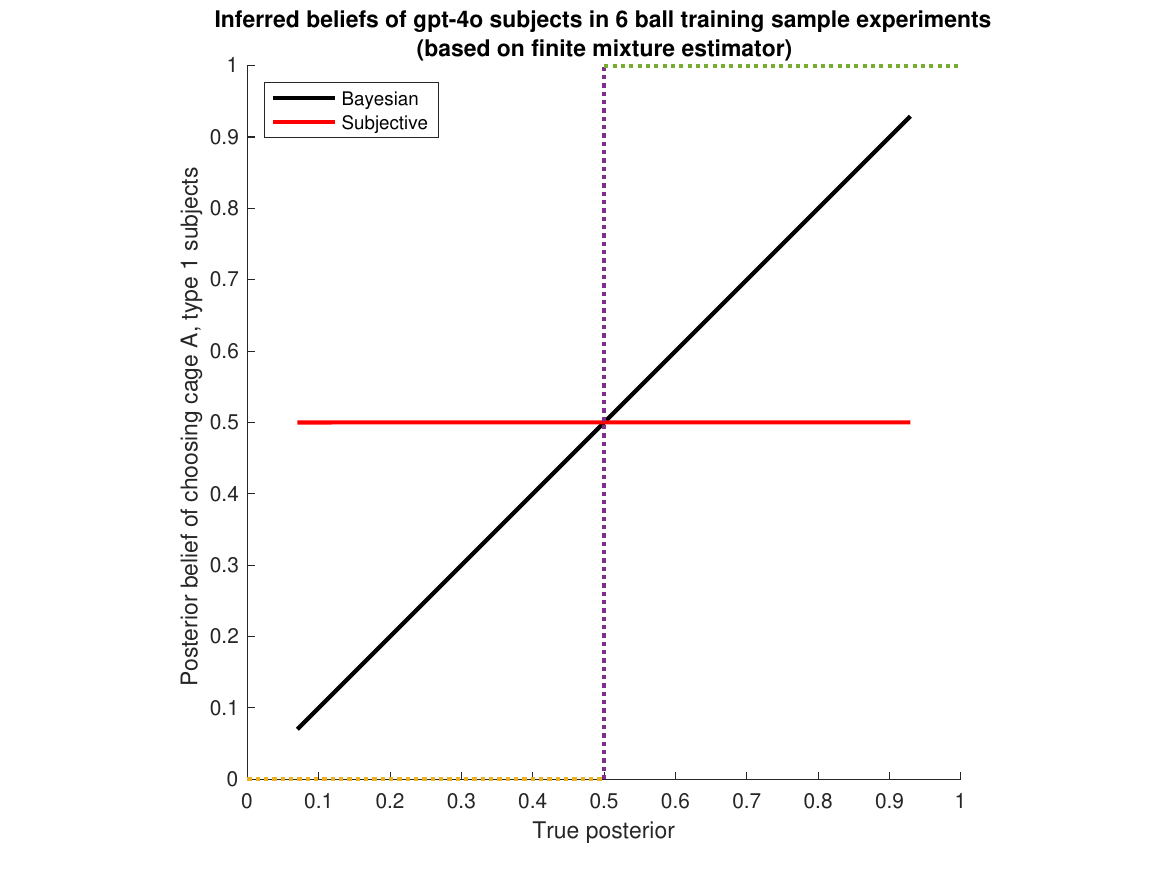}
      \label{fig:beliefs_4o}}
\end{figure}

\begin{figure}[H]
\caption{Predicted vs Actual CCPs in Training and Evaluation Samples}
\centerline{\includegraphics[width=3in, trim=50 170 50 170, clip=true]{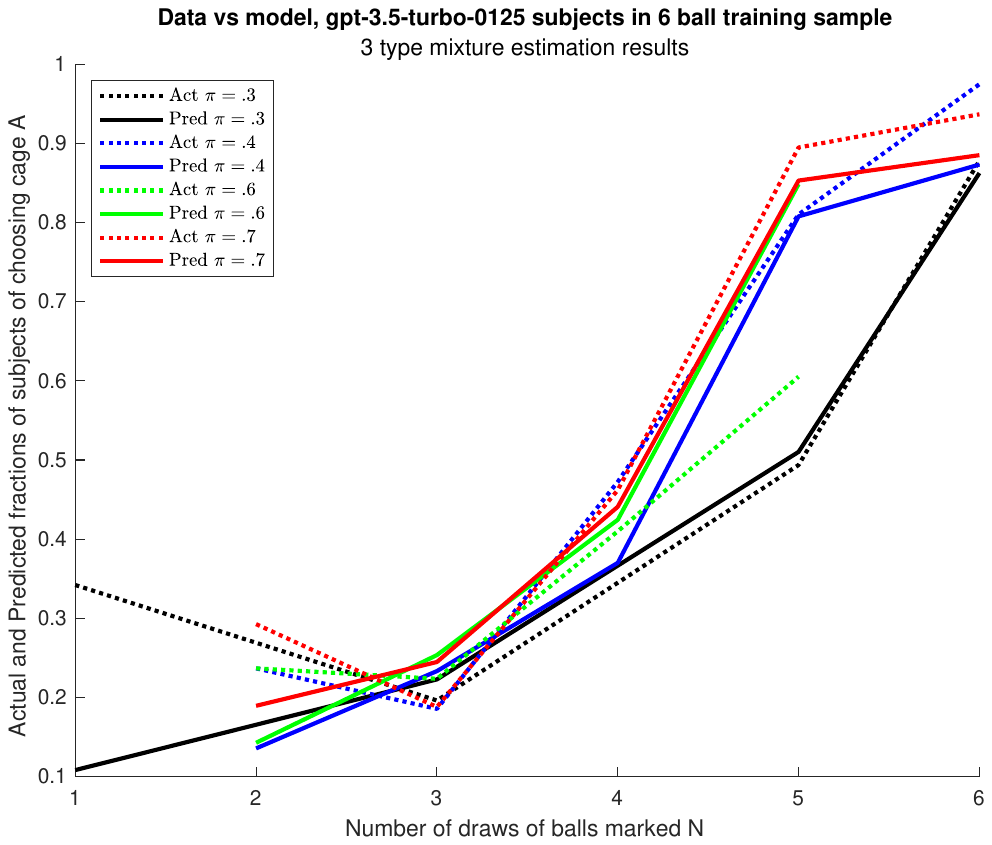}
\includegraphics[width=3in, trim=50 170 50 170, clip=true]{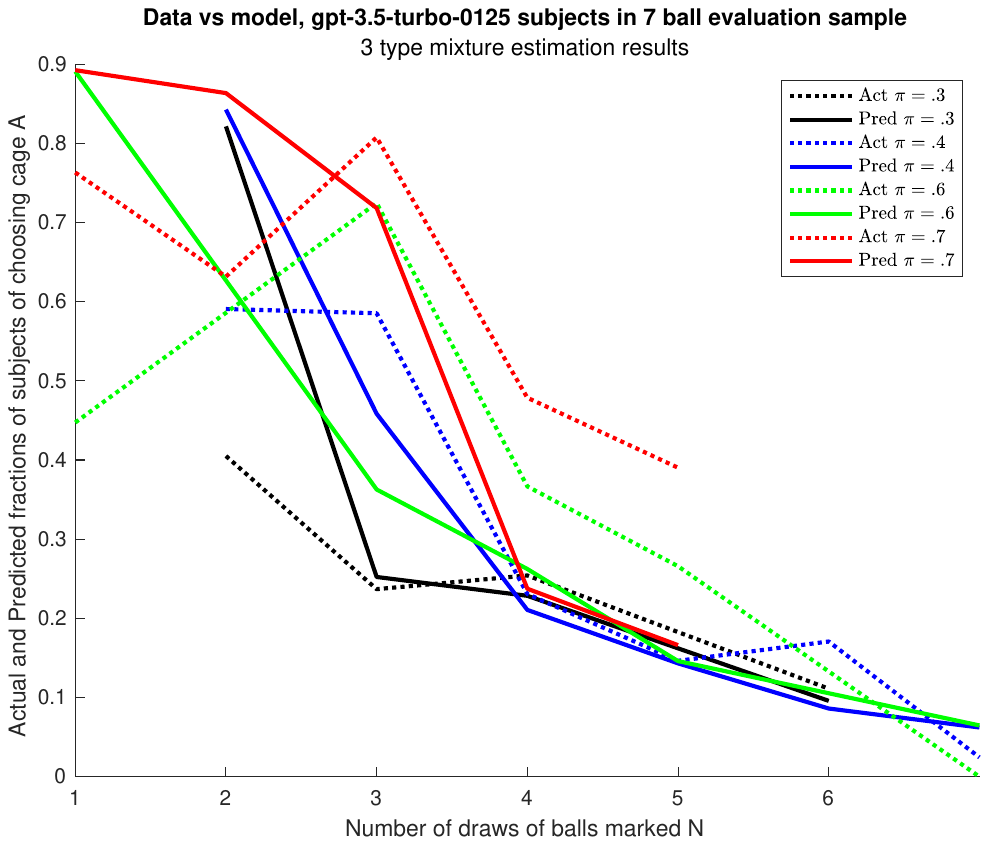}}
\label{fig:wisconsin_gpt3dot5_ccps}
\end{figure}


\label{appendix_figures}

\newpage
\setcounter{figure}{0}
\section{Validation of the Automated GPT Grader}
\label{appendix:grader_validation}

The analysis presented in Section 6 utilizes an automated approach, employing a ``teacher" GPT grader to evaluate the textual reasoning provided by the ``student'' GPT models (GPT-3.5, GPT-4, and GPT-4o). To ascertain the reliability of this method, a validation exercise was conducted. The validation methodology involved a randomly selected subset of 150 textual responses (50 for each student model: GPT-3.5, GPT-4, GPT-4o) from the 6-ball Wisconsin experiment dataset. These responses were independently evaluated by three graders: a human co-author expert in the task, the GPT-o1 model, and the primary GPT-o3-mini grader. All graders utilized the finalized grading prompt detailed in Appendix D. Each response was assessed against the 9 binary grading flags described in Section 6.1, which were subsequently grouped into four conceptual panels: Panel A (Data Read-in), Panel B (Bayes Rule Application), Panel C (Posterior Calculation), and Panel D (Final Decision Consistency). Aggregate panel error rates were calculated, where a panel error was recorded if any underlying flag indicating an error within that panel was triggered for a given response. The error rates reported represent the percentage of the 50 responses for each student model flagged with an error in that panel.

Table \ref{table:grader_validation} presents the comparative panel error rates generated by the three graders on the 50-sample validation subset. The results demonstrate a high degree of concordance between the primary automated grader (GPT-o3-mini), the advanced GPT-o1 grader, and the human expert, particularly for Panels A and B, where error rates are nearly identical. This confirms the automated graders' proficiency in identifying fundamental comprehension and conceptual errors related to Bayes' rule application.

\begin{table}[htbp]
  \centering
  \caption{Comparison of Panel Error Rates (\%) Across Graders (N=50 per student model)}
    \begin{tabular}{llcccc}
    \toprule
    Student Model & Panel   & Error Category                     & GPT-o3-mini & GPT-o1 & Human  \\
          &         &                                    & Grader      & Grader & Grader \\
    \midrule
    \textbf{GPT-3.5}   & Panel A & Data read-in errors                &    0        &   2    &  [2]   \\
                       & Panel B & Bayes Rule application errors      &   76        &  76    & [76]   \\
                       & Panel C & Posterior calculation errors       &   96        &  100   & [98]   \\
                       & Panel D & Final decision inconsistency errors &    6        &  10    & [10]   \\
    \midrule
    \textbf{GPT-4}     & Panel A & Data read-in errors                &    0        &   0    &  [0]   \\
                       & Panel B & Bayes Rule application errors      &    4        &   4    &  [4]   \\
                       & Panel C & Posterior calculation errors       &   64        &  74    & [82]   \\
                       & Panel D & Final decision inconsistency errors &    0        &   0    &  [2]   \\
    \midrule
    \textbf{GPT-4o}    & Panel A & Data read-in errors                &    0        &   0    &  [0]   \\
                       & Panel B & Bayes Rule application errors      &    0        &   0    &  [0]   \\
                       & Panel C & Posterior calculation errors       &   20        &  18    & [22]   \\
                       & Panel D & Final decision inconsistency errors &    8        &  12    & [14]   \\
    \bottomrule
    \end{tabular}%
  \label{table:grader_validation}%
    \caption*{Note: Values represent the percentage of the 50 responses flagged with at least one error within the specified panel by the respective grader.}
\end{table}%

Minor discrepancies arise in Panels C and D, which involve assessing complex numerical calculations and logical consistency. For Panel C (Posterior Calculation), particularly with GPT-4 responses, GPT-o1 aligns more closely with the human grader (74\% error rate) than GPT-o3-mini does (64\% error rate, vs. 82\% for human). This suggests GPT-o1 has a superior, albeit still imperfect, ability to evaluate intricate numerical steps. Similarly, for Panel D (Final Decision Inconsistency), GPT-o1 again tracks human judgment more closely, especially for GPT-4o (12\% vs. 14\% for human, compared to 8\% for o3-mini). These differences likely stem from the challenges LLMs face in precisely evaluating complex fraction comparisons and conversions, a task where GPT-o1 demonstrates marginally better performance. 

Despite these minor variations in evaluating complex numerical reasoning, the overall agreement across graders is substantial. Importantly, the core qualitative findings reported in Section 6 are robustly identified by all graders. This includes the transition from conceptual errors (Panel B) dominating in GPT-3.5 to calculation errors (Panel C) being more prevalent in GPT-4, followed by significant improvement in calculation accuracy (Panel C) for GPT-4o. This validation exercise confirms that GPT-o3-mini serves as a reliable primary grader for the large-scale textual analysis.
\label{appendix_grader_validation}

\end{document}